\documentclass[12pt]{article}
\pdfoutput=1
\usepackage{putex}
\usepackage{comment}
\usepackage{graphicx}
\usepackage{caption}
\usepackage{amsmath}
\usepackage{array}
\usepackage{subcaption}
\usepackage{epstopdf}
\usepackage{enumerate}
\usepackage{cite}
\usepackage{youngtab}
\usepackage{tensor}
\usepackage{slashed}
\usepackage[export]{adjustbox}
\usepackage[aligntableaux=center]{ytableau}
\usepackage[utf8]{inputenc}
\usepackage[
      colorlinks=true,
      linkcolor=blue,
      urlcolor=blue,
      filecolor=black,
      citecolor=red,
      ]{hyperref}

\usepackage{braket}

\usepackage{tikz}
\usepackage{subcaption}

\newcommand{\RNum}[1]{\uppercase\expandafter{\romannumeral #1\relax}}

\newcommand {\be} {\begin {equation}}
\newcommand {\ee} {\end {equation}}

\newcommand {\bes} {\begin {equation*}}
\newcommand {\ees} {\end {equation*}}

\newcommand{\R}{\mathbb{R}}

\newcommand{\ellK}{\mathbb{K}}
\newcommand{\ellE}{\mathbb{E}}

\newcommand{\beq}{\begin{equation}}
\newcommand{\eeq}{\end{equation}}

\def\eqref#1{(\ref{#1})}

\def\ie{\begin{equation}\begin{aligned}}
\def\fe{\end{aligned}\end{equation}}

\numberwithin{equation}{section}

\def\<{\langle}
\def\>{\rangle}

\def\comma{\,,}
\def\period{\,.}
\def\fn#1{\footnote{#1}}
\def\del{\partial}

\def\pmatrix#1#2{\left(\begin{array}{#1}#2\end{array}\right)}
\renewcommand{\eqref}[1]{Eq.~(\ref{#1})}

\def\dbraket#1{\langle\langle #1\rangle\rangle}

\begin{document}

\preprint{PUPT-2630 \\CERN-TH-2021-161}
\institution{PU}{Joseph Henry Laboratories, Princeton University, Princeton, NJ 08544, USA}
\institution{CERN}{Department of Theoretical Physics, CERN, 1211 Meyrin, Switzerland
}

\title{{\LARGE Large Charges on the Wilson Loop in $\mathcal{N}=4$ SYM:\\ Matrix Model and Classical String}}

\authors{Simone Giombi\worksat{\PU}, Shota Komatsu\worksat{\CERN}, Bendeguz Offertaler\worksat{\PU}
}
\abstract{We study the large charge sector of the defect CFT defined by the half-BPS Wilson loop in planar $\mathcal{N}=4$ supersymmetric Yang-Mills theory. Specifically, we consider correlation functions of two large charge insertions and several light insertions in the double-scaling limit where the 't Hooft coupling $\lambda$ and the large charge $J$ are sent to infinity, with the ratio $J/\sqrt{\lambda}$ held fixed. They are holographically dual to the expectation values of light vertex operators on a classical string solution with large angular momentum, which we evaluate in the leading large $J$ limit. We also compute the two-point function of large charge insertions by evaluating the on-shell string action, supplemented by the boundary terms that generalize the one introduced by Drukker, Gross and Ooguri for the Wilson loop without insertions.
For a special class of correlation functions, we reproduce the string results from field theory by using supersymmetric localization. The results are given by correlation functions in an ``emergent'' matrix model whose matrix size is proportional to $J$ and whose spectral curve coincides with that of the classical string. Similar matrix models appeared in the study of extremal correlators in rank-1 $\mathcal{N}=2$ superconformal field theories, but our results hold also for non-extremal cases.} 

\maketitle

\tableofcontents

\section{Introduction and summary}

The study of operators with large quantum numbers has played a central role 
in the development of the gauge/string duality and AdS/CFT correspondence \cite{Berenstein:2002jq, Gubser:2002tv}. Under the duality, operators with large quantum numbers are dual to 
macroscopic string states, which can be quantized by semiclassical methods. For instance, the scaling dimensions of single-trace operators with large charges in the gauge theory are dual to the energies of semiclassical closed string states in AdS. The investigation of such large charge sectors of AdS/CFT also paved the way to the discovery of integrable structures on the gauge and string sides, leading to rather non-trivial dynamical tests of the correspondence (see for instance \cite{Beisert:2010jr} for a review of this extensive topic). More recently, general properties of the large charge expansion in CFTs with global symmetries have been extensively studied using effective field theory methods, see for instance  \cite{Hellerman:2015nra, Monin:2016jmo, Alvarez-Gaume:2016vff, Jafferis:2017zna} and \cite{Gaume:2020bmp} for a recent review. 

In this work, our focus is on the large charge sector of the defect CFT associated with the half-BPS Wilson loop in the $\mathcal{N}=4$ supersymmetric Yang-Mills (SYM) theory. This is a well-known generalization of the usual Wilson loop operator that couples to one of the scalar fields of the theory, and is supported on a circular (or infinite straight line) contour. It preserves a one-dimensional superconformal group $OSp(4^*|4)$ that includes 16 supercharges (half of the superconformal symmetry of the $\mathcal{N}=4$ SYM theory) and the bosonic subgroup $SL(2,\mathbb{R})\times SO(3)\times SO(5)$. Here $SL(2,\mathbb{R})$ is the 1d conformal symmetry preserved by a circle (or line), $SO(3)$ is the symmetry under spacetime rotations transverse to the defect, and $SO(5)$ is the residual R-symmetry that rotates the five scalars that do not couple to the Wilson loop. The correlation functions of operator insertions on the Wilson loop (together with correlation functions involving ``bulk" single-trace operators) define a rather rich defect CFT that is amenable to studies via holography \cite{drukker2006small, giombi2017half, Giombi:2020amn}, localization \cite{giombi2018exact, Giombi:2018hsx, Giombi:2020amn}, conformal bootstrap \cite{liendo2018bootstrapping, Ferrero:2021bsb, Barrat:2021yvp} and integrability \cite{drukker2006small, Drukker:2012de, correa2012exact, Kiryu:2018phb, Grabner:2020nis, Cavaglia:2021bnz}.\footnote{The defect CFT defined by the half-BPS Wilson loop is also of interest in the study of renormalization group flows on defects. It can be thought as the IR fixed point 
of a RG flow which starts in the UV from the ordinary, non-supersymmetric Wilson loop \cite{Polchinski:2011im, Beccaria:2017rbe, Beccaria:2019dws}. See also \cite{Correa:2019rdk, Cuomo:2021rkm, Beccaria:2021rmj} for related work.} 

A Wilson loop operator in the fundamental representation is dual to an open string minimal surface extending in AdS and ending at the boundary on the contour that defines the loop operator. For the half-BPS Wilson loop, the corresponding minimal surface is an open string worldsheet with AdS$_2$ induced geometry. Defect operator insertions on the Wilson loop are dual to fluctuations of the string about the AdS$_2$ geometry, and their correlation functions at strong coupling can be computed holographically by evaluating Witten diagrams in the string sigma model perturbation theory. For instance, the 4-point functions of the ``elementary'' insertions dual to fluctuations of string coordinates in static gauge were computed in \cite{giombi2017half}. Similarly, one can also compute correlation functions of ``composites'' of those excitations (see for instance \cite{giombi2018exact} for some explicit examples). This perturbative approach via AdS$_2$ Witten diagrams is appropriate as long as the quantum numbers of the inserted operators are small compared to the string tension, or $\sqrt{\lambda}$ (where $\lambda=g^2_{\rm YM}N$ is the gauge theory `t Hooft coupling, and in this paper we focus on the planar limit throughout). When the quantum numbers become of order $\sqrt{\lambda}$, the boundary insertions are of the same order as the string action and one then expects a new open string solution to dominate the path-integral. A special class of operator insertions that are ideal for exploring this regime are the chiral primaries corresponding to products of scalar fields transforming in the
symmetric traceless representations of the $SO(5)$ R-symmetry. These operators have protected scaling dimensions, but their two-point and three-point functions are non-trivial functions of the coupling constant that can nevertheless be computed exactly via localization \cite{giombi2018exact, Giombi:2018hsx}. More generally, localization allows to compute the $n$-point functions in a subsector of ``topological'' chiral primaries (to be reviewed below), which have position independent correlation functions. 

\begin{figure}[t]
\centering
\begin{minipage}{0.49\hsize}
\centering
\includegraphics[clip, height=5cm]{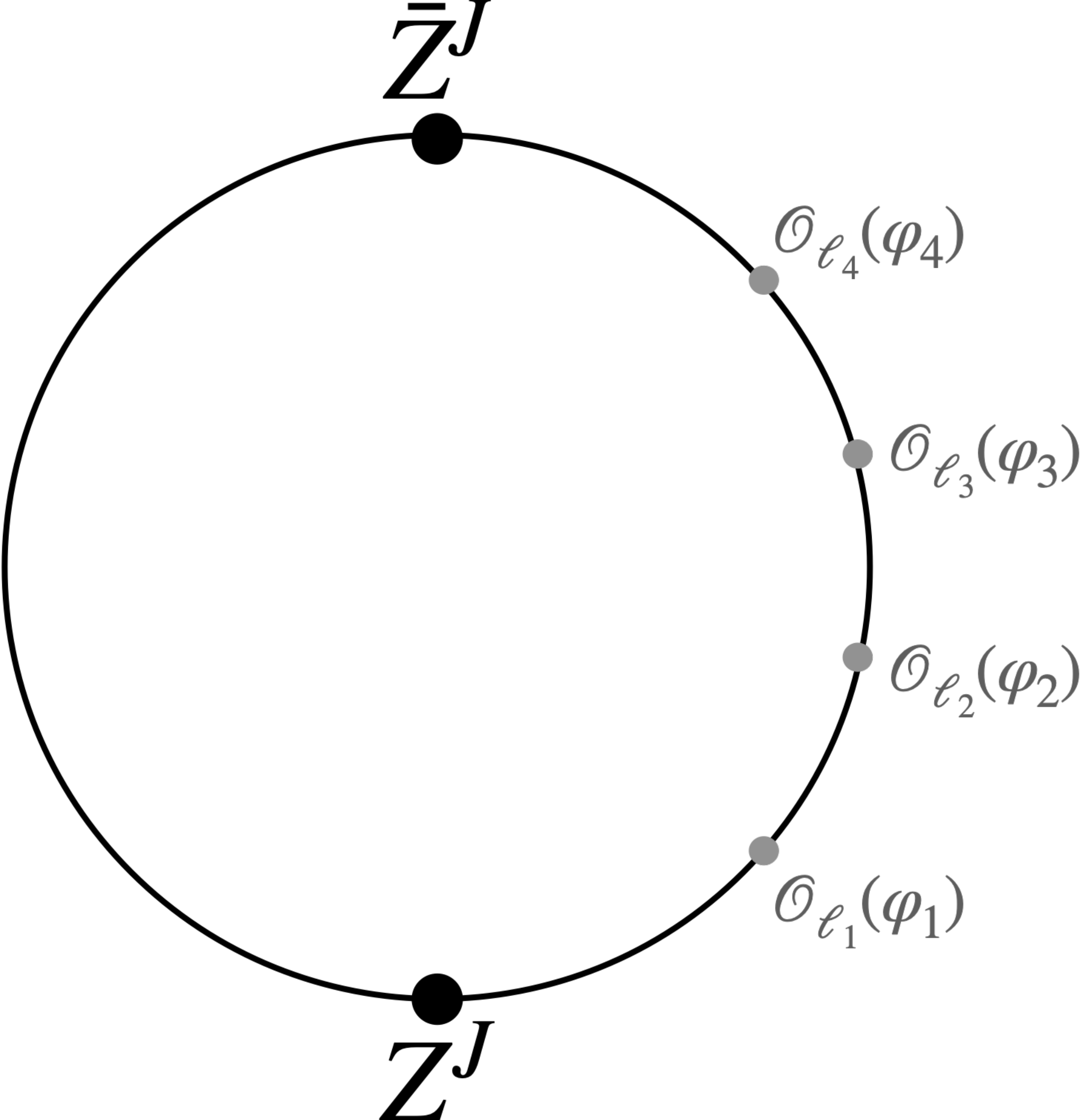}\\
{\bf a.} $\mathcal{N}=4$ SYM
\end{minipage}
\begin{minipage}{0.49\hsize}
\centering
\includegraphics[clip, height=5.2cm]{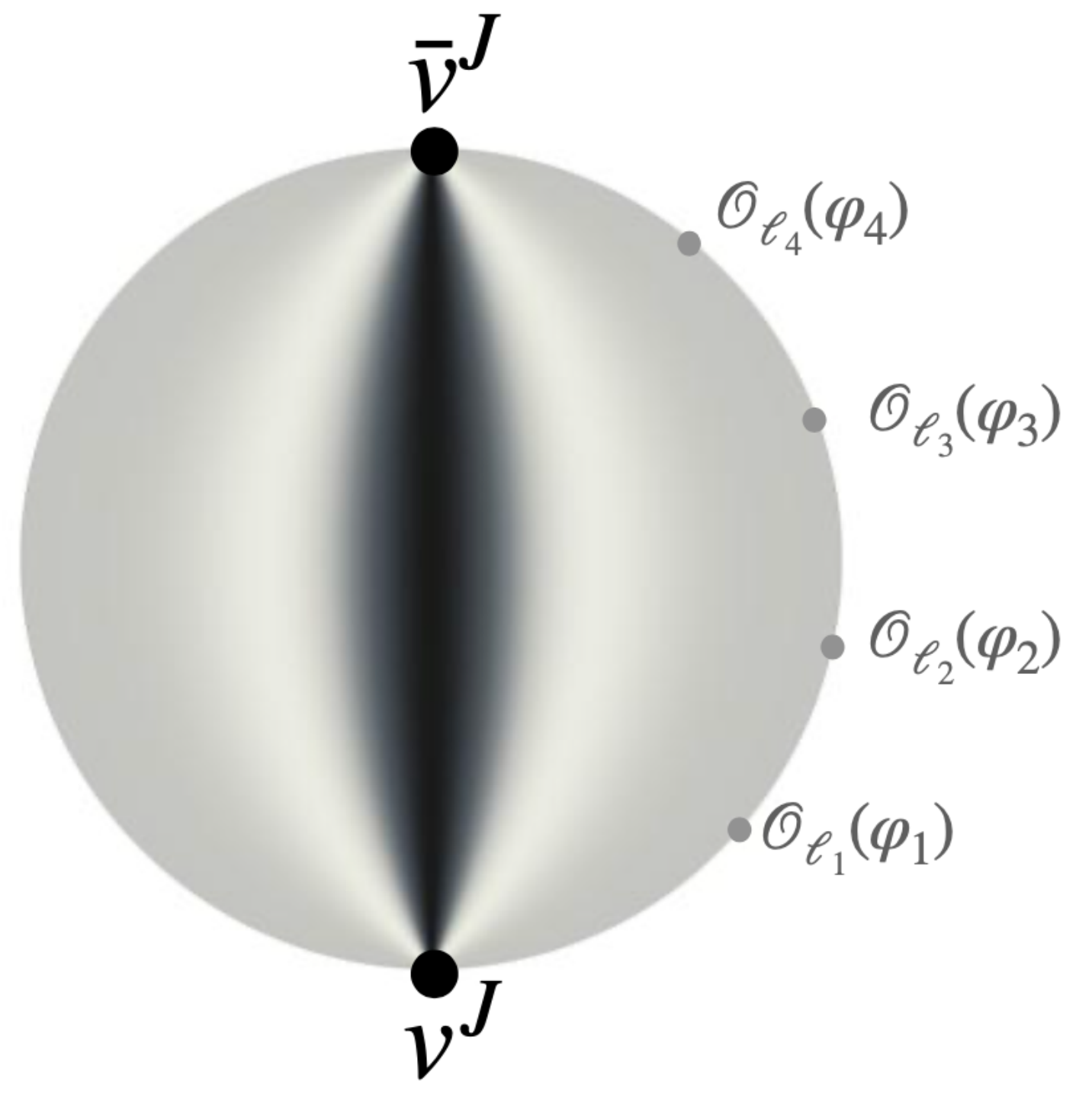}\\
{\bf b.} String
\end{minipage}
\caption{Large charge correlation functions on the gauge theory side and on the string theory side. {\bf a.} Correlation functions of insertions on the half-BPS circular Wilson loop. In this paper, we take two of them ($Z^{J}$ and $\bar{Z}^{J}$) to have large charges $(J\sim \sqrt{\lambda} \gg 1)$ and the others ($\mathcal{O}_{\ell_j}(\varphi_j)$'s) to be light. {\bf b.} At strong coupling, the same correlation functions are described by a non-trivial classical string solution which ends at the Wilson loop on the boundary. In the absence of the insertions, the string worldsheet lies on the Euclidean AdS$_2$ subspace inside AdS$_5\times S^5$. The large charge insertions deform this solution and the string worldsheet has a non-trivial profile around the insertion points. As an illustration of the string solution, the figure above shows a density plot of the worldsheet curvature. The leading large $J$ answer can be computed by evaluating the light vertex operators on this deformed solution.\label{fig1}}
\end{figure}

In the large charge regime, the simplest object to consider in the defect CFT is the insertion of two chiral primaries with charge $J$ such that 
\begin{equation}
J\rightarrow \infty \,,\quad \lambda\rightarrow \infty \,, \quad \frac{J}{\sqrt{\lambda}}~~{\rm fixed}\,.
\label{largeJ-sql}
\end{equation}
On the string theory side, the worldsheet corresponding to this configuration 
was identified first in \cite{drukker2006small} in the limit $\frac{J}{\sqrt{\lambda}}\rightarrow \infty$, and later generalized to finite 
$\frac{J}{\sqrt{\lambda}}$ in \cite{Gromov:2012eu}.\footnote{In \cite{Gromov:2012eu} a general solution 
parametrized by two cusp angles was considered. Here we focus on the case where both angles are zero, which corresponds to insertions on a straight line or circular loop.} We review the solution in detail in Section \ref{sec:correlators from semistring}. Starting from this string worldsheet and inserting additional operators with small quantum numbers, one can then obtain new results on correlation functions of two ``heavy'' (large charge) operators and any number of ``light'' operators, normalized by the heavy-heavy two-point function. See Figure \ref{fig1}. These correlators include in particular the defect CFT OPE data in the heavy-heavy-light configuration. In this paper we focus on the leading order in the large charge limit (\ref{largeJ-sql}), corresponding to the classical analysis on the string side, while subleading corrections corresponding to quantum fluctuations will be discussed in the companion paper \cite{giombi2021}.

A more basic quantity that one can also study is the correlation function of the two heavy operators alone. Since the operators are protected with dimension $\Delta=J$, their two-point function on the Wilson loop has the form
\begin{equation}
\langle O_J O_J\rangle =\frac{n_J(\lambda)}{d^{2J}},
\end{equation}
where $d$ is the distance between the operators, and $n_J(\lambda)$ is a non-trivial function of the coupling constant and charge $J$. This normalization constant has physical meaning because the operators are protected (for instance, for $J=1$ it is related to the Bremsstrahlung function \cite{correa2012exact}). In the large charge limit, \eqref{largeJ-sql}, it is expected to exponentiate as $n_J(\lambda)\sim e^{\sqrt{\lambda}f(J/\sqrt{\lambda})}$, where the non-trivial function in the exponent should correspond to the classical string action. In Section \ref{sec:correlators from semistring} we discuss this holographic two-point calculation in detail (refining and extending earlier attempts in \cite{Miwa2006HolographyOW}). We will propose, in particular, that a suitable modification of the boundary term prescription of \cite{drukker1999wilson} is needed in this case to properly account for the presence of the boundary vertex operators for the large charge insertions. 

A non-trivial test of the string theory predictions can be obtained by comparing to the localization approach, which was developed in \cite{giombi2018exact, Giombi:2018hsx} in the present context of the defect CFT on the Wilson loop. Interestingly, when two chiral primaries with semiclassical charges $J\sim \sqrt{\lambda}$ are inserted, the localization results can be recast in terms of the ``planar'' limit of a matrix model of a $J\times J$ Hermitean matrix.\footnote{This matrix model was obtained first in \cite{Gromov:2012eu} using integrability-based methods, and was rederived in \cite{giombi2018exact} using the localization approach.} Using this matrix model description, in Section \ref{sec:correlators from localization} we obtain the localization prediction for the two-point function of the heavy operators as well as the 
topological $n$-point functions of the two heavy operators and any number of light primaries, working to leading order in the large charge limit. The results are found to be in complete agreement with the string theory predictions we 
obtain in Section \ref{sec:correlators from semistring}. On the string theory side, in addition to the modified boundary term mentioned above, an additional ingredient that is crucial to obtain the correct result for the correlation functions is the averaging over a classical 
modulus of the string solution--- namely, a constant angle $\phi_0$ corresponding to the circle on $S^5$ that is dual to the charge $J$. This averaging is similar to the one 
recently advocated in \cite{yang2021d}, and previously in \cite{Bajnok:2014sza} (see also \cite{Monin:2016jmo} for the analogous effect in the effective field theory description of the large charge expansion of CFTs). 

It is interesting to note that our matrix model shares some features with the matrix model derived in \cite{Grassi:2019txd}, which describes the extremal correlation functions in $\mathcal{N}=2$ SCFTs. In particular, the size of the matrix in both cases is given by the R-charge of the operators. However, there are also notable differences. Firstly, while the matrix model in \cite{Grassi:2019txd} describes a subsector of the rank-1 SCFTs\fn{The generalization of \cite{Grassi:2019txd} to higher-rank theories was discussed in \cite{Beccaria:2020azj}.} e.g.~$SU(2)$ $\mathcal{N}=4$ SYM, our matrix model describes a subsector of $U(N)$ $\mathcal{N}=4$ SYM at large $N$, which is in a completely different parameter regime. Secondly, in the setup discussed in \cite{Grassi:2019txd}, all the physical information is in the two-point function,\fn{This is because all the other extremal correlation functions, including the three-point functions, are automatically determined by the normalization of the two-point function.} and correspondingly, the matrix model only computes such two-point functions. By contrast, in our setup, there are non-extremal higher-point functions which encode non-trivial physical information, and we can study them using our matrix model. Finally and perhaps most importantly, the classical spectral curve of our matrix model, which controls the leading large charge answer, coincides with the spectral curve of the classical string solution describing the holographic dual of the large charge operators on the Wilson loop. This provides a clear physical interpretation of the spectral curve of the large charge matrix model \cite{Gromov:2012eu}. In addition, it suggests that the structure observed in the topological subsector may extend to more general non-SUSY sectors. This optimism comes from the fact that the spectral curve of the classical string is known to control the full non-protected spectrum of semiclassical fluctuations on the worldsheet \cite{Gromov:2008ec,Vicedo:2008jy}.

There are several future directions worth exploring. First, it would be interesting to consider large charge limits of the correlation functions which involve ``bulk operators'', namely operators outside the defect (previous work on some of these observables include \cite{Zarembo:2002ph, giombi2010correlators, giombi2013correlators}). Studying such correlation functions would allow us to analyze the interplay between the large charge limit and the defect crossing equation. Second, the generalization of our results to $1/8$-BPS Wilson loops \cite{drukker2007more, Drukker:2007yx, Drukker:2007qr} should be possible. In this case, the relation to the defect CFT will be lost since general $1/8$-BPS Wilson loops do not preserve the conformal symmetry. Nevertheless, we can analyze the correlation functions on these Wilson loops both from the localization and the (semi-)classical string. Third, the localization analysis can be applied away from the large $N$ limit, in particular to $SU(2)$ $\mathcal{N}=4$ SYM \cite{giombi2018exact}. The large charge limit in this case would be more directly related to large charge limits of rank-1 SCFTs studied in the literature \cite{Hellerman:2017sur,Grassi:2019txd}, and it would be interesting to explore their relation.

The remainder of this paper is organized as follows. In Section \ref{sec:large charges in WL dCFT} we review some basic facts about the defect CFT on the half-BPS Wilson loop and introduce the large charge operators of interest in this paper. In Section \ref{sec:correlators from localization} we review the localization results for the correlation functions on the Wilson loop and the matrix model describing the large charge sector, and then use it to compute the two-point function of the heavy operators and the topological $n$-point functions of two heavy and any number of light operators. In Section \ref{sec:correlators from semistring} we introduce the string solution describing the insertion of two large charges on the Wilson loop, and then discuss the computation of the two-point function of the heavy operators and the correlation functions of the two heavy and any number of light operators. The appendices include some technical details on the ``quasi-momentum'' associated to the matrix model and on the supersymmetries preserved by the string solution. 

\section{Large charges in the Wilson loop defect CFT}\label{sec:large charges in WL dCFT}

This section introduces the defect CFT observables we will analyze in this paper. First, our conventions for $\mathcal{N}=4$ SYM: we work in Euclidean signature and use the standard Cartesian coordinates $x^\mu $, $\mu=1,\ldots,4$, on $\mathbb{R}^4$. We take the gauge group to be $U(N)$, denote the Yang-Mills coupling constant $g_{\rm YM}$, and the 't Hooft coupling $\lambda\equiv g_{\rm YM}^2N$. Below we will also often adopt the  notation $g\equiv \frac{\sqrt{\lambda}}{4\pi}$ commonly used in the integrability literature. 

\paragraph{The Wilson loop defect CFT.}
Our starting point is the half-BPS Wilson loop in $\mathcal{N}=4$ SYM whose contour is a circle and which couples to a single scalar. 
The Wilson loop is
\begin{align}\label{eq:circular WL}
    \mathcal{W}&\equiv\frac{1}{N}\text{Tr P}\text{ exp}\left(\oint \left(iA_\mu(x)\dot x^{\mu}+|\dot x| \Phi_6(x)\right)d\varphi \right)\,,
\end{align}
where $\dot x^{\mu}=\frac{dx^{\mu}}{d\varphi}$ and $\varphi$ is a coordinate parameterizing the loop. Below we will mostly work with the explicit parameterization  $x^\mu(\varphi)=(\cos{\varphi},\sin{\varphi},0,0)$ with $\varphi\in [-\pi,\pi)$ (in which case $|\dot x|=1$). 
The gauge field $A_\mu$ and the scalars $\Phi_I$, $I=1,\ldots,6$, transform in the adjoint of $U(N)$ and the trace is taken in the fundamental. The expectation value of $\mathcal{W}$, as well as its planar limit ($N\to \infty$, $g$ fixed) and supergravity limit ($N\to \infty$, $g\gg 1$), is well known \cite{erickson2000wilson,drukker42exact,pestun2007localization}:
\begin{align}\label{eq:half-BPS WL VEV}
    \braket{\mathcal{W}}_{\mathcal{N}=4\text{ SYM}}=\frac{1}{N}L_{N-1}^1 \left(-\frac{g^2_{\rm YM}}{4}\right)e^{\frac{g^2_{\rm YM}}{8}} &&\overset{N\to \infty}{\rightarrow} &&\frac{1}{2\pi g}I_1(4\pi g) &&\overset{g\gg 1}{\sim} && \frac{e^{4\pi g}}{4\sqrt{2}\pi^2g^{3/2}}.
\end{align}

The Wilson loop preserves an $SL(2,\R)$ subgroup of the $SO(5,1)$ conformal group of $\mathcal{N}=4$ SYM and defines a one dimensional defect CFT in which local adjoint operators $O_i(x)$, $i=1,\ldots,n$ are inserted along the contour \cite{drukker2006small,Cooke:2017qgm,Kim:2017sju,giombi2017half,Beccaria:2017rbe,giombi2018exact,Kiryu:2018phb,Giombi:2018hsx,liendo2018bootstrapping,Beccaria:2019dws,Giombi:2020amn,Grabner:2020nis,Ferrero:2021bsb,Cavaglia:2021bnz}. We will use the ``single bracket'' notation for the unnormalized defect correlators given by 
\begin{align}\label{eq:single bracket}
    \braket{O_1(\varphi_1)\ldots O_n(\varphi_n)}&\equiv \Big\langle\frac{1}{N}\text{Tr P}\left[O_1(\varphi_1)\ldots O_n(\varphi_n)e^{\int (iA_\mu \dot x^\mu +\Phi_6)d\varphi}\right]\Big\rangle_{\mathcal{N}=4\text{ SYM}},
\end{align}
where $O_i(\varphi_i)\equiv O_i(x(\varphi_i))$ and the path ordering in the $\mathcal{N}=4$ SYM correlator acts both on the local operators and on the Wilson loop. It is also convenient to define the ``double bracket'' defect correlators normalized by the Wilson loop expectation value
\begin{align}\label{eq:double bracket}
    \dbraket{O_1(\varphi_1)\ldots O_n(\varphi_n)}&\equiv \frac{\braket{O_1(\varphi_1)\ldots O_n(\varphi_n)}}{\braket{\mathcal{W}}_{\mathcal{N}=4\text{ SYM}}},
\end{align}
which satisfy $\dbraket{1}=1$. 
The defect correlators obey the axioms outlined in Appendix A of \cite{Qiao:2017xif} for correlators in a one dimensional CFT. Since the defect is a circle, the conformal correlators are composed of ratios of the chordal distances, which we will denote
\begin{align}\label{eq:chordal distance}
    d_{ij}\equiv d(\varphi_i,\varphi_j)&\equiv 2\sin\left(\frac{\varphi_i-\varphi_j}{2}\right).
\end{align}

Although our discussion is framed in terms of the circular Wilson loop, the analysis of the half-BPS Wilson line $\mathcal{W}_{\rm line}$ is essentially equivalent.\footnote{One way to explicitly parametrize the half-BPS line is to pick the contour $x^\mu(t)=(t,0,0,0)$ and let
\begin{align}
    \mathcal{W}_{\rm line}&\equiv\frac{1}{N}\text{Tr }\text{P exp}\left(\int_{-\infty}^\infty \left(iA_1(x) +\Phi_6(x)\right)dt\right).
\end{align}} Its expectation value, $\braket{\mathcal{W}_{\rm line}}_{\mathcal{N}=4\text{ SYM}}=1$, differs from \eqref{eq:half-BPS WL VEV} due to a ``conformal anomaly'' \cite{erickson2000wilson,drukker42exact}, but the defect correlators on the circle and line, when normalized by the corresponding Wilson loop operator without insertions, behave in the standard way under conformal transformations. Specifically, under the map $t_i=\tan{\frac{\varphi_i}{2}}$, the correlators are related by
\begin{align}\label{eq:circle to line}
    \frac{\braket{O_1(t_1)\ldots O_n(t_n)\mathcal{W}_{\rm line}}_{\mathcal{N}=4\text{ SYM}}}{\braket{\mathcal{W}_{\rm line}}_{\mathcal{N}=4\text{ SYM}}}&=\left(\frac{d\tan{\frac{\varphi_1}{2}}}{d\varphi_1}\right)^{-\Delta_1}\ldots \left(\frac{d\tan{\frac{\varphi_n}{2}}}{d\varphi_n}\right)^{-\Delta_n}\dbraket{O_1(\varphi_1)\ldots O_n(\varphi_n)}\nonumber\\&=\dbraket{O_1(\varphi_1)\ldots O_n(\varphi_n)}\big\rvert_{d_{ij}\to t_{ij}},
\end{align}
where $t_{ij}\equiv t_i-t_j$. Thus, we may readily switch between correlators on the circle and correlators on the line by exchanging chordal distances with Euclidean distances.

\paragraph{Chiral primaries.}\label{sec:WEUuIIMvLp}
We are interested in the large charge sector of defect correlators of chiral primaries of the form 
\begin{align}\label{eq:chiral primaries}
    \mathcal{O}_J(\varphi,\epsilon)\equiv (\epsilon\cdot \Phi(x(\varphi)))^J,
\end{align}
where $\Phi=(\Phi_1,\ldots,\Phi_5)$ is a vector of the scalar fields not coupled to the Wilson loop in \eqref{eq:circular WL} and $\epsilon=(\epsilon_1,\ldots,\epsilon_5)$ is a complex polarization vector satisfying $\epsilon^2=0$. 
These operators transform in the symmetric traceless representation of $SO(5)\subset SO(6)$ and have protected dimension $\Delta=J$ equal to the R-charge. Conformal symmetry and R-symmetry then fix the two and three-point functions of the chiral primaries, up to the operator normalizations and OPE coefficients:
\begin{align}
    \dbraket{\mathcal{O}_{J_1}(\varphi_1,\epsilon_1)\mathcal{O}_{J_2}(\varphi_2,\epsilon_2)}&=n_{J_1}(g,N)\frac{(\epsilon_1\cdot \epsilon_2)^{J_1}}{d_{21}^{2J}}\delta_{J_1J_2},\label{eq:2-pt function}\\
    \dbraket{\mathcal{O}_{J_1}(\varphi_1,\epsilon_1)\mathcal{O}_{J_2}(\varphi_2,\epsilon_2)\mathcal{O}_{J_3}(\varphi_3,\epsilon_3)}&=c_{J_1J_2J_3}(g,N)\frac{(\epsilon_1\cdot \epsilon_2)^{J_{12|3}}(\epsilon_2\cdot \epsilon_3)^{J_{23|1}}(\epsilon_3\cdot \epsilon_1)^{J_{31|2}}}{d_{21}^{2J_{12|3}}d_{32}^{2J_{23|1}}d_{31}^{2J_{31|2}}}.\label{eq:3-pt function}
\end{align}
Here, $J_{ij|k}\equiv (J_i+J_j-J_k)/2$. The $3$-pt function is zero unless $J_1$, $J_2$ and $J_3$ satisfy the triangle inequality ($J_1+J_2\geq J_3$, plus permutations) and sum to an even integer.

Furthermore, the chiral primaries have a topological sector, which is accessible for finite $N$ and $g$ using localization \cite{giombi2018exact}, in which the correlators are position independent and the OPE is closed. We will discuss the topological sector in greater detail in Section~\ref{sec:correlators from localization}, but we presently define the topological operators to be chiral primaries whose polarization vectors are correlated with their positions in the following way:\fn{This particular definition for the topological operators is equivalent to the one in \cite{giombi2018exact} up to relabelling of scalar flavor indices and flipping the sign of $\Phi_4$.}
\begin{align}\label{eq:topological operator}
    \tilde{\Phi}^J(\varphi)\equiv (\cos{\varphi}\Phi_3(\varphi)-\sin{\varphi}\Phi_4(\varphi)+i\Phi_5(\varphi))^J.
\end{align}
We will usually drop the explicit dependence on $\varphi$. Since $\epsilon(\varphi_i)\cdot \epsilon(\varphi_j)=-\frac{1}{2}d_{ij}^2$, the topological two and three-point correlators are manifestly constant:
\begin{align}
    \dbraket{\tilde{\Phi}^{J_1}\tilde{\Phi}^{J_2}}&=\left(-\frac{1}{2}\right)^{J_1}n_{J_1}\delta_{J_1J_2},\label{eq:topo 2-pt function}\\
    \dbraket{\tilde{\Phi}^{J_1}\tilde{\Phi}^{J_2}\tilde{\Phi}^{J_3}}&=\left(-\frac{1}{2}\right)^{\frac{J_1+J_2+J_3}{2}}c_{J_1J_2J_3}.\label{eq:topo 3-pt function}
\end{align}
In accordance with localization, higher-point correlators are also constant. Given Eqs.~(\ref{eq:2-pt function})-(\ref{eq:3-pt function}), the topological two and three-point functions fully determine the general two and three-point functions, but the same is not true of the higher-point functions because their form is not fixed up to an overall constant by conformal symmetry.

\paragraph{Correlators in the large charge sector.}

In this work, we study the correlators of $2+n$ chiral primaries, two of which have charge $J$ and $n$ of which have charges $\ell_1,\ldots,\ell_{n}$. We analyze the correlators in the \textit{large charge regime}, in which we first take $N\to \infty$ with $g$, $J$, and $\ell_i$ held fixed, and then take $g\to \infty$ with  $\ell_i$ and
\begin{align}\label{eq:J/g}
    \mathcal{J}\equiv \frac{J}{g},
\end{align}
held fixed. Thus, in the planar limit and strongly coupled regime, the two operators with charge $J$ are the same size as the coupling and the operators with charges $\ell_i$ are parametrically smaller. We call the former ``large charges'' or ``heavy operators'' and the latter ``finite charges'' or ``light operators.'' In the present work we focus mostly on the leading behavior, and postpone the analysis of subleading corrections in $1/g$ to \cite{giombi2021}.

In the first half of our analysis, in Section~\ref{sec:correlators from localization}, we start from the planar-exact integral representation of the topological correlators that was derived in \cite{giombi2018exact}, and use it to derive a matrix model that describes the large charge sector. This lets us determine, to leading order, the normalized higher-point correlators
\begin{align}\label{eq:topo higher-point}
    \frac{\dbraket{\tilde{\Phi}^J\tilde{\Phi}^J\prod_{i=1}^n\tilde{\Phi}^{\ell_i}}}{\dbraket{\tilde{\Phi}^J\tilde{\Phi}^J}},
\end{align}
as well as the two-point function
\begin{align}\label{eq:topo two-point}
    \dbraket{\tilde{\Phi}^J\tilde{\Phi}^J}.
\end{align}


In the second half of our analysis, in Section~\ref{sec:correlators from semistring}, we analyze the string dual to the large charge defect correlator
\begin{align}\label{eq:HH correlator}
    \dbraket{Z^J(-\varphi_L)\bar{Z}^J(\varphi_L)}=\frac{2^J n_J}{d^{2J}},
\end{align}
where we have chosen the specific chiral primaries
\begin{align}
    Z\equiv \Phi_4+i\Phi_5, && \text{and} && \bar{Z}\equiv \Phi_4-i\Phi_5,
\end{align}
and where
\begin{align}
    d\equiv d(-\varphi_L,\varphi_L)
\end{align}
is the chordal distance between the large charges. In \eqref{eq:HH correlator}, we have used the rotational symmetry to put the two heavy operators at  $\varphi=\pm\varphi_L$ and assume for simplicity that $\varphi_L\in (0,\pi/2]$, but otherwise let the locations of the insertions be general.\footnote{Because of the conformal symmetry, we could without loss of generality set $\varphi_L=\pi/2$. However, we prefer to keep $\varphi_L$ general in order to explicitly keep track of the conformal form of the correlators.} By evaluating the action and vertex operators of the classical string, we can determine the leading behavior of the two-point function, \eqref{eq:HH correlator}, and of higher correlators with additional insertions of powers of $Z$ and $\bar{Z}$. Finally, we will also be able to use the classical string to reproduce the topological correlators computed by the matrix model. This is because $Z^J$ and $\bar{Z}^J$ become topological in the antipodal configuration $\varphi_L=\pi/2$, since
\begin{align}\label{eq:topological large charges}
    Z^J\left(-\frac{\pi}{2}\right)&=\tilde{\Phi}^J\left(-\frac{\pi}{2}\right), &\bar{Z}^J\left(\frac{\pi}{2}\right)&=(-1)^J\tilde{\Phi}^J\left(\frac{\pi}{2}\right),
\end{align}
and because the light topological operators, $\tilde{\Phi}^\ell$, truncate to linear combinations of powers of $Z$ and $\bar{Z}$ at leading order.


\section{Large charge correlators from localization}\label{sec:correlators from localization}
In this section, we compute correlation functions with large charge insertions in the topological sector. We achieve this by deriving a matrix model that describes the large charge topological sector and applying the matrix model techniques. We will later see in Section \ref{sec:correlators from semistring} that the results obtained from localization are in perfect agreement with the results from the classical string.

\subsection{Matrix model for large charge correlators}\label{sec:matrix model}
\paragraph{Integral representation for the topological sector.} Let us first review the integral representation of the correlation functions of insertions on the half-BPS Wilson loop in $\mathcal{N}=4$ SYM that was derived in \cite{giombi2018exact,Giombi:2018hsx}. 

The correlation functions in the topological sector on the half-BPS Wilson loop at large $N$ are given by the following integral,
\beq\label{eq:integralfirst}
\langle \prod_{k=1}^{n}\tilde{\Phi}^{L_k}\rangle=\oint d\mu \prod_{k=1}^{n}Q_{L_k}(x)\comma
\eeq
where $\tilde{\Phi}^{L}$ is the charge $L$ protected topological insertion defined in \eqref{eq:topological operator}, $d\mu$ is the measure
\beq
d\mu =\frac{1}{4\pi g}\frac{dx(x+x^{-1})}{2\pi i x }e^{2\pi g (x+x^{-1})}\comma
\eeq
and the contour goes counterclockwise once around the origin. We should emphasize that the LHS of \eqref{eq:integralfirst} is a single bracket correlator, as defined in \eqref{eq:single bracket}.

The functions denoted by $Q_{L_k}$ are called {\it Q-functions} and are characterized by the following two important properties:
\begin{enumerate}
\item They are orthogonal under the measure $d\mu$:
\beq \label{eq:Q fn orthogonality}
\oint d\mu \,Q_{J}(x)Q_{M} (x)\propto \delta_{JM}\period
\eeq
\item They are polynomials in $X\equiv g(x-x^{-1})$ with a unit leading coefficient:
\beq \label{eq:Q fn in X basis}
Q_{J}(x)=X^{J}+\cdots\period
\eeq
\end{enumerate}
These two properties uniquely determine the functions $Q_J$, and they can be computed systematically by performing the Gram-Schmidt orthogonalization on the set of monomials $\{1, X, X^{2},\cdots\}$. Applying the Gram-Schmidt procedure, we obtain the simple expression:
\begin{align}
&Q_{J}(x)= \frac{1}{D_J} \left|\begin{array}{cccc}\mathcal{I}_0&\mathcal{I}_1&\cdots &\mathcal{I}_{J}\\\mathcal{I}_1&\mathcal{I}_2&\cdots &\mathcal{I}_{J+1}\\\vdots &\vdots &\ddots &\vdots\\\mathcal{I}_{J-1}&\mathcal{I}_{J}&\cdots &\mathcal{I}_{2J-1}\\1&X&\cdots &X^{J}\end{array}\right|\comma\label{eq:expQ}\\
&D_{J}\equiv \det \left(\mathcal{I}_{j+k-2}\right)_{1\leq j,k\leq J}\qquad \qquad \mathcal{I}_j\equiv \oint d\mu \left(g (x-x^{-1})\right)^{j}\period
\end{align}
One important identity that follows from the orthogonalization is the expression for the two-point function
\beq\label{eq:twopt}
\langle \tilde{\Phi}^{J}\tilde{\Phi}^{M}\rangle=\oint d\mu \,\, Q_{J}(x)Q_{M}(x)= \frac{D_{J+1}}{D_J}\delta_{JM}\period
\eeq
The relations, \eqref{eq:expQ} and \eqref{eq:twopt}, hold for any choice of the measure $d\mu$ as long as the $Q$'s satisfy the aforementioned two properties. This will be important in the analysis of higher-point functions, as we will see shortly.  

\paragraph{Matrix model.}Another identity that we use in this section is the matrix model expression for the determinant $D_J$. This can be derived by expanding the determinant and expressing each component as an integral (see Section 7 of \cite{giombi2018exact}). The result reads
\beq\label{eq:matrixmodel}
D_J =\frac{g^{J (J-1)}}{J!}\left(\prod_{k=1}^{J}\oint d\mu (x_k)\right) \prod_{i<j} (x_i-x_j)^2 \left(1+\frac{1}{x_i x_j}\right)^{2}\period
\eeq
Alternatively we can write it in terms of $X_k\equiv g (x_k-x_k^{-1})$ as\fn{If one instead performs a change of variables $x_i =i e^{i\theta_i}$, the interaction term  in \eqref{eq:matrixmodel} becomes
\beq
(x_i-x_j)^2\left(1+\frac{1}{x_i x_j}\right)^2=-16 \sin^2 \left(\frac{\theta_i-\theta_j}{2}\right)\sin^2 \left(\frac{\theta_i+\theta_j}{2}\right)\comma
\eeq
which coincides with the Vandermonde factor for the matrix model of the orthogonal group $SO(2N)$.}
\beq
D_J =\frac{1}{J!}\left(\prod_{k=1}^{J}\oint \underbrace{\frac{dX_k}{8g^2\pi^2 i}e^{2\pi \sqrt{X_k^2+4g^2}}}_{=d\mu(x_k)}\right) \prod_{i<j} (X_i-X_j)^2 \comma
\eeq
where the integration contours are around the branch cut of $\sqrt{X_k^2+4g^2}$. Setting aside the unusual choice of contour, this can be regarded as the eigenvalue integral for the partition function of a $J\times J$ Hermitian matrix model with potential $e^{2\pi \sqrt{M^2+4g^2}}/(8g^2\pi^2 i)$:
\beq
D_J=\frac{Z_J}{J!}\qquad\text{with}\qquad  Z_J\equiv \oint [dM]\frac{e^{2\pi \sqrt{M^2+4g^2}}}{8g^2\pi^2 i}\period
\eeq

We can also express the two-point function, \eqref{eq:twopt}, directly in terms of this matrix model. For this purpose, we first rewrite $D_{J+1}$ as
\beq\nonumber
D_{J+1}=\frac{1}{(J+1)!}\left(\oint \frac{dY}{8g^2\pi^2i}e^{2\pi \sqrt{Y^2+4g^2}}\right)\left[\left(\prod_{k=1}^{J}\oint \frac{dX_k}{8g^2\pi^2 i}e^{2\pi \sqrt{X_k^2+4g^2}}(Y-X_k)^2\right) \prod_{i<j} (X_i-X_j)^2\right] \period
\eeq
The terms in the square bracket can be identified with the expectation value of the square of the determinant operator in the matrix model, $\det (Y-M)^2$. Dividing this by $D_J$, we obtain
\beq\label{eq:2ptdirectly}
\langle \tilde{\Phi}^{J}\tilde{\Phi}^{J}\rangle=\frac{D_{J+1}}{D_J}=\frac{1}{J+1}\oint d\mu(y)\langle \det (Y-M)^2\rangle_{J}\comma
\eeq
where $Y\equiv g (y-y^{-1})$ and $\braket{\bullet}_J$ denotes the expectation value in the matrix model:
\begin{align}
    \braket{f(M)}_J\equiv \frac{1}{Z_J}\oint [dM]\frac{e^{2\pi \sqrt{M^2+4g^2}}}{8g^2\pi^2 i}f(M).
\end{align}

Let us make several comments about this matrix model. First, \eqref{eq:matrixmodel} was derived initially from integrability in \cite{Gromov:2012eu} in the computation of the generalized cusp anomalous dimension. It was later re-derived in \cite{giombi2018exact} from localization and was shown to control the two-point function in the topological sector as well. 

Second, it has several similarities with the large-charge matrix model in \cite{Grassi:2019txd}, which computes the extremal correlation functions in rank-1 $\mathcal{N}=2$ SCFTs. In both cases, the matrix model was derived by applying the Gram-Schmidt orthogonalization to the localization results and the sizes of the matrices are related to the charges of the operators. Note however that the parameter regimes described by the two models are quite different. The matrix model in \cite{Grassi:2019txd} is for the rank-1 SCFTs, a canonical example being $\mathcal{N}=4$ SYM with the $SU(2)$ gauge group. By contrast, our matrix model is for the large $N$ limit of $\mathcal{N}=4$ SYM. Another difference is that our model can be applied to the non-extremal correlation functions, as we will see below.

 Third, since the measure is given by
$d\mu \sim e^{2\pi g (x+x^{-1})}$
the standard large $N$ limit in the matrix model corresponds to the limit in which $J$ is large but $J/g$ is kept fixed. In terms of the 't Hooft coupling $\lambda$, this corresponds to the limit
\beq
J\sim \sqrt{\lambda} \gg 1\comma
\eeq
which is precisely the strong coupling limit in which the theory is described by classical strings. In other words, the large $J$ expansion of the matrix model gives the semi-classical quantization of the classical string.
\paragraph{Higher-point functions.}
Let us now use the formalism reviewed above to derive a matrix model representation for the correlation functions of two large-charge insertions ($\tilde{\Phi}^{J}$) and several light insertions ($\tilde{\Phi}^{\ell_k}$):
\beq
\langle \tilde{\Phi}^{J}\tilde{\Phi}^{J}\prod_{j=1}^{n}\tilde{\Phi}^{\ell_j}\rangle=\oint d\mu \, Q_{J} (x) Q_{J}(x) \prod_{j=1}^{n}Q_{\ell_j}(x)\period
\eeq
To compute this correlation function, we first deform the measure by exponentiating the light operators:
\beq
d\tilde{\mu}\equiv d\mu \exp \left[t \prod_{j=1}^{n}Q_{\ell_j}(x)\right]\period
\eeq
Next, we construct the orthogonal polynomials of $X$ under this measure, which we denote by $\tilde{Q}_{J}(x,t)$. Using the relations reviewed in the previous section, we can express the two-point function of the $\tilde{Q}_J$'s as a ratio of (deformed) determinants,
\beq\label{eq:deformedtwo}
\oint d\tilde{\mu}\,\, \tilde{Q}_J(x,t)\tilde{Q}_J(x,t)=\frac{\tilde{D}_{J+1}}{\tilde{D}_J}\comma
\eeq
with
\beq
\tilde{D}_J \equiv \det \left(\tilde{\mathcal{I}}_{j+k-2}\right)_{1\leq j,k\leq J}\comma\qquad \tilde{\mathcal{I}}_j\equiv \oint d\tilde{\mu}\left(g(x-x^{-1})\right)^{j}\period
\eeq

Then we differentiate the left hand side of \eqref{eq:deformedtwo} with respect to $t$ and set $t$ to zero. Since both the measure and $\tilde{Q}$ depend on $t$, the differentiation produces three different terms, 
\beq
\oint d\mu \, Q_{J} (x) Q_{J}(x) \prod_{j=1}^{n}Q_{\ell_j}(x) \comma\qquad \oint d\mu \left.\frac{d\tilde{Q}_{J}}{dt}\right|_{t=0}Q_{J}\comma\qquad \oint d\mu \,Q_{J}\,\left.\frac{d\tilde{Q}_{J}}{dt}\right|_{t=0}\period
\eeq
However, the latter two terms identically vanish for the following reason: Since the leading coefficient of the polynomial $\tilde{Q}_J$ is $1$, the derivative $d\tilde{Q}_J/dt$ is always a polynomial of lower degree. Thus one can express $d\tilde{Q}_J/dt|_{t=0}$ as a linear combination of $Q_M$ with $M<L$. We can then use the orthogonality of the $Q$-functions, \eqref{eq:twopt}, to conclude that the latter two integrals vanish.

In summary, we arrive at the equality
\beq
\begin{aligned}
\langle \Phi^{J}\Phi^{J}\prod_{j=1}^{n}\Phi^{\ell_j}\rangle&=\oint d\mu Q_{J} (x) Q_{J}(x) \prod_{j=1}^{n}Q_{\ell_j}(x)=\left.\frac{d}{dt}\left[\oint d\tilde{\mu}\,\,\tilde{Q}_J(x,t)\tilde{Q}_{J}(x,t)\right]\right|_{t=0}\\
&=\left.\frac{d}{dt}\frac{\tilde{D}_{J+1}}{\tilde{D}_{J}}\right|_{t=0}\period
\end{aligned}
\eeq
If we divide this correlation function by the two-point function of the heavy insertions, we arrive at the simpler formula
\beq
\frac{\langle \tilde{\Phi}^{J}\tilde{\Phi}^{J}\prod_{j=1}^{n}\tilde{\Phi}^{\ell_j}\rangle}{\langle \tilde{\Phi}^{J}\tilde{\Phi}^{J}\rangle}=\left.\frac{d\log \tilde{D}_{J+1}}{dt}-\frac{d\log \tilde{D}_{J}}{dt}\right|_{t=0}\period
\eeq

Now the final step is to translate this into the matrix model. As with $D_J$, the deformed determinant admits a matrix model representation \eqref{eq:matrixmodel} with $d\mu$ replaced with $d\tilde{\mu}$. By differentiating it with respect to $t$ and setting it to zero, we get 
\beq
\left.\frac{d\tilde{D}_J}{dt}\right|_{t=0}=\frac{g^{J (J-1)}}{J!}\left(\prod_{k=1}^{J}\oint d\mu (x_k)\right) \prod_{i<j} (x_i-x_j)^2 \left(1+\frac{1}{x_i x_j}\right)^{2} \sum_{k=1}^{J}\left(\prod_{j=1}^{n}Q_{\ell_j}(x_k)\right)
\eeq
In the matrix model language,  after dividing by $D_J$, this corresponds to the expectation value of the ``single-trace operator'' ${\rm Tr}\left[\prod_{j=1}^{n}Q_{\ell_j}(M)\right]$. Therefore, we arrive at the relation
\beq\label{eq:higherpointmatrix}
\frac{\langle \tilde{\Phi}^{J}\tilde{\Phi}^{J}\prod_{j=1}^{n}\tilde{\Phi}^{\ell_j}\rangle}{\langle \tilde{\Phi}^{J}\tilde{\Phi}^{J}\rangle}=\left<{\rm Tr}\left[\prod_{j=1}^{n}Q_{\ell_j}(M)\right]\right>_{J+1}-\left<{\rm Tr}\left[\prod_{j=1}^{n}Q_{\ell_j}(M)\right]\right>_{J}\period
\eeq
Note that this formula naturally reduces to the original integral representation \eqref{eq:integralfirst} when $J=0$ because $\langle \bullet \rangle_{J=1}$ is a single-integral with the measure $d\mu$, and $\langle \bullet \rangle_{J=0}=0$.

\subsection{Large \texorpdfstring{$J$}{J} analysis}\label{sec:large J analysis}
\paragraph{Higher-point functions.} To evaluate \eqref{eq:higherpointmatrix} in the large $J$ limit, it is convenient to use the quasi-momentum discussed in \cite{Gromov:2012eu,giombi2018exact}:
\beq\label{eq:sumexpressionp}
\begin{aligned}
p (x)&\equiv \frac{1+x^2}{1-x^{2}}\sum_{k=1}^{J}\frac{1}{g(x-x^{-1})-g(x_k-x_k^{-1})}\\
&=\frac{x^2}{g(1-x^{2})}\sum_{k=1}^{J}\left(\frac{1}{x-x_k}+\frac{1}{x+\frac{1}{x_k}}-\frac{1}{x}\right)\period
\end{aligned}
\eeq
To motivate this definition, let us consider the one form $p(x)du$, where $u$ is the Zhukovsky variable often used in the integrability literature:
\beq
u(x)\equiv g\left(x+\frac{1}{x}\right)\period
\eeq
An important property of this one form is that it has poles with residue $-1$ at $x=x_k$ and $x=-x_k^{-1}$ and poles with residue $J$ at $x=0$ and $x=\infty$. In addition, it is related to the resolvent in the matrix model that we introduced earlier. Using the standard definition, the resolvent of the matrix model is
\beq\label{eq:defofRX}
R(X)\equiv \sum_{k=1}^{J}\frac{1}{X-X_k}\comma
\eeq
where we recall that the $X_k$'s are the eigenvalues of the matrix $M$.
Now, by explicit computation, we can check
\beq\label{resolventpu}
R(X)dX=-p(x)du(x)\comma
\eeq
where $X\equiv g (x-x^{-1})$.

Using \eqref{eq:defofRX}, we can express the expectation value of the ``single-trace'' operators as
\beq\label{eq:integralsingletr0}
\left<{\rm Tr}\left[\prod_{j=1}^{n}Q_{\ell_j}(M)\right]\right>_{J} =\oint_{\mathcal{C}} \frac{dX}{2\pi i }\left<R(X)\right>_J\prod_{j=1}^{n}Q_{\ell_j}(X)\comma
\eeq
where $\mathcal{C}$ is a contour which encircles all the eigenvalues $X_k$'s counterclockwise. To proceed, it is convenient to rewrite the above expression in terms of the $x$, rather than the $X$, variables. To do so, we need to remember that the $x$-plane is a double cover of the $X$-plane and each eigenvalue $X_k$ has two images $x_k$ and $-x_k^{-1}$ in the $x$-plane. Taking this into account, we obtain the expression 
\beq\label{eq:integralsingletr}
\left<{\rm Tr}\left[\prod_{j=1}^{n}Q_{\ell_j}(M)\right]\right>_{J} =-\oint_{\mathcal{C}_{+}\cup \mathcal{C}_{-}} \frac{du(x)}{4\pi i }\left<p(x)\right>\prod_{j=1}^{n}Q_{\ell_j}(x)\comma
\eeq
 where the contours $\mathcal{C}_{+}$ and $\mathcal{C}_{-}$ encircle $x_k$'s and $-x_k^{-1}$'s counterclockwise respectively. 
Note that we used the symmetry of $Q_{\ell_j}(x)$
 \beq
 Q_{\ell_j}(x)=Q_{\ell_j}(-x^{-1})\comma
 \eeq
 and the relation \eqref{resolventpu} to get \eqref{eq:integralsingletr}. The expression \eqref{eq:integralsingletr} is valid at finite $J$. If we take the large $J$ limit, we can approximate $\langle p(x)\rangle$ with its classical value $p_{\rm cl}(x)$, which was computed\fn{Precisely speaking, the matrix model studied in \cite{Gromov:2012eu} is different from the one analyzed here. However the large $J$ behavior of the quasi-momentum in the two matrix models was shown to be the same in \cite{giombi2018exact}. } in \cite{Gromov:2012eu}. We then get
 \beq\label{eq:integrallimit}
\left<{\rm Tr}\left[\prod_{j=1}^{n}Q_{\ell_j}(M)\right]\right>_{J} \overset{J\to \infty}{=}-\oint_{\mathcal{C}_{+}\cup \mathcal{C}_{-}} \frac{du(x)}{4\pi i }p_{\rm cl}(x)\prod_{j=1}^{n}Q_{\ell_j}(x)\period
\eeq
The classical limit of the quasi-momentum $p_{\rm cl}$ is given in terms of elliptic functions \cite{Gromov:2012eu}, but we do not need its explicit form in this paper. Instead we use its following properties (see also Figure \ref{fig:quasi1}):
\begin{figure}[t]
\centering
\includegraphics[clip,height=5cm]{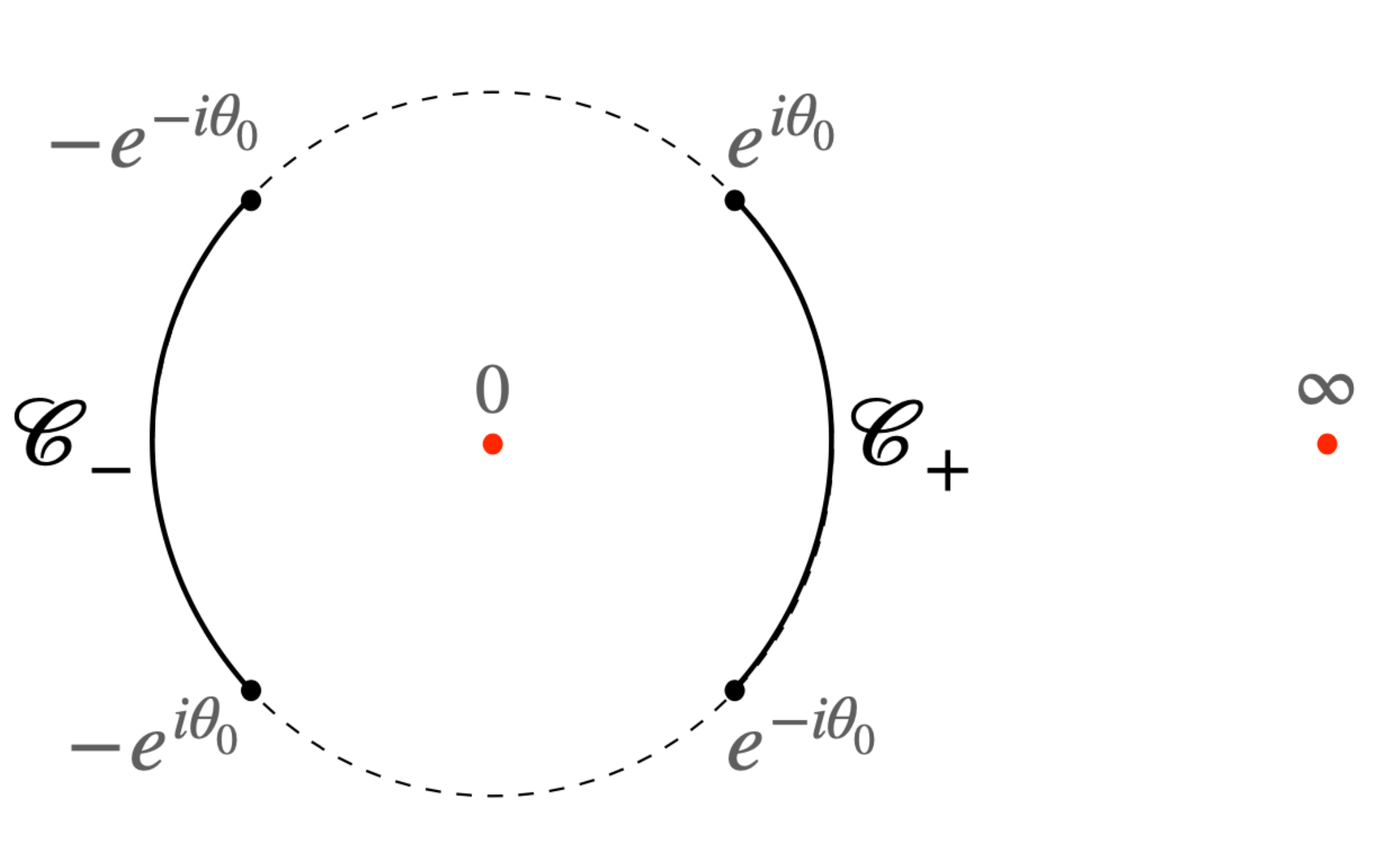}
\caption{Analytic structure of $p_{\rm cl}(x)$. It has two branch cuts $\mathcal{C}_{\pm}$, around which the integral $\oint p_{\rm cl} (x)du$ gives $-J$, as well as two poles at $x=0$ and $x=\infty$, around which $\oint p_{\rm cl}(x)dx$ gives $J$. In addition, $p_{\rm cl}=\pi$ at the {\it right} branch points, $e^{\pm i\theta}$, and $p_{\rm cl}=-\pi$ at the {\it left} branch points, $-e^{\pm i\theta}$.}\label{fig:quasi1}
\end{figure}
\begin{itemize}
\item In the $x$-plane, $p_{\rm cl}(x)$ has two (square-root) branch cuts along a unit circle; one along the arc $\mathcal{C}_{+}=[e^{-i\theta_0}, e^{i\theta_0}]$ and the other along the arc $\mathcal{C}_{-}=[-e^{-i\theta_0},-e^{i\theta_0}]$.
\item The parameter $\theta_0$ determining the size of the branch cuts is related to the charge $J$ of the heavy operators by  
\begin{align}\label{eq:thetaandcharge}
    J=4g \left[\ellK(\sin^2\theta_0)-\ellE(\sin^2\theta_0)\right].
\end{align}
Here $\ellK$ and $\ellE$ are the complete elliptic integrals of the first and second kind. $J$ is a monotonic function of $\theta_0$ satisfying $J=0$ when $\theta_0=0$ and $J\to \infty$ when $\theta_0\to \pi/2$.
As will be explained in Section \ref{sec:correlators from semistring}, the angle $\theta_0$ also controls the extension of the classical string solution in one of the $S^5$ directions.
\item On the branch cuts, $p_{\rm cl}$ satisfies 
\beq
p_{\rm cl}(x+i0)+p_{\rm cl}(x-i0)=\pm 2\pi \qquad \text{on }\mathcal{C}_{\pm}\period
\eeq In particular, at the branch points, $p_{\rm cl}(x)=\pm \pi$. The integrals around the branch cuts and at $x=0,\infty$ are given by
\beq\label{eq:periodcpm}
J=\oint_{x=0}p_{\rm cl}(x)du=\oint_{x=\infty}p_{\rm cl}(x)du=-\oint_{\mathcal{C}_{+}}p_{\rm cl}(x)du=-\oint_{\mathcal{C}_{-}}p_{\rm cl}(x)du\period
\eeq
\item $p_{\rm cl}(x)=-p_{\rm cl}(-x^{-1})$
\item The derivative $\del_Jp_{\rm cl}(x) du$ admits a simple expression (see Appendix \ref{ap:quasimomentum} for the derivation):
\beq\label{eq:derpcl}
\del_J p_{\rm cl}(x)du=\frac{(x+x^{-1})dx}{\sqrt{(x-e^{-i\theta_0})(x-e^{i\theta_0})(x+e^{-i\theta_0})(x+e^{i\theta_0})}}\period
\eeq
\end{itemize}
Let us also note that, as shown in \cite{Gromov:2012eu}, $p_{\rm cl}(x)$ coincides with the quasi-momentum of the {\it classical string solution}\fn{For the definition of the quasi-momentum in the classical string, see e.g.~\cite{Kazakov:2004qf,Dorey:2006zj}.} to be discussed in the next section, which is a holographic dual of the large charge operator on the Wilson loop. This relation is intriguing for several reasons: First it provides a clear physical interpretation of the spectral curve of our large charge matrix model. (By contrast, an analogous interpretation for the large charge matrix model for $\mathcal{N}=2$ SCFTs has not yet been established at the present time). Second the quasi-momentum of the classical string (and therefore $p_{\rm cl}$ discussed here) is expected to control observables beyond the topological subsector. For instance, in the case of closed string solutions, one can also describe the spectrum of general semi-classical fluctuations using the quasi-momentum as was demonstrated in \cite{Gromov:2008ec,Vicedo:2008jy}. We plan to revisit this (in the context of correlation functions) in the upcoming paper \cite{giombi2021}.

To evaluate \eqref{eq:integrallimit}, we deform the contour and bring it around $x=0$ and $x=\infty$. Using the invariance of the integrand under $x\to -x^{-1}$, we then arrive at
 \beq\label{eq:finallargeL}
 \left<{\rm Tr}\left[\prod_{j=1}^{n}Q_{\ell_j}(M)\right]\right>_{J} \overset{J\to \infty}{=}2\oint_{x=0} \frac{du(x)}{4\pi i }p_{\rm cl}(x)\prod_{j=1}^{n}Q_{\ell_j}(x)\comma
 \eeq
 where the contour encircles $x=0$ counterclockwise and the factor of $2$ comes from the contribution at infinity. Plugging this expression into the large $J$ limit of the formula for the higher-point function \eqref{eq:higherpointmatrix}, we obtain
 \beq\label{eq:integralQtoperform}
 \begin{aligned}
 \frac{\langle \tilde{\Phi}^{J}\tilde{\Phi}^{J}\prod_{j=1}^{n}\tilde{\Phi}^{\ell_j}\rangle}{\langle \tilde{\Phi}^{J}\tilde{\Phi}^{J}\rangle}&\overset{J\to\infty}{=}\oint_{x=0} \frac{du(x)}{2\pi i }\del_Jp_{\rm cl}(x)\prod_{j=1}^{n}Q_{\ell_j}(x)\\
 &=\oint_{x=0}\frac{dx (x+x^{-1})}{2\pi i}\frac{\prod_{j=1}^{n}Q_{\ell_j}(x)}{\sqrt{(x-e^{-i\theta_0})(x-e^{i\theta_0})(x+e^{-i\theta_0})(x+e^{i\theta_0})}}\period
 \end{aligned}
 \eeq
 The final step is to substitute $Q_{\ell_j}$ with its strong coupling limit determined in \cite{giombi2018exact},
 \beq\label{eq:Qsmall1}
 Q_{\ell}(x)\overset{g\to \infty}{=}(-i)^{\ell}\left(\frac{g}{2\pi}\right)^{\ell/2}H_{\ell}\left(i\sqrt{\frac{\pi g}{2}}(x-x^{-1})\right)\comma
 \eeq
 where $H_{\ell}$ is the Hermite polynomial. As can be seen from this expression, the leading contribution at large $g$ comes from the highest power in the Hermite polynomial. Therefore, for the purpose of computing the leading answer, we can simplify \eqref{eq:Qsmall1} to
 \beq
 Q_{\ell}(x)\overset{g\to \infty}{=}g^{\ell} (x-x^{-1})^{\ell}\period
 \eeq
 Performing the integral in \eqref{eq:integralQtoperform} analytically, we find that the result is nonzero only when the total length of light operators
 \beq
 \ell_{\rm tot}\equiv \sum_{j=1}^{n}\ell_j\comma
 \eeq
  is even, and it is given by
 \beq\label{eq:topo higher-point from matrix model}
  \frac{\langle \tilde{\Phi}^{J}\tilde{\Phi}^{J}\prod_{j=1}^{n}\tilde{\Phi}^{\ell_j}\rangle}{\langle \tilde{\Phi}^{J}\tilde{\Phi}^{J}\rangle}\overset{g\to\infty}{=}(-g^2\sin^2\theta_0)^{\ell_{\rm tot}/2}\pmatrix{c}{\ell_{\rm tot}\\\ell_{\rm tot}/2}\period
 \eeq
 We will later see that this agrees precisely with the result from the classical string.
  \paragraph{Two-point function.} We can also compute the two-point function, $\langle \tilde{\Phi}^{J}\tilde{\Phi}^{J} \rangle$, using the matrix model techniques. 
  
For this purpose, we start with the ``planar'' approximation of the expression \eqref{eq:2ptdirectly},
\beq\label{eq:largeLtwopnt}
 \langle \tilde{\Phi}^{J}\tilde{\Phi}^{J}\rangle=\frac{1}{J+1}\oint d\mu(y)\left<\det (Y-M)^2\right>_{J}\overset{J\to \infty}{\sim}\frac{1}{J+1}\oint d\mu(y)\,e^{2\left<{\rm Tr}\log (Y-M)\right>_J}\period
 \eeq
 The exponent can be computed using the matrix model techniques. Namely, we have
 \beq
 \begin{aligned}
 &\left<{\rm Tr}\log (Y-M)\right>_{J}\overset{J\to\infty}{=}-\oint_{\mathcal{C}_{+}\cup\mathcal{C}_{-}}\frac{du(x)}{4\pi i }p_{\rm cl}(x)\log \left[g(y-x)(1+1/xy)\right] \period
 \end{aligned}
 \eeq
To evaluate this integral, we rewrite the logarithmic term as\footnote{$\Lambda$ is an artificial cut-off that we introduce solely to write the logarithm as an integral of rational functions with simple poles. This makes the contour deformation analysis easier.}
\beq
\log \left[g(y-x)(1+1/xy)\right] =\lim_{\Lambda\to\infty}\left[\log (g \Lambda)+\int_{\Lambda}^{y}dy^{\prime}\left(\frac{1}{y^{\prime}-x}+\frac{1}{y^{\prime}+1/x}-\frac{1}{y^{\prime}}\right)\right]\period
\eeq
We then get
 \beq
 \begin{aligned}
 &\left<{\rm Tr}\log (Y-M)\right>_{J}\overset{J\to\infty}{=}\\
 &\lim_{\Lambda\to\infty}\left[J\log (g\Lambda)-\int^{y}_{\Lambda}dy^{\prime}\oint_{\mathcal{C}_{+}\cup\mathcal{C}_{-}}\frac{du(x)}{4\pi i }p_{\rm cl}(x) \left(\frac{1}{y^{\prime}-x}+\frac{1}{y^{\prime}+1/x}-\frac{1}{y^{\prime}}\right)\right]\period
 \end{aligned}
 \eeq
 Here we used \eqref{eq:periodcpm} to evaluate $\oint p_{\rm cl}(x)du$.
 Next we deform the contour of $x$ and pick up the residues of poles at $x=y^{\prime},-1/y^{\prime}, 0,\infty$. As a result we obtain
 \beq
 \left<{\rm Tr}\log (Y-M)\right>_{L\to \infty}=\lim_{\Lambda\to\infty}\left[J\log (g\Lambda)-\int^{u(y)}_{\Lambda}p_{\rm cl}(y^{\prime})du(y^{\prime})\right]\period
 \eeq
 
 Now, inserting this expression to the original integral \eqref{eq:largeLtwopnt}, we find that the integral of $y$ can be evaluated at the saddle point. The saddle point equation receives a contribution from $e^{2\pi g (y+1/y)}=e^{2\pi u(y)}$ in the measure and gives
 \beq
 p_{\rm cl}(y)=\pi\period
 \eeq
 This condition is satisfied at the right branch points of $p_{\rm cl}$ (i.e., $y=e^{\pm i\theta_0}$). We thus obtain the following expression for the two-point function in the large $J$ limit:
 \beq\label{eq:twopointdirectintegral}
 \langle \tilde{\Phi}^{J}\tilde{\Phi}^{J}\rangle\sim \exp \left(2J\log (g\Lambda)+2\pi u(e^{i\theta_0})-2\int_{\Lambda}^{u(e^{i\theta_0})}p_{\rm cl}(x)du\right)\period
 \eeq
 Here we took $y=e^{i\theta_0}$; the other choice, $e^{-i\theta_0}$, will give the same answer. 
 
\begin{figure}[t]
\centering
\includegraphics[clip,height=4.5cm]{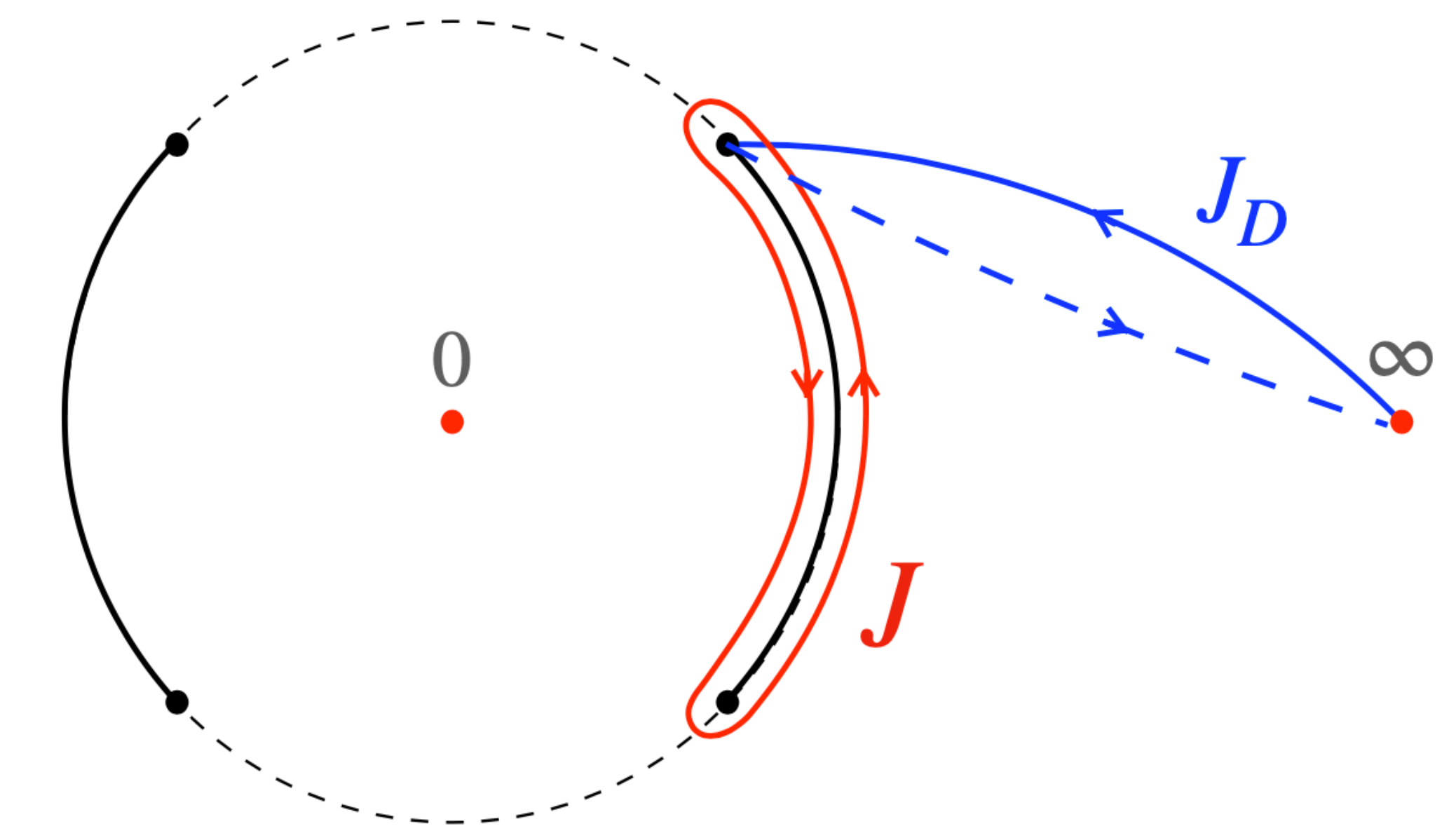}
\caption{The period integrals on the spectral curve. The charge of the operator, $J$, is given by an integral of $-p_{\rm cl}du$ around a branch cut. On the other hand, the logarithm of the two-point function, $\log \langle \tilde{\Phi}^{J}\tilde{\Phi}^{J}\rangle$, is given (up to terms that come from regularization) by an integral on a ``dual'' cycle denoted by $J_{D}$.}\label{fig:quasi2}
\end{figure}
 Before proceeding, let us make a few comments. The last term in \eqref{eq:twopointdirectintegral} can be rewritten as an integral from $x=\Lambda$ on the first sheet (denoted by $\Lambda_+$) to $x=-\Lambda$ on the second sheet (denoted by $\Lambda_-$):
 \beq
 -2\int_{\Lambda}^{u(e^{i\theta_0})}p_{\rm cl}(x)du=\int^{\Lambda_{+}}_{\Lambda_{-}}p_{\rm cl}(x)du\period
 \eeq
 On the other hand, the left hand side of \eqref{eq:twopointdirectintegral} is given by a ratio of matrix integrals $D_{J+1}/D_J$, and it becomes $\exp \left(\del_J\mathcal{F}_0\right)$ in the large $J$ limit, where $\mathcal{F}_0$ is the ``genus-0'' free energy
 \beq
 \log D_J \overset{J\to\infty}{\sim} \mathcal{F}_0\period
 \eeq 
 Note that $J$ is also given by an integral of $p_{\rm cl}(x)du$ albeit on a different contour $\mathcal{C}_{\pm}$. Therefore \eqref{eq:twopointdirectintegral} can be regarded as a relation between period integrals along two ``dual'' cycles\fn{The extra terms in $J_D$ can be viewed as a regularization which makes the period integral finite.}:
 \beq
 \begin{aligned}
&\del_J \mathcal{F}_{0}=J_D\comma\\
&J=-\oint_{\mathcal{C}_{\pm}}p_{\rm cl}(x)du\comma\qquad J_D\equiv \int^{\Lambda_{+}}_{\Lambda_{-}}p_{\rm cl}(x)du+2J\log (g\Lambda)+2\pi u(e^{i\theta_0})\period
\end{aligned}
 \eeq
 See also Figure \ref{fig:quasi2}.
 This is an analog of the relations known in conventional large $N$ matrix models and topological strings, which relate the genus $0$ free energy with the prepotential of period integrals \cite{Eynard:2014zxa,Eynard:2007kz}.
 
 Now, to evaluate \eqref{eq:twopointdirectintegral}, we first differentiate the exponent with respect to $J$. This gives two contributions: one is proportional to the deformation of the branch point $\del_{J}u(y^{\ast})$ while the other comes from the explicit $J$-dependence in $p_{\rm cl}$ and $J\log(g\Lambda)$. The first contribution is proportional to $\pi -p_{\rm cl}$, which vanishes at the saddle point. Therefore, we only need the second contribution, which gives
 \beq
 \begin{aligned}
\del_{J} \log \langle \tilde{\Phi}^{J}\tilde{\Phi}^{J}\rangle&\sim 2\log (g\Lambda)-2\int^{e^{i\theta_0}}_{\Lambda}\del_Jp_{\rm cl}(x)du\\
&=2\log (g\Lambda)-2\int^{e^{i\theta_0}}_{\Lambda}dx \frac{x+x^{-1}}{\sqrt{(x-e^{i\theta_0})(x-e^{-i\theta_0})(x+e^{i\theta_0})(x+e^{-i\theta_0})}}\period
\end{aligned}
 \eeq
 This integral can be evaluated analytically, and after sending the cut-off $\Lambda$ to infinity, we get
 \beq \label{eq:d/dJ log Phi^JPhi^J}
 \del_{J}\log \langle \tilde{\Phi}^{J}\tilde{\Phi}^{J}\rangle\overset{J\to\infty}{=}\log (-g^2\sin^2{\theta_0}).
 \eeq
 
 It is useful to pause at this point and switch to the compact notation
 \begin{align}\label{eq:c and sin(theta_0)}
     c\equiv \sin{\theta_0}.
 \end{align}
Because of \eqref{eq:thetaandcharge}, $c\in[0,1)$ is a monotonic function of $J$ defined implicitly by the relation
\begin{align}\label{eq:J and c-sqr}
    \frac{J}{4g}=\frac{\mathcal{J}}{4}=\ellK(c^2)-\ellE(c^2)\period
\end{align}
It satisfies $c=0$ when $\mathcal{J}=0$ and $c\to 1$ when $\mathcal{J}\to \infty$. This parametrization of the large charge is convenient in the remainder of the localization analysis in this section, and especially in the dual string analysis in Section~\ref{sec:correlators from semistring}. When we wish to emphasize its $J$-dependence or let $J$ vary, we will write $c(J)$ instead of $c$.

Returning to the computation of $\braket{\tilde{\Phi^J}\tilde{\Phi}^J}$, it follows from \eqref{eq:J and c-sqr} that
 \begin{align}\label{eq:dJ/dc^2}
      \frac{dJ}{dc^2}=\frac{2g\mathbb{E}(c^2)}{1-c^2}.
 \end{align}
This lets us integrate the $c^2$-dependent piece of the RHS of \eqref{eq:d/dJ log Phi^JPhi^J} with respect to $J$:
\begin{align}\label{eq:cjintegral}
\int^{J}_{0}dJ^{\prime}\log (c^2 (J^{\prime}))&=2g\int^{c^2}_0 dx \frac{\mathbb{E}(x)}{1-x}\log x \\
&=4g\left[\mathbb{K}(c^2)-\mathbb{E}(c^2)\right]\log (c^2)+8g\left[\mathbb{E}(c^2)-\mathbb{E}(0)\right]\nonumber\\
&=\log (c^{2J})+8g \left[\mathbb{E}(c^2)-\mathbb{E}(0)\right]\period\nonumber
 \end{align}
Meanwhile, the integral of the LHS of \eqref{eq:d/dJ log Phi^JPhi^J} yields $\log\braket{\tilde{\Phi}^J\tilde{\Phi}^J}-\log\braket{1}$. Putting everything together, we finally obtain the two-point correlator,
 \beq\label{eq:finaltop2}
 \dbraket{\tilde{\Phi}^J\tilde{\Phi}^J}=\frac{n_J}{(-2)^J}=(-g^2c^2)^{J}e^{8g\left(\ellE(c^2)-\ellE(0)\right)}.
 \eeq
We will be able to rederive \eqref{eq:finaltop2} from the classical action of the dual string in Section~\ref{sec:correlators from semistring}. 

\subsection{Large charge correlators from the Bremsstrahlung function}\label{sec:bremsstrahlung}
The large charge behavior of the topological two-point correlator can alternatively be determined from the ratio of incrementally shifted two-point functions, 
\begin{align}\label{eq:R_J}
    R_J\equiv \frac{\dbraket{\tilde{\Phi}^J\tilde{\Phi}^J}}{\dbraket{\tilde{\Phi}^{J-1}\tilde{\Phi}^{J-1}}},
\end{align}
which can in turn be related to the Bremsstrahlung function, $B_J$.\footnote{The Bremsstrahlung function $B_0$ was introduced in \cite{correa2012exact} as a quantity related to the cusp anomalous dimension of the Wilson loop and the radiation emitted by an accelerating quark, and was computed using localization. The generalized Bremsstrahlung function, $B_J$, was defined and computed using integrability in \cite{Gromov:2012eu} and computed using localization in \cite{giombi2018exact}.} This lets us take advantage of the leading large charge expression for $B_J$ determined in \cite{Gromov:2012eu} and the first subleading correction determined in \cite{Sizov:2013joa}. We will derive the relation between $R_J$ and $B_J$ using the localization framework developed in \cite{giombi2018exact}, which we therefore first briefly summarize.

Firstly, the half-BPS Wilson loop defined in \eqref{eq:circular WL} can be generalized to a $1/8-$BPS Wilson loop \cite{Drukker:2007yx,Drukker:2007qr}, $\mathcal{W}_{1/8}$, whose contour $x_\mu(\varphi)$ is any closed curve on the two-sphere $x_1^2+x_2^2+x_3^2=1$ and which couples to the scalars as $\vec{\Phi}\times \vec{x}\cdot d\vec{x}$, where $\vec{x}\equiv (x_1,x_2,x_3)$ and $\vec{\Phi}\equiv (\Phi_3,-\Phi_4,\Phi_6)$.\footnote{We pick this combination of scalar fields to couple to the $1/8$-BPS Wilson loop in order to match our conventions for the half-BPS Wilson loop and topological operators in Section~\ref{sec:large charges in WL dCFT}.} The expectation value of $\mathcal{W}_{1/8}$ is found by replacing $g\to g\frac{A(4\pi-A)}{4\pi^2}$ in \eqref{eq:half-BPS WL VEV}, where $A$ is the area of one of the two regions of the two-sphere demarcated by the contour. The $1/8$-BPS loop reduces to the half-BPS loop when $\vec{x}=(\cos{\varphi},\sin{\varphi},0)$, in which case $A=2\pi$.

Secondly, taking $L$ derivatives of the $1/8$-BPS Wilson loop with respect to $A$ inserts $L$ copies of the unit charge topological operator $\tilde{\Phi}\equiv x_1(\varphi)\Phi_3-x_2(\varphi)\Phi_4+x_3(\varphi)\Phi_6+i\Phi_5$\footnote{More generally, we may equivalently define the topological operator to be $\vec{x}(\varphi)\cdot \vec{\Phi}+i\hat{n}\cdot \vec{\Phi}^\perp$, where $\vec{\Phi}^\perp\equiv (\Phi_1,\Phi_2,\Phi_5)$ and $\hat{n}$ is any unit $3$-vector. We use the freedom to pick $\hat{n}=(0,0,1)$ so that this topological operator reduces to the topological operator defined in \eqref{eq:topological operator} when the $1/8$-BPS Wilson loop reduces to the half-BPS Wilson loop.} along the loop:
\begin{align}
    \partial_A^L \langle\mathcal{W}_{1/8}\rangle_{\mathcal{N}=4\text{ SYM}}&=\langle \mathcal{W}_{1/8}[\underbrace{\tilde{\Phi}\cdots \tilde{\Phi}}_{L}]\rangle_{\mathcal{N}=4\text{ SYM}}.
\end{align}
For this subsection, we will use $\langle \ldots\rangle_{1/8}\equiv \langle \mathcal{W}_{1/8}[\ldots]\rangle_{\mathcal{N}=4\text{ SYM}}$ to denote the topological defect correlators on the $1/8$-BPS loop. These reduce to the topological correlators on the half-BPS Wilson loop when $A=2\pi$. Namely, $\braket{\ldots}_{1/8}\rvert_{A=2\pi}=\braket{\ldots}$. We also use the notation $\underbrace{\tilde{\Phi}\cdots \tilde{\Phi}}_{L}$ to denote $L$ non-coincident copies of $\tilde{\Phi}$, which is different from the charge $L$ topological operator $\tilde{\Phi}^{L}$ that can be thought of as  the normal-ordered insertion of $L$ coincident copies of $\tilde{\Phi}$.

Thirdly, the higher charge topological operators on the half-BPS Wilson loop can be defined by imposing two properties that are directly parallel to the two properties in Eqs.~(\ref{eq:Q fn orthogonality})-(\ref{eq:Q fn in X basis}) defining the $Q$-functions. Namely,
\begin{align}\label{eq:topo op orthogonality}
    \braket{\tilde{\Phi}^J\tilde{\Phi}^M}_{1/8}\propto \delta_{JM},
\end{align}
and $\tilde{\Phi}^J$ is a linear combination of the $\underbrace{\tilde{\Phi}\cdots \tilde{\Phi}}_{L}$ operators for $L\leq J$ with a unit leading coefficient:
\beq
\tilde{\Phi}^J=\underbrace{\tilde{\Phi}\cdots \tilde{\Phi}}_J+\ldots\period
\eeq
In this way the higher charge topological operators define an orthogonal basis that can be constructed from non-coincident insertions of the unit topological operators using the Gram-Schmidt procedure.

Finally, the topological correlators are closed under the OPE. As a special case, the OPE between a charge $J$ insertion and a unit charge is simply
\begin{align}\label{eq:W5ZpWcKNwu}
    \tilde{\Phi}^J\,\tilde{\Phi}&=\tilde{\Phi}^{J+1}+\frac{\braket{\tilde{\Phi}^J\,\tilde{\Phi}^J}_{1/8}}{\braket{\tilde{\Phi}^{J-1}\,\tilde{\Phi}^{J-1}}_{1/8}}\tilde{\Phi}^{J-1}\period
\end{align}
The prefactors on the RHS of \eqref{eq:W5ZpWcKNwu} can be determined by taking the expectation value after multiplying both sides above by $\tilde{\Phi}^{J+1}$ or $\tilde{\Phi}^{J-1}$, and also noting
that
\begin{align}\label{eq:HRDL6AIYo2}
    \braket{\tilde{\Phi}\,\tilde{\Phi}^J\,\tilde{\Phi}^{J+1}}_{1/8}=\braket{\tilde{\Phi}^{J+1}\,\tilde{\Phi}^{J+1}}_{1/8}.
\end{align}
This is a special case of the statement that the extremal topological correlators, for which the largest charge is equal to the sum of all the smaller charges, reduce to two-point functions. Namely,
\begin{align}\label{eq:extremal correlators}
    \braket{\tilde{\Phi}^{J_1}\tilde{\Phi}^{J_2}\ldots \tilde{\Phi}^{J_n}}_{1/8}&=\braket{\tilde{\Phi}^{J_1}\tilde{\Phi}^{J_1}}_{1/8}, && \text{if }J_1=J_2+\ldots+J_n,
\end{align}
as was shown in \cite{giombi2018exact}.

We have now laid all of the groundwork needed to relate the topological two-point function to the Bremsstrahlung function. We begin with the expression derived in \cite{giombi2018exact} (see also \cite{correa2012exact}) for the Bremsstrahlung function as the second area derivative of the two-point function:
\begin{align}\label{eq:UQFQ77bj6W}
    B_J= -\partial_A^2 \log{\braket{\tilde{\Phi}^J\tilde{\Phi}^J}}_{1/8}\biggr\rvert_{A=2\pi}.
\end{align}
Each area derivative in \eqref{eq:UQFQ77bj6W} can either act on the Wilson loop to insert a factor of $\tilde{\Phi}$ or on one of the topological operators. In particular, when we expand
\beq
\tilde{\Phi}^J=\underbrace{\tilde{\Phi}\cdots \tilde{\Phi}}_{J}+c_J(A)\underbrace{\tilde{\Phi}\cdots \tilde{\Phi}}_{J-1}+\ldots\comma
\eeq
then all the coefficients other than the leading one depend on the area. We have explicitly labelled the next-to-leading coefficient, $c_J(A)$, since it plays an important role in what follows. \eqref{eq:UQFQ77bj6W} therefore evaluates to
\begin{align}\label{eq:MQGcbvJ8bR}
    -B_J&=\frac{\braket{\tilde{\Phi}^J\tilde{\Phi}^J\tilde{\Phi}\tilde{\Phi}}}{\braket{\tilde{\Phi}^J\tilde{\Phi}^J}}+2\frac{dc_J}{dA}\biggr\rvert_{A=2\pi}.
\end{align}
Various other terms on the RHS drop out due to the orthogonality in \eqref{eq:topo op orthogonality}. The next step is to deduce a more transparent form for $\frac{dc_J}{dA}$. Taking one area derivative of $\braket{\tilde{\Phi}^J\tilde{\Phi}^{J-1}}=0$ yields
\begin{align}\label{eq:j0lrzagESx}
    \frac{dc_J}{dA}\biggr\rvert_{A=2\pi}&=-\frac{\braket{\tilde{\Phi}^J\tilde{\Phi}^{J-1}\tilde{\Phi}}}{\braket{\tilde{\Phi}^{J-1}\tilde{\Phi}^{J-1}}}=-R_J.
\end{align}
To get to the second equality, we used \eqref{eq:HRDL6AIYo2} and recalled the definition of $R_J$ in \eqref{eq:R_J}.

Finally, applying the OPE in \eqref{eq:W5ZpWcKNwu} to both copies of $\tilde{\Phi}^J\tilde{\Phi}$ in the four point correlator on the RHS of \eqref{eq:MQGcbvJ8bR}, we can write it in terms of $R_J$ as well:
\begin{align}\label{eq:d4F40tIvE9}
    \frac{\braket{\tilde{\Phi}^J\tilde{\Phi}^J\tilde{\Phi}\tilde{\Phi}}}{\braket{\tilde{\Phi}^J\tilde{\Phi}^J}}&=R_{J+1}+R_J.
\end{align}
Therefore, Eqs.~(\ref{eq:MQGcbvJ8bR}), (\ref{eq:j0lrzagESx}) and (\ref{eq:d4F40tIvE9}) yield
\begin{align}\label{eq:H9UAvjl3DA}
    -B_J=R_{J+1}-R_J,
\end{align}
which is the desired relation for $R_J$ in terms of $B_J$. It holds for any $g$ and $J$. 

We are in particular interested in \eqref{eq:H9UAvjl3DA} in the large charge regime. If we expand $R_J$ and $B_J$ in $1/g$, 
\begin{align}
    R_J&\equiv g^2 r_2\left(J/g\right)+g r_1(J/g)+\ldots,\label{eq:RLODwsvWLK}\\
    B_J&\equiv g b_1(J/g)+b_0(J/g)+\ldots,\label{eq:vIPg0qUwzK}
\end{align}
then we can use \eqref{eq:H9UAvjl3DA} to relate $r_2$ and $r_1$ to $b_1$ and $b_0$. The latter two were determined in \cite{Gromov:2012eu,Sizov:2013joa}:
\begin{align}\label{eq:RYMw4WV3JZ}
    b_1(J/g)&=\frac{1-c^2}{2\ellE(c^2)},& b_0(J/g)&=\frac{1}{2}b_1'(J/g).
\end{align}
Here, we have written $b_1$ in terms of the parameter $c^2$ introduced in \eqref{eq:J and c-sqr}, and have found that $b_0$ simplifies when written in terms of $b_1$.\footnote{To get this result, we set $r=1$ and $\sin^2{\psi}=c^2$ in Eqs. (5.11) and (5.12) of \cite{Sizov:2013joa}.} Substituting Eqs.~(\ref{eq:RLODwsvWLK}) and (\ref{eq:vIPg0qUwzK}) into \eqref{eq:H9UAvjl3DA} and matching the two sides order by order yields the following pair of differential equations:
\begin{align}\label{eq:wdrewswe}
    r_2'(x)&=-b_1(x), &\frac{1}{2}r_2''(x)+r_1'(x)&=-b_0(x).
\end{align}
Eqs.~(\ref{eq:RYMw4WV3JZ}) and (\ref{eq:wdrewswe}) together imply $r_1'(x)=0$. Furthermore, writing $r_2'(J/g)=g\frac{dr_2}{dc^2}\frac{dc^2}{dJ}$ and noting from \eqref{eq:dJ/dc^2} that $b_1(J/g)=g\frac{dc^2}{dJ}$, we find $\frac{dr_2}{dc^2}=-g^2$. Finally, imposing the initial conditions $r_2(0)=r_1(0)=0$,\footnote{These initial conditions can be justified as follows. As we'll discuss in Section~\ref{sec:small and large J/g}, the small $\mathcal{J}$ expansion of the large charge correlators matches the large $J$ expansion of the strongly coupled finite charge correlators. Thus, we can partly determine the behavior of $R_J$ as $\mathcal{J}\to 0$ using the expression for the finite charge correlator determined in \cite{giombi2018exact}, which is given in \eqref{eq:n_ell finite charge}. From
\begin{align}
    R_J=-\frac{n_J}{2n_{J-1}}=-\frac{gJ}{\pi}\left(1-\frac{3J}{8\pi g}+O(1/g^2)\right)=g^2\left(-\frac{\mathcal{J}}{\pi}+\frac{3\mathcal{J}^2}{8\pi^2 g}+O(\mathcal{J}^3)\right)+g O(\mathcal{J}^2)+O(g^0).
\end{align}
we explicitly see that $r_2\to 0$ and $r_1\to 0$ as $\mathcal{J}\to 0$.
} we arrive at the following particularly simple form for $R_J$:
\begin{align}\label{eq:RqOAARYZO3}
    R_J&=-g^2c^2+O(g^0).
\end{align}

\eqref{eq:RqOAARYZO3} lets us determine the large charge limit of the two-point correlator. Keeping only the leading order term in $R_J$, we find
\begin{align}
    \dbraket{\tilde{\Phi}^J\tilde{\Phi}^J}&=\prod_{j=1}^J R(j)=(-g^2)^J \text{exp}\left(\sum_{j=1}^J \log{c^2(j)}\right).
\end{align}
To leading order, we may approximate the sum by the integral computed in \eqref{eq:cjintegral}. This reproduces the result in the previous subsection, \eqref{eq:finaltop2}.

Returning to \eqref{eq:d4F40tIvE9}, 
\eqref{eq:RqOAARYZO3} also lets us determine the four-point correlator:
\begin{align}\label{eq:zvANrCh5vU}
    \frac{\braket{\tilde{\Phi}^J\tilde{\Phi}^J\tilde{\Phi}\tilde{\Phi}}}{\braket{\tilde{\Phi}^J\tilde{\Phi}^J}}&=2R_J-B_J=-2g^2c^2-\frac{g}{2}\frac{1-c^2}{\ellE(c^2)}+O(g^0).
\end{align}
We will see in Section~\ref{sec:correlators from semistring} that, in the dual string calculation, $2R_J$ is the classical contribution to the four point function. The other term, $-B_J$, corresponds to the contribution from semiclassical fluctuations, which we will discuss in \cite{giombi2021}.

Finally, \eqref{eq:RqOAARYZO3} also lets us determine the normalized extremal OPE coefficient,
\begin{align}\label{eq:extremal OPE coeff}
    \frac{c_{J+\ell,J,\ell}^2}{n_{J+\ell}n_Jn_\ell}&\equiv\frac{\dbraket{\tilde{\Phi}^{J+\ell}\tilde{\Phi}^J\Phi^\ell}^2}{\dbraket{\Phi^{J+\ell}\Phi^{J+\ell}}\dbraket{\Phi^J\Phi^J}\dbraket{\Phi^\ell\Phi^\ell}},
\end{align}
to next-to-leading order. We first note from \eqref{eq:dJ/dc^2} how $c$ changes when $J$ changes by a finite amount:
\begin{align}\label{eq:C(J+j)}
    c^2(J+j)&=c^2+\frac{j}{2g}\frac{1-c^2}{\ellE(c^2)}+O(j^2/J^2).
\end{align}
Therefore,
\begin{align}\label{eq:J+ell over J}
    \frac{\braket{\tilde{\Phi}^{J+\ell}\tilde{\Phi}^{J+\ell}}}{\braket{\tilde{\Phi}^J\tilde{\Phi}^J}}=\prod_{j=1}^\ell R_{J+j}=(-g^2c^2)^\ell \left(1+\frac{\ell(\ell+1)}{4g}\frac{1-c^2}{c^2\ellE(c^2)}+O(1/g^2)\right).
\end{align}
Noting the strong coupling expansion of $n_\ell$ for finite $\ell$ \cite{giombi2018exact},
\begin{align}\label{eq:n_ell finite charge}
    n_\ell(g)&=\left(\frac{2g}{\pi}\right)^\ell \ell! \left(1-\frac{3}{16\pi g}\ell(\ell+1)+O(1/g^2)\right),
\end{align}
the normalized extremal OPE coefficient becomes
\begin{align}\label{eq:extremal OPE coeff subleading}
    \frac{c_{J+\ell,J,\ell}^2}{n_{J+\ell}n_Jn_\ell}&=\frac{\left(g\pi c^2\right)^\ell}{\ell!}\left(1+\frac{\ell(\ell+1)}{4g}\left[\frac{3}{4\pi}+\frac{1-c^2}{c^2\ellE(c^2)}\right]+O(1/g^2)\right).
\end{align}

\section{Large charge correlators from the classical string}\label{sec:correlators from semistring}

Now we turn to the analysis of the large charge correlators using the classical string solution in AdS$_5\times S^5$ that is holographically dual to the Wilson operator with $Z^J$ and $\bar{Z}^J$ inserted. 

The classical string was discussed previously in a few closely related contexts. The string with $\mathcal{J}\to \infty$ was first identified in \cite{drukker2006small}, and its spectrum of fluctuations in the BMN limit was matched to the anomalous dimensions, computed in weakly coupled gauge theory, of ``words'' inserted on the Wilson loop that are composed of many copies of $Z$ interspersed with a few copies of an orthogonal scalar. Then \cite{Miwa2006HolographyOW} used the same classical string to explore the $\mathcal{J}\to \infty$ limit of the two-point function in \eqref{eq:HH correlator}. 
Finally, the string with general $\mathcal{J}$ is a special case of a more general string solution identified in Appendix E of \cite{Gromov:2012eu}, which was used to determine the classical limit of the cusp anomalous dimension of the Wilson line with cusps simultaneously in both the ambient and internal contours at the locations of the $Z^J$ and $\bar{Z}^J$ insertions. (In our analysis the Wilson loop has insertions but no cusps.)

We will first review the general $\mathcal{J}$ string solution and, as an additional check building on the previous works, we show in Appendix~\ref{sec:semistring susy} that the supersymmetries of the dual string match those of the Wilson loop with insertions by extending the discussion in Sec.~4.2 of \cite{drukker2006small} to the case with general $\mathcal{J}$. The main new results of our analysis are the evaluation of the classical action of the string solution (including the identification of the relevant boundary terms), which gives the holographic prediction for the two-point function of the large charge insertions, and the calculation of higher-point correlators with two heavy charges using the classical string. 


\subsection{Identifying the dual string solution}\label{eq:identify semistring}
In this section, generalizing Section 4.1 in \cite{drukker2006small}, we identify the classical string dual to the Wilson operator with $Z^J$ and $\bar{Z}^J$ inserted and $\mathcal{J}$ finite. We begin by writing the metric in AdS$_5\times S^5$ using global coordinates on AdS$_5$ and embedding coordinates on $S^5$:
\begin{align}
    ds^2&=ds^2_{\text{AdS}_5}+ ds^2_{S^5},
\end{align}
where
\begin{equation}
\begin{aligned}
&ds^2_{\text{AdS}_5}=d\rho^2-\cosh^2{\rho}dt^2+\sinh^2{\rho}\left(d\psi_1^2+\sin^2{\psi_1}\left(d\psi_2^2+\sin^2{\psi_2}d\psi^2\right)\right)\\
&ds^2_{S^5}=d\Theta_Id\Theta_I.
\end{aligned}
\end{equation}
Here, $\rho\in [0,\infty)$ is the bulk coordinate of AdS$_5$ such that $\rho=0$ is the middle of AdS$_5$ and $\rho\to \infty$ is the conformal boundary, $\mathbb{R}\times S^3$; $t$ is the time coordinate along $\mathbb{R}$; and $\psi_1,\psi_2\in [0,\pi]$ and $\psi\in [0,2\pi)$ are spherical coordinates on $S^3$. Furthermore, the coordinates $\Theta_I$, $I=1,\ldots 6$, satisfy $\Theta_I\Theta_I=1$ and embed $S^5$ in $\mathbb{R}^6$.

We let the contour of the half-BPS Wilson operator on the boundary consist of the two antiparallel, antipodal lines that lie at $\rho\to\infty, \psi_1=\psi_2=\frac{\pi}{2}$ and $\psi=0,\pi$, such that the line at $\psi=0$ runs in the positive $t$ direction and the line at $\psi=\pi$ runs in the negative $t$ direction. Furthermore, we put $Z^J$ in the infinite past, $t\to -\infty$, and $\bar{Z}^J$ in the infinite future, $t\to \infty$. 
Given this symmetric choice of the Wilson contours in global coordinates, we may restrict our attention to the AdS$_3$ subspace of AdS$_5$ with $\psi_1=\psi_2=\frac{\pi}{2}$. We could also fix $\psi=0,\pi$ and restrict to an AdS$_2$ subspace but it is convenient to keep track of $\psi$ for when we ultimately change to Poincar\'e coordinates and map this configuration to the circular Wilson loop with two insertions at $\varphi=\pm \varphi_L$, as in \eqref{eq:HH correlator}.

The coordinate $\Theta_I$ is dual to $\Phi_I$. Since the Wilson loop couples only to $\Phi_6$ and the insertions $Z^J$ and $\bar{Z}^J$ are composed of only $\Phi_4$ and $\Phi_5$, it follows that the dual string also lies in the $S^2$ subspace of $S^5$ defined by $\Theta_4^2+\Theta_5^2+\Theta_6^2=1$, $\Theta_1=\Theta_2=\Theta_3=0$. We introduce the coordinates $\theta\in[0,\pi]$ and $\phi\in [0,2\pi)$ to parametrize this two-sphere such that $\theta$ is the polar angle from the north pole, $\Theta_6=1$, and $\phi$ is the azimuthal angle that rotates parallel to the $\Theta_4$-$\Theta_5$ plane:
\begin{align}
    \Theta_4&=\sin{\theta}\cos{\phi},&\Theta_5&=\sin{\theta}\sin{\phi},&\Theta_6&=\cos{\theta}.
\end{align}

Therefore, restricting to the AdS$_3\times S^2$ subspace that contains the dual string, we may write the relevant parts of the metric as
\begin{align}
    ds^2&=ds^2_{\text{AdS}_3}+ds^2_{S^2}, \label{eq:rB4vz3h29T}\\ ds^2_{\text{AdS}_3}&=d\rho^2-\cosh^2{\rho}dt^2+\sinh^2{\rho}d\psi^2, &ds^2_{S^2}&=d\theta^2+\sin^2{\theta}d\phi^2.\label{eq:metric on AdS3 and S2}
\end{align}
It is convenient to extend $\rho$ to negative values and identify $(-\rho,t,\psi)=(\rho,t,\psi+\pi)$.

\begin{figure}[t!]
\centering
\begin{minipage}{0.49\hsize}
\centering
\includegraphics[clip, height=10cm]{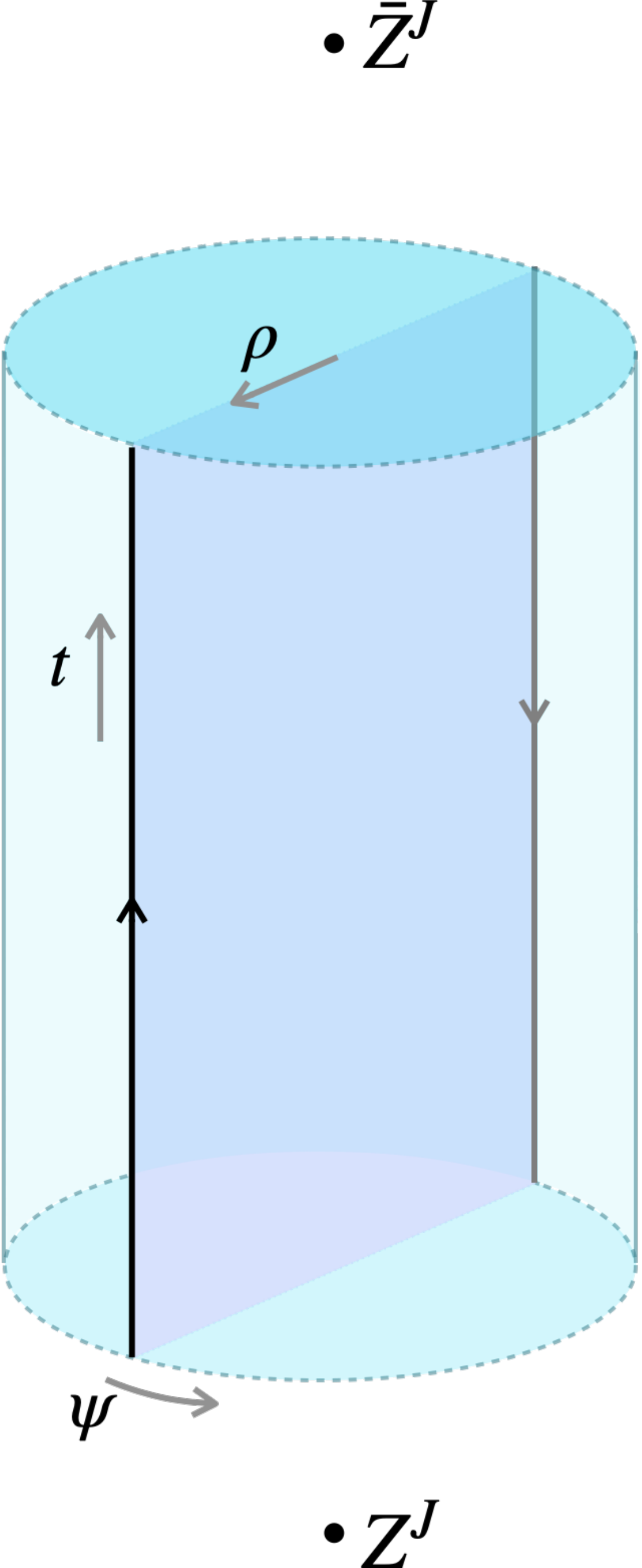}\\
\vspace{0.5cm}
{\bf a.} Classical string solution in AdS$_5$
\end{minipage}
\begin{minipage}{0.49\hsize}
\centering
\vspace{2cm} 
\includegraphics[clip, height=6cm]{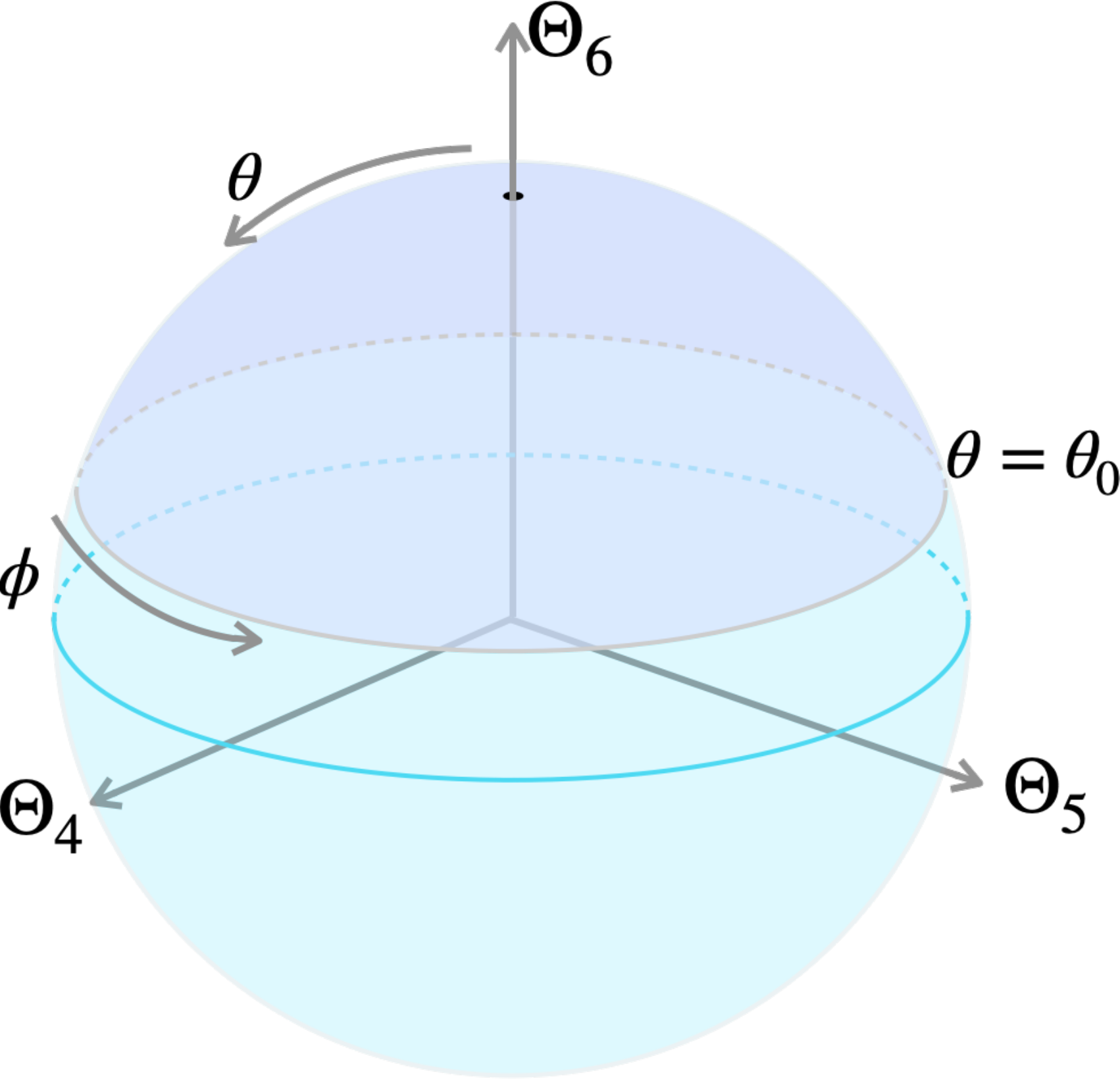}\\
\vspace{2.5cm} 

{\bf b.} Classical string solution in $S^5$
\end{minipage}
\caption{The classical string solution in AdS$_5\times S^5$ dual to the Wilson loop operator with $Z^J$ and $\bar{Z}^J$. {\bf a.} In global AdS$_5$ coordinates, the string forms a strip stretching between the antipodal, antiparallel lines forming the contour of the Wilson operator. The insertions, $Z^J$ and $\bar{Z}^J$, are located at the infinite past and future, respectively. {\bf b.} In $S^5$, the string wraps part of the upper half of the two-sphere $\Theta_4^2+\Theta_5^2+\Theta_6^2=1$. When the string is at the boundary of AdS, $\rho\to \pm \infty$, it is at the north pole of the sphere, $\theta=0$. When the string is in the center of AdS, $\rho=0$, it is at the maximum polar angle on the sphere, $\theta=\theta_0$. Furthermore, translations in $t$ in AdS are accompanied by rotations in $\phi$ on the sphere.} 
\label{fig:semiclassical string}
\end{figure}

The string moving in the background in \eqref{eq:rB4vz3h29T} is governed by the Nambu-Goto action (recall that we use the notation $g\equiv \frac{\sqrt{\lambda}}{4\pi}$),
\begin{align}\label{eq:Nambu-Goto}
    S_{\rm NG}[\Psi]&=-2g\int d^2\sigma \sqrt{-h},
\end{align}
where $\Psi$ is shorthand for the all the coordinates of the string, $2g$ is the string tension, $\sigma^\alpha$, $\alpha=0,1$, are the worldsheet coordinates, and $h$ denotes the determinant of $h_{\alpha\beta}$, the metric induced on the worldsheet. Following \cite{drukker2006small}, we identify the classical string dual to the Wilson operator with $Z^J$ and $\bar{Z}^J$ in the large charge limit as the solution extremizing \eqref{eq:Nambu-Goto} with the following properties:
\begin{enumerate}[1.]
    \item The string is incident on the antiparallel lines at $\psi=0,\pi$ on the boundary of AdS$_3$ and is at  $\theta=0$, the north pole of $S^2$, when it reaches the boundary.
    \item As the string moves away from the antiparallel lines on the boundary, it simultaneously moves away from the north pole in $S^2$ and reaches a maximum angle, $\theta_0\in [0,\pi/2)$, when it reaches the middle of AdS$_3$.
    \item The string rotates in $S^2$ in the $\phi$ direction with angular momentum $J$.
\end{enumerate}

A family of string solutions satisfying the equations of motion and having the first two properties is given by:
\begin{equation}
\begin{aligned}
    &\rho=\sigma^1, \qquad t=\sigma^0, \qquad \psi=0,\\ 
    &\sin{\theta}=\frac{c}{\cosh{\sigma^1}}, \qquad \phi=\phi_0+\sigma^0.
\label{eq:semistring global coords 1}
\end{aligned}
\end{equation}
Here, the worldsheet coordinates, $\sigma^1\equiv \rho$ and $\sigma^0\equiv t$, span $\mathbb{R}$, the parameter $c\in[0,1)$ is related to the maximum value of $\theta$ by
\begin{align}
    c=\sin{\theta_0}\,,
\end{align}
and $\phi_0\in[0,2\pi)$ is a modulus whose significance will become clear in the next section. The string is depicted in Figure~\ref{fig:semiclassical string}. While we have presented the string solution using the Nambu-Goto description, one may also use the conformal gauge and the Polyakov action, as in \cite{drukker2006small, Gromov:2012eu}. The solution in the conformal gauge is related to \eqref{eq:semistring global coords 1} by a change of variable of the $\sigma^2$ coordinate (involving elliptic functions) with $\sigma^0=t$ unchanged. We find it more convenient to work with the Nambu-Goto action and the solution \eqref{eq:semistring global coords 1}.

Finally, to complete the identification of the classical string dual to the Wilson loop operator with insertions, we require that it have the correct angular momentum in accordance with the third property above. This fixes the value of the parameter $c$ in terms of the charge $J$. Noting the angular momentum density $\Pi_\phi\equiv -2g\frac{\partial \sqrt{-h}}{\partial (\partial_0 \phi)}=-2g\sin^2{\theta}\sqrt{-h}h^{0\alpha }\partial_\alpha\phi$, and setting the anguluar momentum of the string equal to $J$, we find
\begin{align}\label{eq:s4VcJNwmrV}
    J&=\int_{-\infty}^\infty d\rho \Pi_\phi=2g \int_{-\infty}^\infty d\rho \frac{c^2\sech^2{\rho}}{\sqrt{\cosh^2{\rho}-c^2}}=4g(\ellK(c^2)-\ellE(c^2)),
\end{align}
which matches \eqref{eq:J and c-sqr} derived from the matrix model side. This tells us that the maximum polar angle of the classical string is the same as the angle parametrizing the branch cuts of $p_{\rm cl}$ (see Figure~\ref{fig:quasi1}), which justifies our use of the same symbol, $\theta_0$, for both. Therefore, we may also identify the parameter $c$ in the string solution in \eqref{eq:semistring global coords 1} with the parameter introduced in \eqref{eq:c and sin(theta_0)}. In fact, as shown in \cite{Gromov:2008ec}, the classical quasi-momentum of the matrix model coincides with that obtained from the string solution.  

To connect our discussion above with previous work, we note that the string solution found in \cite{drukker2006small} is given by \eqref{eq:semistring global coords 1} when $\mathcal{J}\to \infty$ or $c=1$. The classical string we consider is a simple generalization of the $c=1$ string that preserves the same supersymmetries, as we check explicitly in Appendix~\ref{sec:semistring susy}. It is in turn a special case of the string considered in \cite{Gromov:2012eu} when the Wilson loop on the boundary does not have cusps in either the ambient contour or the internal contour at the locations of the two insertions.\footnote{Specifically, to match the string solution defined in \eqref{eq:semistring global coords 1} to the one in \cite{Gromov:2012eu}, the parameters $\ell_\theta$, $\ell_\phi$, $\kappa$ and $\gamma$ in Appendix E.1 of \cite{Gromov:2012eu} should be set to $\ell_\theta=\ell_\phi=0$ and $\kappa=\gamma=\frac{1}{c}$.}

\begin{figure}[t]
\centering
\begin{minipage}{0.49\hsize}
\centering
\includegraphics[clip, height=5cm]{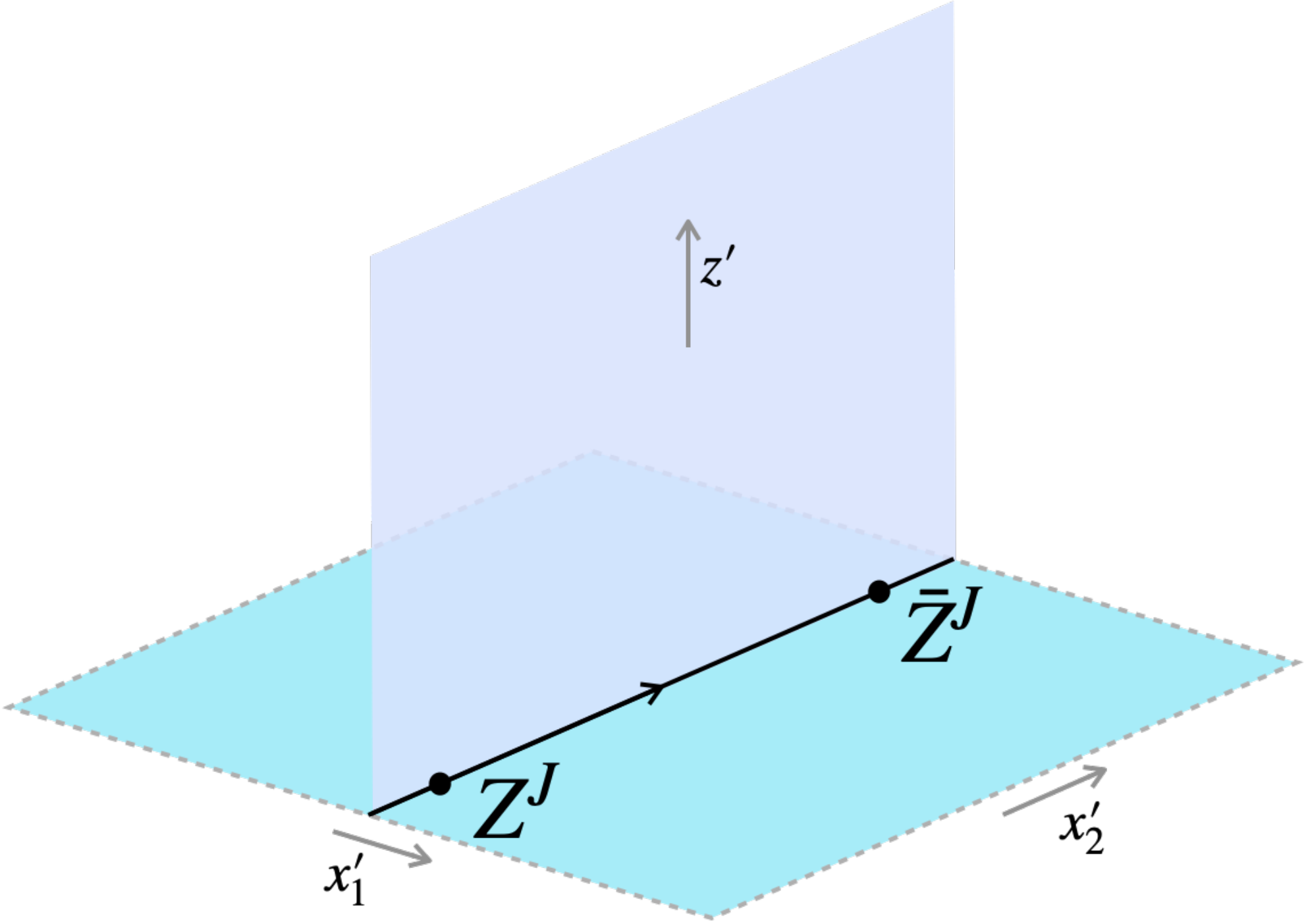}\\
{\bf a.} Half-plane
\end{minipage}
\begin{minipage}{0.49\hsize}
\centering
\vspace{1.3cm}
\includegraphics[clip, height=3.7cm]{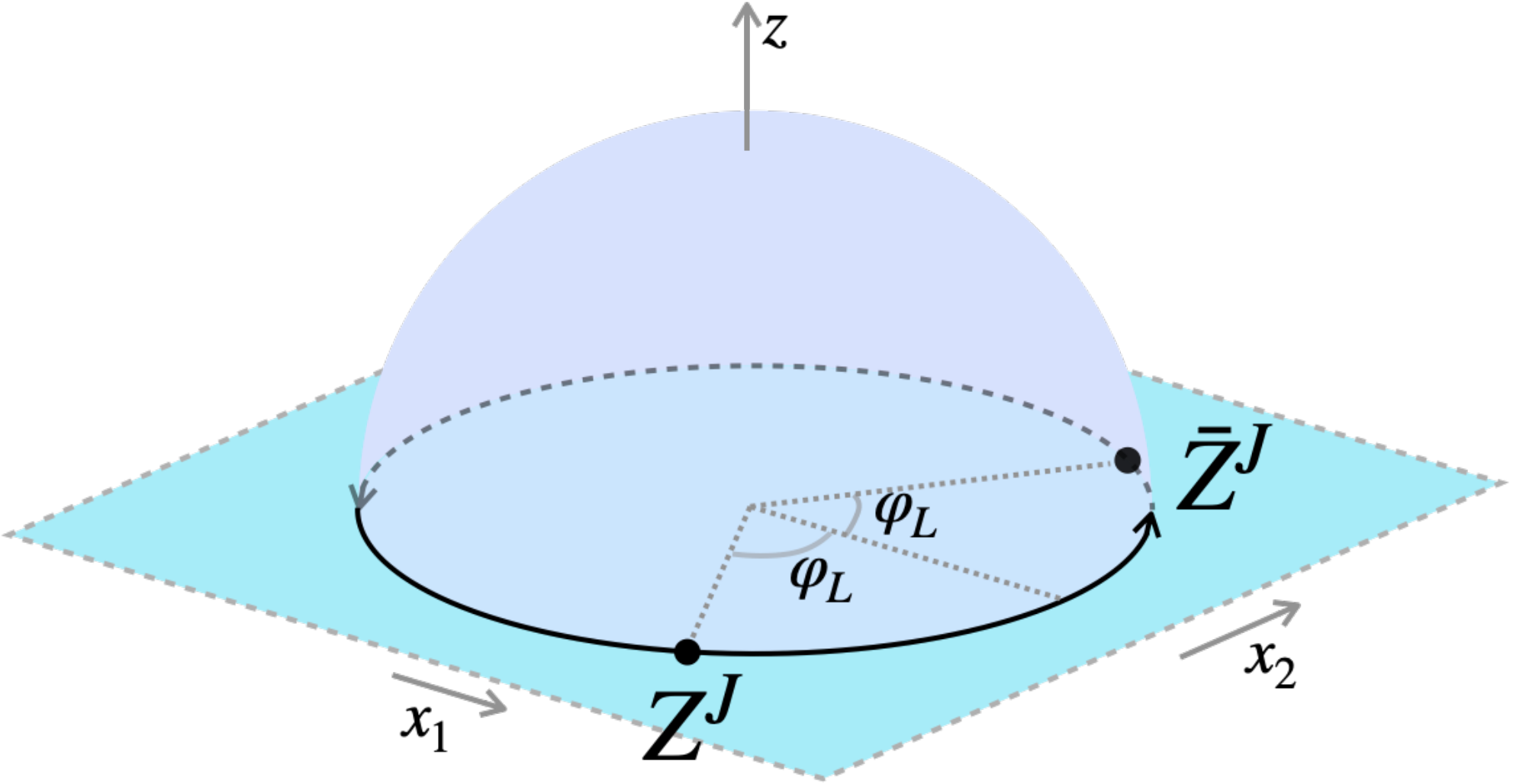}\\
{\bf b.} Hemisphere
\end{minipage}
\caption{The classical string solution in AdS global coordinates in Figure~\ref{fig:semiclassical string} can be mapped to: {\bf a.} a half-plane in Poincar\'e coordinates using \eqref{eq:global to Poincare}, in which case the string is dual to the Wilson line $x_1'=0$ with $Z^J$ and $\bar{Z}^J$ at $x_2'=\mp 1$; or to {\bf b.} the hemisphere in Poincar\'e coordinates using \eqref{eq:Poincare to Poincare}, in which case the classical string is dual to the Wilson loop $x_1^2+x_2^2=1$ with $Z^J$ and $\bar{Z}^J$ at $\varphi=\mp \varphi_L$.}
\label{fig:half-plane and hemisphere}
\end{figure}

To close this section, we transform the solution in \eqref{eq:semistring global coords 1} into a form more suitable for computing correlations functions on the Wilson loop. We begin by continuing from Lorentzian to Euclidean signature, letting $t=-i\tau$ where $\tau\in \mathbb{R}$ is Euclidean time. Then $\phi=\phi_0-i\tau$ in (\ref{eq:semistring global coords 1}), and the Euclidean action is $S_{\rm NG}=2g\int d^2\sigma \sqrt{h}$. The Euclidean worldsheet coordinates will be denoted by $\sigma^1=\rho$ and $\sigma^2\equiv i\sigma^0=\tau$. Secondly, as an intermediate step, we map the solid cylinder in global coordinates to the half-volume in Poincar\'e coordinates using the transformation
\begin{align}\label{eq:global to Poincare}
    z'&=\frac{1}{\cosh{\rho}\cosh{\tau}+\sinh{\rho}\cos{\psi}},&x_1'&=z'\sinh{\rho}\sin{\psi}, &x_2'&=z'\sinh{\tau}\cosh{\rho}.
\end{align}
This maps the Wilson contour to the line $x_1'=0$ on the boundary $z'=0$ and sends $Z^J$ to $x_2'=-1$ and $\bar{Z}^J$ to $x_2'=1$. It maps the dual string to the half-plane $x_1'=0$. See Figure~\ref{fig:half-plane and hemisphere} a. 

Next, we shift the string in the $x_1'$ direction by $a\equiv\cot\left(\frac{\varphi_L}{2}\right)$ and then invert, rescale and recenter to get the transformation
\begin{align}\label{eq:Poincare to Poincare}
    z&=\frac{2az'}{{z'}^2+(x_1'+a)^2+{x_2'}^2}, &x_1&=\frac{a^2-{x_1'}^2-{x_2'}^2-{z'}^2}{{z'}^2+(x_1'+a)^2+{x_2'}^2}, &x_2&=\frac{2ax_2'}{{z'}^2+(x_1'+a)^2+{x_2'}^2}.
\end{align}
This maps the Wilson contour to the circle $(x_1,x_2)=(\cos{\varphi},\sin{\varphi})$ on the boundary $z=0$ and sends $Z^J$ to $\varphi=-\varphi_L$ and $\bar{Z}^J$ to $\varphi=\varphi_L$. It maps the dual string to the hemisphere $x_1^2+x_2^2+z^2=1$. See Figure~\ref{fig:half-plane and hemisphere} b. 

Therefore, the composition of Eqs.~(\ref{eq:global to Poincare}) and (\ref{eq:Poincare to Poincare}) produces the classical string in EAdS$_3\times S^2\subset \text{EAdS}_5\times S^5$ that we can use to study the large charge correlator in Eq.~(\ref{eq:HH correlator}). The solution in these coordinates takes the explicit form
\begin{equation}
\begin{aligned}
&z=\frac{\sin{\varphi_L}}{\Delta},\quad x_1=\frac{\cos{\varphi_L}\cosh{\tau}\cosh{\rho}+\sinh{\rho}}{\Delta},\quad 
x_2=\frac{\sin{\varphi_L}\sinh{\tau}\cosh{\rho}}{\Delta},\\ &\sin{\theta}=\frac{c}{\cosh{\rho}}, \quad \phi=\phi_0-i\tau\,,
\quad \Delta=\cosh{\tau}\cosh{\rho}+\cos{\varphi_L}\sinh{\rho}.
\label{eq:semistring MY coords 1}
\end{aligned}
\end{equation}
When $c=1$, this solution is essentially equivalent to the one in \cite{Miwa2006HolographyOW}.

The string in \eqref{eq:semistring MY coords 1} is incident on $Z^J$ at $\varphi=-\varphi_L$ as $\tau\to -\infty$, on $\bar{Z}^J$ at $\varphi=\varphi_L$ as $\tau\to\infty$,  and on the unit circle as $\rho\to \pm \infty$. To approach the specific point $\varphi$ on the circle from the worldsheet, we fix $\tau$ and send $\rho\to \eta \infty$, where $\eta=+1$ for $\varphi\in[-\varphi_L,\varphi_L]$ and $\eta=-1$ otherwise and $\tau$ is determined by the pair of implicit equations
\begin{align}\label{eq:rho,tau to varphi}
   \frac{\cosh{\tau}\pm \sinh{\tau}}{\cosh{\tau}+\eta\cos{\varphi_L}}=2\sin^2\left(\frac{\varphi\pm \varphi_L}{2}\right)/\sin^2{\varphi_L}.
\end{align}
This follows from \eqref{eq:semistring MY coords 1} and we have expressed it in terms of the chordal distances from $\varphi$ to the insertions.

A note about notation: Going forward, we will continue to write $d^2\sigma\equiv d\rho d\tau$ for the measure on the worldsheet and use $\alpha$ and $\beta$ as free or summed worldsheet indices, but we will label particular components of worldsheet tensors using $\rho,\tau$ instead of $1,2$.

\subsection{Two-point function}\label{sec:two-pt from semistring}

The two-point correlator in \eqref{eq:HH correlator} can be computed using a semiclassical evaluation of the string path integral of the appropriate vertex operators weighted by the action:
\begin{align}\label{eq:string path integral}
    \dbraket{Z^J(-\varphi_L)\bar{Z}^J(\varphi_L)}&=\frac{\int D\Psi e^{-S[\Psi]}\text{ }v^J(-\varphi_L;\Psi) \bar{v}^J(\varphi_L;\Psi) }{\int D\Psi \text{ }e^{-S[\Psi]}}.
\end{align}
Here $\int D\Psi$ is the path integral over all the fields in the superstring sigma model, which are collectively denoted by $\Psi$. Furthermore, $v$ and $\bar{v}$ are the vertex operators dual to $Z$ and $\bar{Z}$, which are determined by sending the insertions of $\Theta_4\pm i \Theta_5=\sin{\theta}e^{\pm i\phi}$ to the boundary of the world sheet:
\begin{align}
    v(\varphi;\Psi)&\equiv 2g\lim_{\rho\to \eta\infty}\frac{\sin{\theta}e^{i\phi}}{z}, &
    \bar{v}(\varphi;\Psi)&\equiv2g\lim_{\rho\to \eta \infty}\frac{\sin{\theta}e^{-i\phi}}{z}.\label{eq:v and vbar}
\end{align}
The limits are taken in accordance with \eqref{eq:rho,tau to varphi}.
We have defined the vertex operators with the usual factor of $1/z$ (recall that the unit charge chiral primaries have unit conformal dimension) that appears in the AdS/CFT dictionary when extrapolating bulk points to the boundary. The vertex operators also include the natural string tension factor $2g=\frac{\sqrt{\lambda}}{2\pi}$.  And, finally, $S[\Psi]$ in \eqref{eq:string path integral} is the complete action of the string, which is the sum of the Nambu-Goto action and a boundary action:
\begin{align}\label{eq:total action}
    S[\Psi]\equiv S_{\rm NG}[\Psi]+S_{\rm bdy}[\Psi].
\end{align}
Identifying $S_{\rm bdy}[\Psi]$ is the focus of the next section.

In the large charge regime, the path integral in \eqref{eq:string path integral} is dominated by the classical string solution given in \eqref{eq:semistring MY coords 1}, which we shall schematically denote $\Psi_{\rm cl}(c,\phi_0,\varphi_L)$ (or $\Psi_{\rm cl}(c)$ when we only want to emphasize the $c$-dependence). The action and the vertex operators in \eqref{eq:string path integral} can therefore be evaluated on this solution. Let us adopt the shorthand
\begin{align}\label{eq:semiclassical c-string action}
    S_{\rm cl}(c)&\equiv S[\Psi_{\rm cl}(c,\phi_0,\varphi_L)]
\end{align}
for the classical action of the string, and
\begin{align}
    v_{\rm cl}(\varphi)&\equiv v(\varphi;\Psi_{\rm cl}(c,\phi_0,\varphi_L)), &\bar{v}_{\rm cl}(\varphi)&\equiv \bar{v}(\varphi;\Psi_{\rm cl}(c,\phi_0,\varphi_L))
\end{align}
for the vertex operators evaluated on the classical string solution. The action $S_{\rm cl}(c)$ does not depend on $\phi_0$ or $\varphi_L$, since changing $\phi_0$ is implemented by an isometry on $S^2$ and changing $\varphi_L$ is implemented by an isometry on EAdS$_3$. 
By contrast, the vertex operators, $v_{\rm cl}$ and $\bar{v}_{\rm cl}$, do depend on $c$, $\phi_0$ and $\varphi_L$ as well as on $\varphi$, but we drop the explicit dependence for ease of notation. The two-point function in the large charge limit therefore becomes
\begin{align}\label{eq:tJY12zm78k}
\dbraket{Z^J(-\varphi_L)\bar{Z}^J(\varphi_L)}&=\frac{1}{2\pi}\int_{0}^{2\pi}d\phi_0\text{ }v^J_{\text{cl}}(-\varphi_L) \bar{v}_{\text{cl}}^J(\varphi_L)e^{-S_{\rm cl}(c)+S_{\rm cl}(0)}.
\end{align}
The term $S_{\rm cl}(0)$ in the exponent comes from the denominator in \eqref{eq:string path integral}. To understand and apply \eqref{eq:tJY12zm78k}, we need to understand the origin of the integral over $\phi_0$, determine the classical vertex operators, $v_{\rm cl}$ and $\bar{v}_{\rm cl}$, and evaluate the classical action $S_{\rm cl}(c)-S_{\rm cl}(0)$.

We begin with the integral over $\phi_0$. Recall that moduli of classical solutions play an important role in relating the observables in semiclassical eigenstates of a quantum mechanical system to the observables along trajectories of the corresponding classical system. The quantum mechanical counterpart of a family of classical trajectories labelled by a modulus is a family of coherent states rather than eigenstates. Since a particular eigenstate can be realized by a suitable linear combination of the coherent states, one should suitably average the expectation values of the classical trajectories over the moduli to reproduce the quantum expectation value. See Sections 2-3 of \cite{Bajnok:2014sza} and also \cite{yang2021d} for discussions. In our case, the semiclassical string with a given $\phi_0$ is dual to a coherent state of average charge $J$, and integrating over $\phi_0$ in \eqref{eq:tJY12zm78k} ensures that the resulting expectation value on the LHS is for an eigenstate of charge $J$.

Next, we note the explicit form of the vertex operators on the classical string:
\begin{equation}
\begin{aligned}
    &v_{\rm cl}(\varphi)=2g\lim_{\rho \to \eta\infty}\frac{\Delta}{\sin{\varphi_L}}\frac{c}{\cosh{\rho}}e^{i\phi_0+\tau}=2gc e^{i\phi_0}\frac{d}{d(\varphi,\varphi_L)^2}\\
   & \bar{v}_{\rm cl}(\varphi)=2g\lim_{\rho\to \eta \infty}\frac{\Delta}{\sin{\varphi_L}}\frac{c}{\cosh{\rho}}e^{-i\phi_0-\tau}
=2gce^{-i\phi_0}\frac{d}{d(\varphi,-\varphi_L)^2}\,,
\label{eq:v cl}
\end{aligned}
\end{equation}
where $\Delta$ was defined in \eqref{eq:semistring MY coords 1}. Above we used \eqref{eq:rho,tau to varphi} to evaluate the limit in terms of $\varphi$ instead of $\tau$ and wrote the result in terms of the chordal distances introduced in \eqref{eq:chordal distance}. Thus, the contribution of the large charge vertex operators to the RHS of \eqref{eq:tJY12zm78k} is\footnote{One may also define the large charge vertex operators to be $v^J\equiv \lim_{\tau \to -\infty}(2g\sin{\theta}e^{i\phi}/z)^J$ and $\bar{v}^J\equiv \lim_{\tau \to \infty}(2g\sin{\theta}e^{-i\phi}/z)^J$, so that we send $\tau \to \mp\infty$ without sending $\rho\to \pm \infty$ first. The order of limits does not matter and the resulting classical vertex operators reproduce $v_{\rm cl}^J(-\varphi_L)$ and $\bar{v}_{\rm cl}^J(\varphi_L)$, as given by \eqref{eq:v cl}.}
\begin{align}\label{eq:ikbvY1zQYE}
    v^J_{\rm cl}(-\varphi_L)\bar{v}^J_{\rm cl}(\varphi_L)&=\frac{(2gc)^{2J}}{d^{2J}}.
\end{align}
This has the correct position dependence for a conformal two-point correlator, as in \eqref{eq:HH correlator}. Furthermore, the factors of $e^{i\phi_0}$ and $e^{-i\phi_0}$ from the two vertex operators evidently cancel and the integral over $\phi_0$ in \eqref{eq:tJY12zm78k} simply yields $1$. Stated differently, given the explicit forms of the vertex operators in \eqref{eq:v cl}, the integral over $\phi_0$ ensures that the RHS of \eqref{eq:tJY12zm78k} is nonzero only if equal numbers of $v_{\rm cl}$ and $\bar{v}_{\rm cl}$ are inserted on the boundary of the worldsheet. This mirrors the requirement from R-symmetry that the LHS is nonzero only if equal numbers of $Z$ and $\bar{Z}$ are inserted along the Wilson loop. The role of the integration over $\phi_0$ will be slightly less trivial when we consider higher-point functions below.

The last step in evaluating \eqref{eq:tJY12zm78k} is to determine the value of the action on the classical string solution, $S_{\rm cl}(c)$. For simplicity we compute the action for the case of antipodal insertions, which is equal to the action for the case with general insertions. Setting $\varphi_L=\pi/2$, the form of the string in EAdS$_3$ simplifies to
\begin{align}\label{eq:semistring MY antipodal}
    z&=\frac{1}{\cosh{\tau}\cosh{\rho}}, &x_1&=\frac{\tanh{\rho}}{\cosh{\tau}}, &x_2&=\tanh{\tau}.
\end{align}
We also note the explicit form of the metric induced on the worldsheet,
\begin{align}
    h_{\rho\rho}&=\frac{\cosh^4{\rho}-c^2}{\cosh^4{\rho}-c^2\cosh^2{\rho}},& h_{\tau\tau}&=\cosh^2{\rho}-\frac{c^2}{\cosh^2{\rho}}, &h_{\rho\tau}=h_{\tau\rho}=0.
\end{align}
This follows equally from \eqref{eq:semistring MY antipodal}, \eqref{eq:semistring MY coords 1} or the Euclidean continuation of \eqref{eq:semistring global coords 1}.

\paragraph{Computing the action.}

As is familiar from the study of classical strings dual to Wilson loops without insertions \cite{Maldacena:1998im,Rey:1998ik,drukker1999wilson,erickson2000wilson}, the string action is not simply the Nambu-Goto or Polyakov action since the area of the minimal surface diverges as it approaches the Wilson contour on the boundary. One possible remedy in that context is to regularize the area by shifting the contour from $z=0$ to $z=\epsilon$, remove the divergent piece from the area that is proportional to the length of the contour as $\epsilon\to 0^+$, and interpret it as a renormalization of the mass of the particle moving around the loop \cite{Maldacena:1998im,Rey:1998ik}. An alternative approach is to take the Legendre transform of the Nambu-Goto action with respect to the bulk radial coordinate \cite{drukker1999wilson}. This adds a boundary term to the string action that does not change the equations of motion or the extremal surface but does change the value of the action and makes it well-defined (and, when applicable, in agreement with gauge theory). We adopt this second approach for the computation of the action of the classical string dual to the Wilson loop with insertions, but must modify the precise prescription in \cite{drukker1999wilson} to reflect the modified boundary conditions of the string arising from the insertions.

Because of its important role in what follows\footnote{There also appears to be a subtle issue with a minus sign in the original discussion \cite{drukker1999wilson}. We thank Nadav Drukker for correspondence on this detail.} let us briefly review the boundary term proposed in \cite{drukker1999wilson} for the classical string dual to a smooth Wilson loop without insertions. We call this the DGO boundary term, in reference to the authors' initials. We begin by combining the inverse bulk coordinate $u\equiv 1/z$ with the embedding coordinates $\Theta_I$ for $S^5$ into $u_I\equiv u \Theta_I$. This satisfies $u_Iu_I=u^2$. In these coordinates, the metric on EAdS$_5\times S^5$ can be written
\begin{align}
    ds^2&=u^2 dx^\mu dx_\mu + \frac{du^2}{u^2}+d\Theta_Id\Theta_I=u^2 dx^\mu dx_\mu + \frac{du_I du_I}{u^2}.
\end{align}
Then the induced metric on the worldsheet is $h_{\alpha\beta}=u^2 \partial_\alpha x^\mu \partial_\beta x_\mu+u^{-2}\partial_\alpha u_I \partial_\beta u_I$, and we define the canonical momenta conjugate to $u_I$ and $u$ to be
\begin{align}
    \Pi_I^\alpha &\equiv 2g\frac{\partial \sqrt{h}}{\partial (\partial_\alpha u_I)}=\frac{2g}{u^2}\sqrt{h}h^{\alpha\beta}\partial_\beta u_I,
    &\Pi_u^\alpha &\equiv 2g\frac{\partial \sqrt{h}}{\partial (\partial_\alpha u)}=\frac{2g}{u^2}\sqrt{h}h^{\alpha\beta}\partial_\beta u.
\end{align}
In this language, the DGO prescription is to take the Legendre transform with respect to $u$ or, equivalently, with respect to the $u_I$ by adding to the Nambu-Goto action the following boundary action:
\begin{align}\label{eq:DGO bdy}
    S_{\rm DGO}[\Psi]&=-\int d^2\sigma \partial_\alpha (u \Pi_u^\alpha)=-\int d^2\sigma \partial_\alpha (u_I \Pi_I^\alpha ).
\end{align}
The variation of the Nambu-Goto action plus the DGO action under $u_I\to u_I+\delta u_I$ about a classical solution is then
\begin{align}\label{eq:NG+DGO boundary variation}
    \delta S_{\rm NG}[\Psi_{\rm classical}]+ \delta S_{\rm DGO}[\Psi_{\rm classical}]=-\int d^2\sigma \partial_\alpha \left(u_I \delta \Pi^\alpha_I\right).
\end{align}
For the variation to be zero, we must send $\delta \Pi_I^\alpha\to 0$ at the boundary of the worldsheet, which means the complete action is to be viewed as a functional of the momentum $\Pi_I$ instead of the coordinates $u_I$ at the boundary. The DGO boundary term can therefore be thought of as imposing Neumann, rather than Dirichlet, boundary conditions on $u_I$.\footnote{This in turn is equivalent to imposing Dirichlet conditions on the $\Theta^I$, see \cite{drukker1999wilson}, which is just the familiar statement that the coupling to the scalars in the Wilson loop sets the boundary values of the $S^5$ coordinates of the dual string.} 

As a simple check of \eqref{eq:DGO bdy}, we compute the action of the string with $c=0$, which is dual to the half-BPS Wilson loop without insertions. In this case $\sqrt{h}=\cosh{\rho}$, $\partial_\rho(u\Pi^\rho_u)=2g\cosh{\rho}$, and $\partial_\tau(u\Pi^\tau_u)=2g\sech^2{\tau}\sech{\rho}$, which reproduces the well known result
\begin{align}\label{eq:action without insertions}
    S_{\rm NG}[\Psi_{\rm cl}(0)]+ S_{\rm DGO}[\Psi_{\rm cl}(0)]=-2g \int d\tau d\rho \sech{\rho}\sech^2{\tau}=-4\pi g.
\end{align}
This is just the string tension, $2g$, times the regularized area of EAdS$_2$, $-2\pi$, and matches the leading behavior of the gauge theory result in the supergravity regime given in \eqref{eq:half-BPS WL VEV}.

Our characterization of the boundary term in \eqref{eq:DGO bdy} closely follows the one in \cite{drukker1999wilson} except that the original discussion was framed in terms of $z$ and $Y_I\equiv z\Theta_I$, and the momenta $\Pi^\alpha_z\equiv \frac{2g}{z^2}\sqrt{h}h^{\alpha\beta}\partial_\beta z$ and  $\Pi^\alpha_{Y_I}\equiv \frac{2g}{z^2}\sqrt{h}h^{\alpha\beta}\partial_\beta Y_I$. But since $z\Pi_z^\alpha=-u\Pi^\alpha_u$ and $Y_I\Pi^\alpha_{Y_I}=-u_I\Pi^\alpha_I$, the DGO boundary term expressed in terms of the $z$ and $Y_I$ coordinates does not have the minus sign in \eqref{eq:DGO bdy}. Therefore, in contrast with \eqref{eq:NG+DGO boundary variation}, the variation about a classical solution of the Nambu-Goto action plus DGO boundary term under $Y^I\to Y^I+\delta Y^I$ does not vanish when the momenta $\Pi^\alpha_{Y_I}$ conjugate to $Y^I$ are fixed at the boundary. For this reason, the DGO term should be viewed as a Legendre transform in $u$, not $z$, or in $u_I$, not $Y^I$. 

Finally, we are ready to propose a precise form for $S_{\rm bdy}[\Psi]$ and evaluate the action. We begin with a few motivating observations. First, the vertex operators in \eqref{eq:v and vbar} are naturally written in terms of $u_4$ and $u_5$ as $v\equiv 2g\lim_{\rho\to \eta \infty} (u_4+iu_5)$ and $\bar{v}\equiv 2g \lim_{\rho\to \eta \infty} (u_4-iu_5)$. Secondly, the limiting behavior of $u_4\pm iu_5$ evaluated on the classical solution, \eqref{eq:semistring MY antipodal}, as the string approaches the insertions is
\begin{align}\label{eq:SALaUlK8RH}
    \lim_{\tau\to -\infty} u_4+iu_5 &= \frac{ce^{i\phi_0}}{2}, &
    \lim_{\tau\to \infty} u_4-iu_5 &= \frac{c e^{-i\phi_0}}{2}.
\end{align}
These limits are taken with $\rho$ fixed but general. Finally, the limiting behavior of the momenta conjugate to $u_4\pm iu_5$, $\Pi^\alpha_{4\pm i 5}\equiv  \frac{1}{2}(\Pi^\alpha_4 \mp i \Pi^\alpha_5)$, evaluated on the classical solution is
\begin{align}\label{eq:WLZy2rz2F1}
    \lim_{\tau \to -\infty}\Pi_{4-i5}&=0, &\lim_{\tau \to \infty}\Pi_{4+i5}&=0.
\end{align}
By contrast, $u_4+iu_5$ and $u_4-iu_5$ diverge as $\tau\to \infty$ and $\tau\to -\infty$, respectively and $\Pi_{4-i5}$ and $\Pi_{4+i5}$ approach non-trivial functions of $\rho$ as $\tau\to \infty$ and $\tau\to -\infty$, respectively.\footnote{Specifically, $\lim_{\tau\to \infty}\Pi_{4-i5}^\alpha=-\lim_{\tau\to -\infty}\Pi_{4+i5}^\alpha=4gc/\cosh^2{\rho}/\sqrt{\cosh^2{\rho}-c^2}\delta^{\tau\alpha}$.}

Eqs.~(\ref{eq:SALaUlK8RH}) and (\ref{eq:WLZy2rz2F1}) indicate that, as compared to the general prescription in \cite{drukker1999wilson}, inserting $Z^J$ and $\bar{Z}^J$ on the Wilson loop changes the boundary condition of $u_4+iu_5$ at $\tau\to -\infty$ and of $u_4-iu_5$ at $\tau\to \infty$ from Neumann to Dirichlet. We therefore propose that the correct boundary action is the DGO action in \eqref{eq:DGO bdy}, except with the Legendre transformation on $u_4\pm iu_5$ at $\tau=\mp \infty$ removed:
\begin{align}\label{eq:Sbdy}
    S_{\rm bdy}[\Psi]&\equiv S_{\rm DGO}[\Psi]-\int d\rho \lim_{\tau\to -\infty} (u_4+iu_5) \Pi^\tau_{4+i5}+\int d\rho \lim_{\tau\to \infty}(u_4-iu_5)\Pi^\tau_{4-i5}.  
\end{align}
Like the Nambu-Goto action, this boundary action is not finite by itself. The bulk and boundary actions should be added as in \eqref{eq:total action} with IR cutoffs imposed on both $\rho$ and $\tau$, and the sum then approaches a finite value as the cutoffs are removed. 

The evaluation of the complete action is now simple, especially using the coordinates in \eqref{eq:semistring MY antipodal}. Firstly, we note
\begin{align}\label{eq:PQfmteL6tD}
(u_4+iu_5)\Pi^\tau_{4+i5}&=\frac{\tanh{\tau}-1}{2}\Pi_\phi,&(u_4-iu_5)\Pi^\tau_{4-i5}&=\frac{\tanh{\tau}+1}{2}\Pi_\phi,
\end{align}
where $\Pi_\phi$ is the angular momentum density introduced in \eqref{eq:s4VcJNwmrV}. The second two terms in \eqref{eq:Sbdy} therefore evaluate to 
\begin{align}\label{eq:Dirichlet bdy terms}
    -\int d\rho \lim_{\tau\to -\infty} (u_4+iu_5) \Pi^\tau_{4+i5}=\int d\rho \lim_{\tau\to \infty}(u_4-iu_5)\Pi^\tau_{4-i5}=\int d\rho \Pi_\phi&=J.
\end{align}
Furthermore, one can check that the $\rho$ term in the DGO Lagrangian satisfies $\partial_\rho(u \Pi^\rho_u)=2g\sqrt{h}$ and therefore cancels with the Nambu-Goto Lagrangian, while the $\tau$ term simplifies to $\partial_\tau(u\Pi^\tau_u)=2g\sech^2{\tau}/\sqrt{\cosh^2{\rho}-c^2}$. Therefore,
\begin{align}\label{eq:semistring NG+DGO action}
    S_{\rm NG}[\Psi_{\rm cl}(c)]+S_{\rm DGO}[\Psi_{\rm cl}(c)]&=-2g \int_{-\infty}^\infty \frac{d\tau}{\cosh^2{\tau}}\int_{-\infty}^\infty \frac{d\rho}{\sqrt{\cosh^2{\rho}-c^2}}\nonumber\\&=-8g \ellK(c^2).
\end{align}
Combined with \eqref{eq:Dirichlet bdy terms} and recalling \eqref{eq:s4VcJNwmrV}, we determine the complete action of the classical string to be
\begin{align}\label{eq:JHXsNz95Yl}
    S_{\rm cl}(c)=S_{\rm NG}+S_{\rm bdy}= -8g \ellK(c^2)+2J=-8g\ellE(c^2).
\end{align}
Then, from \eqref{eq:tJY12zm78k} and including the contribution of the vertex operators \eqref{eq:ikbvY1zQYE}, we obtain the two-point function
\begin{align}\label{eq:final HH 2-pt}
\dbraket{Z^J(-\varphi_L)\bar{Z}^J(\varphi_L)}&=\frac{(2gc)^{2J}}{d^{2J}}e^{8g\ellE(c^2)-8g\ellE(0)}.
\end{align}
Note that if we put $Z^J$ and $\bar{Z}^J$ in the topological configuration by sending $\varphi_L\to \pi/2$, as in \eqref{eq:topological large charges}, we find
\begin{align}\label{eq:HH topological final}
    \dbraket{\tilde{\Phi}^J\tilde{\Phi}^J}&=(-g^2c^2)^Je^{8g \ellE(c^2)-8g\ellE(0)}.
\end{align}
This is in precise agreement with the localization result in \eqref{eq:finaltop2}. 

We should emphasize that although we have for simplicity written the boundary action \eqref{eq:Sbdy} using the global coordinates $\rho$ and $\tau$, the complete action is independent of the choice of worldsheet coordinates and IR regularization.\footnote{This can be seen as follows. Firstly, the sum of the Nambu-Goto and DGO actions is manifestly worldsheet reparametrization invariant and one can also readily check that it is finite without need for regularization. Secondly, the first boundary term in \eqref{eq:Sbdy} can be written in a coordinate independent way as follows (the treatment of the second boundary term is analogous):
\begin{align}\label{eq:coord invariant boundary term}
    -\int d\rho \lim_{\tau \to -\infty} (u_4+iu_5)\Pi^\tau_{4+i5}=-\lim_{\Lambda\to \infty}\int_{\eta_\Lambda} d\lambda \sqrt{\gamma} n_\alpha P^\alpha.
\end{align}
Here, $\eta_\Lambda$ is a family of curves that approach $Z^J$ as $\Lambda\to \infty$, $\lambda$ is a coordinate along $\eta_\Lambda$ that we may for concreteness take to run from $-1$ to $1$, $\gamma_{\alpha\beta}\equiv\frac{d\eta^\alpha}{d\lambda}\frac{d\eta^\beta}{d\lambda}h_{\alpha\beta}$ is the induced metric on $\eta_\Lambda$, $n_\alpha$ is a unit normal vector (i.e., $\frac{d\eta^\alpha}{d\lambda}n_\alpha=0$ and $n_{\alpha}n_\beta h^{\alpha\beta}=1$), and finally $P^\alpha\equiv (u_4+iu_5)\Pi^\alpha_{4+i5}/\sqrt{h}$. The point, then, is that the limit on the RHS of \eqref{eq:coord invariant boundary term} is independent of the particular choice of $\eta_\Lambda(\lambda)\equiv(\rho_\Lambda(\lambda),\tau_\Lambda(\lambda))$ as long as $\rho_\Lambda(+1)\to \infty$, $\rho_\Lambda(-1)\to -\infty$ and $\text{max}_{\lambda\in[-1,1]}\tau_\Lambda(\lambda)\to -\infty$ as $\Lambda\to \infty$, which is what we mean when we say that $\eta_\Lambda$ ``approaches $Z^J$'' as $\Lambda\to \infty$.}

Before proceeding to the computation of the higher-point functions, we close this section with some comments about the evaluation of the string action. We begin by noting, since $d\ellE(c^2)/dc=(\ellK(c^2)-\ellE(c^2))/c$, that the two-point function is extremized with respect to $c$: 
\begin{align}\label{eq:c-extremization}
    \frac{\partial}{\partial c}\left(2J\log{c}+S_{\rm cl}(c)\right)\biggr\rvert_{c=c(J)}=0.
\end{align}
Here, we treat $c$ and $J$ as independent variables when taking the partial derivative, and $c(J)$ is the particular value of $c$ satisfying \eqref{eq:s4VcJNwmrV}. If we view the vertex operators in \eqref{eq:string path integral} as additional boundary terms in a ``total'' action, 
\begin{align}
    S_{\rm tot}[\Psi]&\equiv -J\log \left(v(-\varphi_L;\Psi)\bar{v}(\varphi_L;\Psi)\right)+S[\Psi],
\end{align}
then \eqref{eq:c-extremization} is equivalent to 
\begin{align}\label{eq:c-extremization 2}
    \frac{\partial S_{\rm tot}[\Psi_{\rm cl}(c)]}{\partial c}\biggr\rvert_{c=c(J)}=0\period
\end{align}
This tells us that the total action is invariant when we vary $\theta$ by $\delta\theta=\delta c/\sqrt{\cosh^2{\rho}-c(J)^2}$ about the classical string solution, which is equivalent to varying $c$ by $\delta c$ about $c(J)$, as we would expect from a purportedly extremal solution. By contrast, the Nambu-Goto action plus the DGO action without the additional two terms in \eqref{eq:Sbdy}, which evaluates to \eqref{eq:semistring NG+DGO action} on the classical string, does not satisfy this extremization property. Therefore, we may say that modifying the boundary conditions of $u_4+iu_5$ and $u_4-iu_5$ to be Dirichlet instead of Neumann at $\tau\to -\infty$ and $\tau\to \infty$, respectively--- and implementing the necessary Legendre transforms with the boundary action given in \eqref{eq:Sbdy}--- fixes the $\theta$ variational problem, which would otherwise not be well defined.\footnote{Incidentally, \eqref{eq:c-extremization} uniquely determines $S_{\rm cl}(c)$. It allows to completely bypass the detailed discussion of the boundary action and uses only the fact that the $c$ dependence of the vertex operators goes like $c^{2J}$ and that the variation of $S_{\rm tot}[\Psi]$ about the classical string solution should vanish, including when we vary $\theta$ by varying $c$.} 

Requiring the $\theta$ variational problem to be well defined does not uniquely determine $S_{\rm bdy}[\Psi]$. For example, the boundary terms in the DGO action implementing the Legendre transforms of $u_1$, $u_2$, and $u_3$ at $\tau\to \pm \infty$ are all zero because $\Theta_1=\Theta_2=\Theta_3=0$ on the string solution. And given \eqref{eq:PQfmteL6tD}, the same is true for the two terms implementing the Legendre transforms of $u_4+iu_5$ at $\tau\to \infty$ and of $u_4-iu_5$ at $\tau\to -\infty$. Any of these terms may therefore be dropped without changing the value of the action. Thus, our findings do not rule out the possibility that the correct boundary prescription is to also impose Dirichlet conditions on $u_1$, $u_2$, and $u_3$ at $\tau \to \pm \infty$, and/or perhaps on $u_4+iu_5$ at $\tau\to\infty$ and $u_4-iu_5$ at $\tau\to -\infty$. However, we consider the boundary conditions we identified and the boundary action given in \eqref{eq:Sbdy} to be the most plausible. Our proposal could be further tested by applying it to a more general string solution in which the behavior of the $S^5$ directions orthogonal to the large charge directions is not trivial. One possibility is to study the semiclassical string dual to the quarter-BPS Wilson loop \cite{Drukker:2006ga} with $Z^J$ and $\bar{Z}^J$ additionally inserted.

Next, it is instructive to compare the behavior of \eqref{eq:final HH 2-pt} in the limit $\mathcal{J}\to \infty$ to the results in \cite{Miwa2006HolographyOW}.  Noting that $2\mathcal{J}\log{c}\to 0$ and $\ellE(c^2)\to 1$ as $\mathcal{J}\to \infty$, we find that in the large $\mathcal{J}$ limit
\begin{align}\label{eq:large J/g limit}
\dbraket{Z^J(-\varphi_L)\bar{Z}^J(\varphi_L)}&=\frac{(2g)^{2J}}{d^{2J}}\text{exp}\left[g\left(8-4\pi  + O(\mathcal{J}e^{-\frac{\mathcal{J}}{2}})\right)+O(g^0)\right].
\end{align}
Eq. (18) in \cite{Miwa2006HolographyOW} differs from our result above by an extra $2J$ in the exponent.\footnote{Specifically, Eq. (18) in \cite{Miwa2006HolographyOW} gives the following value for the total action (including the contributions from the vertex operators):
\begin{align}\label{eq:MW action}
    S_{\rm tot}&=2J\log{\frac{2\ell}{\epsilon}}-2J-\frac{2R^2}{\pi \alpha'}.
\end{align}
Identifying $2\ell\to d$ as the distance between the insertions, $\frac{R^2}{2\pi \alpha'}\to 2g$ as the string tension and $\epsilon$ as an IR regulator as $z\to 0$ that \cite{Miwa2006HolographyOW} imposed, we find that the above expression differs from the negative log of \eqref{eq:large J/g limit} by $2J+4\pi g+2J\log{\frac{\epsilon}{2g}}$. The $4\pi g$ and $2J\log{\frac{\epsilon}{2g}}$ terms in the difference arise because, unlike \cite{Miwa2006HolographyOW}, we normalize the Wilson loop with insertions by the Wilson loop without insertions (see Eqs.~(\ref{eq:double bracket}) and (\ref{eq:string path integral})) and we include a factor of $2g/z$ in the vertex operators as we approach the boundary (see \eqref{eq:v and vbar}). Accounting for these differences in convention, the substantive difference between the two results is therefore simply $2J$.} This is because the analysis in \cite{Miwa2006HolographyOW} used the DGO boundary action in \eqref{eq:DGO bdy} instead of the modified boundary action in \eqref{eq:Sbdy}. We have seen that this difference in choice of boundary term indeed changes the action by $2J$.

Finally, it is interesting to compare our analysis of the classical string dual to large charges $Z^J$ and $\bar{Z}^J$ inserted \textit{on} the Wilson loop to the string calculation of the correlation function of the Wilson loop and a large charge single-trace local operator ${\rm tr} Z^J$ inserted \textit{away from} the Wilson loop \cite{Zarembo:2002ph, Enari:2012pq, giombi2013correlators}. The classical string solution in that case is topologically a cone whose boundary is the disjoint union of a point incident on the local insertion and a circle incident on the Wilson loop. Our analysis is most directly similar to the one in \cite{Enari:2012pq}, where the action is also computed with a DGO boundary term added to the Nambu-Goto action.\footnote{\cite{Zarembo:2002ph,giombi2013correlators} used different, equivalent methods to regulate the semiclassical action.} In that paper, the DGO term is written not as a bulk integral of a total derivative like in \eqref{eq:DGO bdy}, which would give contributions from both boundaries, but rather as a boundary integral over only the boundary incident on the Wilson loop. Given our discussion above, we interpret this to mean that the $u_I$ coordinates satisfy Neumann conditions on the boundary incident on the Wilson loop, as in the DGO prescription, but Dirichlet conditions on the boundary incident on the local insertion.\footnote{More precisely, we would say that the combination of the $u_I$ coordinates dual to $Z$ satisfies the Dirichlet condition on the insertion. The other linear combinations should still satisfy Neumann conditions but the boundary terms implementing their Legendre transforms evaluate to zero and can effectively be dropped.} Moreover, if we instead write the DGO term in \cite{Enari:2012pq} as a bulk integral of a total derivative or equivalently add the DGO term also for the boundary incident on the local operator, then we find the action changes by $J$, just like in \eqref{eq:Dirichlet bdy terms}. Thus, the semiclassical string analysis of $Z^J$ and $\bar{Z}^J$ inserted on the Wilson loop is very similar to the analysis of ${\rm tr}Z^J$ inserted away from the Wilson loop, except that splitting the boundary of the string worldsheet into a part incident on the local insertions and a part incident on the Wilson loop proper is more subtle in the former case.

\subsection{Higher-point functions}\label{sec:higher-pt from semistring}

The procedure we used to compute the large charge limit of the two-point function can be straightforwardly extended to higher-point functions. For every extra insertion of $Z^\ell(\varphi)$ in the Wilson loop correlator, we simply add an extra factor of $v_{\rm cl}^\ell(\varphi)$ in the integral over $\phi_0$ in \eqref{eq:tJY12zm78k}, and likewise for $\bar{Z}^\ell(\varphi)$ and $\bar{v}^\ell_{\rm cl}(\varphi)$. Thus, in the large charge limit,
\begin{align}\label{eq:HH LLL correlator}
    &\frac{\dbraket{Z^J(-\varphi_L)\bar{Z}^J(\varphi_L)\prod_{i=1}^n Z^{\ell_i}(\varphi_i)\prod_{j=1}^{m}\bar{Z}^{\bar{\ell}_j}(\bar{\varphi}_j)}}{\dbraket{Z^J(-\varphi_L)\bar{Z}^J(\varphi_L)}}=\frac{1}{2\pi}\int_0^{2\pi}d\phi_0\prod_{i=1}^n v_{\rm cl}^{\ell_i}(\varphi_i)\prod_{j=1}^{m} \bar{v}_{\rm cl}^{\bar{\ell}_j}(\bar{\varphi}_j)\nonumber\\&=\frac{(2gc)^{2\ell_{\rm tot}}d^{2\ell_{\rm tot}}}{\prod_{i=1}^n d(\varphi_i,\varphi_L)^{2\ell_i} \prod_{j=1}^m d(\bar{\varphi}_j,-\varphi_L)^{2\bar{\ell}_j}}\delta_{\ell_{\rm tot},\bar{\ell}_{\rm tot}}.
\end{align}
Here, we let $\ell_{\rm tot}\equiv \sum_{i=1}^n \ell_i$ and $\bar{\ell}_{\rm tot}\equiv \sum_{j=1}^m \bar{\ell}_j$.

We could also be interested in higher-point functions in which the light operators are not just finite powers of $Z$ and $\bar{Z}$. However, the orthogonal scalars $\Phi_1$, $\Phi_2$ and $\Phi_3$ are effectively ``invisible'' in the large charge limit, as is clear from the perspective of the dual string: since $\Theta_i=0$ for $i=1,2,3$ on the classical string solution, the corresponding classical vertex operators, $\lim_{\rm bulk\rightarrow bdy}\Theta_i/z$, are trivially zero. This means that correlators involving light operators of the form $(\epsilon\cdot \Phi)^\ell$, where $\epsilon\cdot \Phi=\epsilon_1\Phi_1+\epsilon_2\Phi_2+\epsilon_3 \Phi_3+\epsilon_Z Z+\epsilon_{\bar{Z}}\bar{Z}$, reduce to linear combinations of the correlators in \eqref{eq:HH LLL correlator} after the truncation $\epsilon\cdot \Phi\rightarrow \epsilon_Z Z+\epsilon_{\bar{Z}}\bar{Z}$. Alternatively, we can compute correlators of the more general light operators directly using the classical string by replacing $\epsilon\cdot \Phi$ in the Wilson loop correlator with $\epsilon_Z v_{\rm cl}+\epsilon_{\bar{Z}}\bar{v}_{\rm cl}$ in the integral over $\phi_0$. (The correlators resulting from such truncation may of course be zero, like when $\epsilon_Z=\epsilon_{\bar{Z}}=0$, in which case we need to go beyond the classical analysis to determine the leading non-trivial contribution. For instance, to determine the four-point correlator $\dbraket{Z^J\bar{Z}^J \Phi_1\Phi_1}$, we need to study fluctuations of the semiclassical string, which will be discussed in \cite{giombi2021}.) 

As a special case, we will compute the higher-point topological correlator in \eqref{eq:topo higher-point}. Firstly, we again set $\varphi_L=\pi/2$ to put the large charges in the topological configuration. The light vertex operators then simplify to $v_{\rm cl}(\varphi)=2gce^{i\phi_0}/(1-\sin{\varphi})$ and $\bar{v}_{\rm cl}(\varphi)=2gce^{-i\phi_0}/(1+\sin{\varphi})$.  Furthermore, since $\tilde{\Phi}^\ell(\varphi)=\left(\cos{\varphi}\Phi_3+\left(\frac{1-\sin{\varphi}}{2}\right)Z-\left(\frac{1+\sin{\varphi}}{2}\right)\bar{Z}\right)^\ell$, the light vertex operators for the topological operators on the classical string are manifestly position independent:
\begin{align}\label{eq:topologicla vertex op}
    \tilde{\Phi}^\ell\to \left(\frac{1-\sin{\varphi}}{2}v_{\rm cl}(\varphi)+\frac{1+\sin{\varphi}}{2}\bar{v}_{\rm cl}(\varphi)\right)^\ell=(gc)^\ell\left(e^{i\phi_0}-e^{-i\phi_0}\right)^\ell.
\end{align}
Therefore, \eqref{eq:topo higher-point} becomes
\begin{align}
    \frac{\dbraket{\tilde{\Phi}^J\tilde{\Phi}^J\prod_{i=1}^n \tilde{\Phi}^{\ell_i}}}{\dbraket{\tilde{\Phi}^J\tilde{\Phi}^J}}&=\frac{(gc)^{\ell_{\rm tot}}}{2\pi}\int_0^{2\pi}d\phi_0 (e^{i\phi_0}-e^{-i\phi_0})^{\ell_{\rm tot}}.
\end{align}
Here, we again have defined $\ell_{\rm tot}\equiv \sum_{i=1}^n \ell_i$. If we expand the integrand using the binomial theorem, the only term that does not integrate to zero has equal numbers of $e^{i\phi_0}$ and $e^{-i\phi_0}$. Thus, the integral counts the number of ways of grouping $\ell_{\rm tot}$ objects into two equal halves, and the correlator simplifies to
\begin{align}\label{eq:17AVeb3vDx}
    \frac{\dbraket{\tilde{\Phi}^J\tilde{\Phi}^J\prod_{i=1}^n \tilde{\Phi}^{\ell_i}}}{\dbraket{\tilde{\Phi}^J\tilde{\Phi}^J}}&=(-g^2c^2)^{\ell_{\rm tot}/2}\binom{\ell_{\rm tot}}{\ell_{\rm tot}/2}.
\end{align}
This matches the localization result in \eqref{eq:topo higher-point from matrix model}.

We can also extend the analysis to correlators involving two large operators with unequal charges $J$ and $J+\ell$, where $\ell$ is held fixed in the large charge limit. The string path integral is dominated by the same saddle point, $\Psi\to \Psi_{\rm cl}(c(J))$, and therefore the classical action and the vertex operators are the same as before.\footnote{We could also alternatively localize the path integral to $\Psi\to \Psi_{\rm cl}(c(J'))$ for any $J'=J+O(g^0)$, including $J'=J+\ell$. The resulting string correlators, now with vertex operators $v_{\rm cl},\bar{v}_{\rm cl}\propto c(J')$ and action $S_{\rm cl}(c(J'))$, have the same leading behavior in the large charge limit. This follows from \eqref{eq:c-extremization 2}. The analogous statement for the one-dimensional Laplace integral $\int e^{-Nf(x)}g(x)dx$, with $N$ large and $f$ minimized at $x_0$, is that the leading term $e^{-Nf(x_0)}g(x_0)$ is invariant under $x_0\to x_0+a/N$ because $f'(x_0)=0$. See also the next footnote.} The only difference is that the classical vertex operator for the $J+\ell$ charge has $\ell$ extra copies of $v_{\rm cl}(-\varphi_L)$. For instance, the generalization of \eqref{eq:HH LLL correlator} is 
\begin{align}\label{eq:HH LLL correlator 3}
    &\frac{\dbraket{Z^{J+\ell}(-\varphi_L)\bar{Z}^J(\varphi_L)\prod_{i=1}^n Z^{\ell_i}(\varphi_i)\prod_{j=1}^{m}\bar{Z}^{\bar{\ell}_j}(\bar{\varphi}_j)}}{\dbraket{Z^J(-\varphi_L)\bar{Z}^J(\varphi_L)}}=\frac{1}{2\pi}\int_0^{2\pi}d\phi_0v^\ell_{\rm cl}\left(-\frac{\varphi_L}{2}\right)\prod_{i=1}^n v_{\rm cl}^{\ell_i}(\varphi_i)\prod_{j=1}^{m} \bar{v}_{\rm cl}^{\bar{\ell}_j}(\bar{\varphi}_j)\nonumber\\&=\frac{(2gc)^{\ell_{\rm tot}+\bar{\ell}_{\rm tot}+\ell}d^{\ell_{\rm tot}+\bar{\ell}_{\rm tot}-\ell}}{\prod_{i=1}^n d(\varphi_i,\varphi_L)^{2\ell_i} \prod_{j=1}^m d(\bar{\varphi}_j,-\varphi_L)^{2\bar{\ell}_j}}\delta_{\ell_{\rm tot}+\ell,\bar{\ell}_{\rm tot}}.
\end{align}
Furthermore, to determine the corresponding topological correlators, we again set $\varphi_L=\pi/2$, replace $\tilde{\Phi}^{J+\ell}\to v_{\rm cl}(-\frac{\pi}{2})^{J+\ell}=(gc)^{J+\ell}e^{i(J+\ell)\phi_0}$ and $\tilde{\Phi}^J\to (-1)^J \bar{v}_{\rm}(\frac{\pi}{2})^J=(-gc)^Je^{-iJ\phi_0}$ for the heavy operators and use \eqref{eq:topologicla vertex op} for the light operators. This yields
\begin{align}\label{eq:unequal HH correlator}
    \frac{\dbraket{\tilde{\Phi}^{J+\ell}\tilde{\Phi}^J\prod_{i=1}^n \tilde{\Phi}^{\ell_i}}}{\dbraket{\tilde{\Phi}^{J}\tilde{\Phi}^{J}}}&=\frac{(gc)^{\ell_{\rm tot}+\ell}}{2\pi}\int_0^{2\pi}d\phi_0 e^{i\ell \phi_0}(e^{i\phi_0}-e^{-i\phi_0})^{\ell_{\rm tot}}=(-g^2c^2)^{\frac{\ell_{\rm tot}+\ell}{2}}\binom{\ell_{\rm tot}}{(\ell_{\rm tot}-\ell)/2},
\end{align}
which indeed reduces to \eqref{eq:17AVeb3vDx} when $\ell=0$. One consequence of Eqs.~(\ref{eq:17AVeb3vDx}) and (\ref{eq:unequal HH correlator}) is that $\tilde{\Phi}^{\ell_1}\tilde{\Phi}^{\ell_2}$ and $\tilde{\Phi}^{\ell_1+\ell_2}$ are equal in the large charge limit, but $\tilde{\Phi}^{J+\ell}$ and $\tilde{\Phi}^J\tilde{\Phi}^\ell$ are not.

As a simple check of \eqref{eq:unequal HH correlator}, we note that when $\ell=\ell_{\rm tot}$ it reproduces the leading behavior of $\braket{\tilde{\Phi}^{J+\ell}\tilde{\Phi}^{J+\ell}}/\braket{\tilde{\Phi}^J\tilde{\Phi}^J}$, which is given in \eqref{eq:J+ell over J}.\footnote{The leading term in \eqref{eq:J+ell over J} can also be deduced using only results from the dual string. Firstly, given \eqref{eq:C(J+j)}, we see that $c(J)^\ell=c(J+\ell)^\ell(1+O(1/g))$ for finite $\ell$. Also, it follows from \eqref{eq:c-extremization} that $c(J)^{2J}e^{-S_{\rm cl }(c(J))}=c(J+\ell)^{2J}e^{-S_{\rm cl}(c(J+\ell))}(1+O(1/g))$. Applying these to \eqref{eq:HH topological final} implies $\braket{\tilde{\Phi}^{J+\ell}\tilde{\Phi}^{J+\ell}}=(-g^2c(J)^2)^\ell\braket{\tilde{\Phi}^{J}\tilde{\Phi}^{J}}(1+O(1/g))$, which is the desired result.} This is in accordance with the property of extremal correlators stated in \eqref{eq:extremal correlators}. It also follows that, if we use \eqref{eq:unequal HH correlator} to determine the normalized extremal OPE coefficient defined in \eqref{eq:extremal OPE coeff}, then we reproduce the leading order term, $(g\pi c^2)^\ell/\ell!$, in \eqref{eq:extremal OPE coeff subleading}. This normalized OPE coefficient can alternatively be extracted from a simple conformal block calculation, as we'll see next.

\paragraph{CFT data from a four-point function.}
The final correlator we consider is the non-topological correlator with two equal large charges and two equal light charges. This is a special case of \eqref{eq:HH LLL correlator} and is given by
\begin{align}\label{eq:four-point function}
    \frac{\dbraket{Z^J(-\varphi_L)\bar{Z}^J(\varphi_L)Z^\ell(\varphi)\bar{Z}^\ell(\bar{\varphi})}}{\dbraket{Z^J(-\varphi_L)\bar{Z}^J(\varphi_L)}}&=(2gc)^{2\ell}\frac{d(-\varphi_L,\varphi_L)^{2\ell}}{d(\varphi,\varphi_L)^{2\ell}d(\bar{\varphi},-\varphi_L)^{2\ell}}.
\end{align}
This four point function contains some simple large charge CFT data. We choose to study the conformal block expansion in the $Z^\ell Z^J\to \bar{Z}^\ell \bar{Z}^J$ channel, so that the smallest exchanged operators also have large charges. (By contrast, the smallest operators exchanged in the $Z^\ell \bar{Z}^\ell \to Z^J \bar{Z}^J$ channel are light). 

Therefore, let us map $\varphi\to \varphi_1$, $\bar{\varphi}\to \varphi_3$, $-\varphi_L\to \varphi_2$ and $\varphi_L\to \varphi_4$, and assume for simplicity that $-\pi<\varphi_1<\varphi_2<\varphi_3<\varphi_4<\pi$. The four point function can be written
\begin{align}\label{eq:four-point line}
    \dbraket{Z(\varphi_1)Z^J(\varphi_2)\bar{Z}(\varphi_3)\bar{Z}^J(\varphi_4)}&=\frac{(2gc)^{2(J+\ell)}e^{-S_{\rm cl}(c)}}{(d_{21}d_{43})^{J+\ell}}\left(\frac{d_{31}}{d_{42}}\right)^{J-\ell}\frac{\chi^{J+\ell}}{(1-\chi)^{2\ell}},
\end{align}
where we have introduced the conformally invariant cross ratio, $\chi$:
\begin{align}\label{eq:cross ratio}
    \chi&\equiv \frac{d_{21}d_{43}}{d_{31}d_{42}}, &\frac{\chi}{1-\chi}&=\frac{d_{21}d_{43}}{d_{41}d_{32}}, & \frac{1}{1-\chi}&=\frac{d_{31}d_{42}}{d_{41}d_{32}}.
\end{align}
(\ref{eq:four-point line}) is in a form that is convenient for comparing with the conformal block expansion \cite{Dolan:2011dv},
\begin{align}\label{eq:conformal block expansion}
    &\dbraket{Z(\varphi_1)Z^J(\varphi_2)\bar{Z}(\varphi_3)\bar{Z}^J(\varphi_4)}\\&\hspace{2cm}=\frac{1}{(d_{21}d_{43})^{J+\ell}} \left(\frac{d_{31}}{d_{42}}\right)^{J-\ell}\sum_{n} p_n\chi^{\Delta_n} {_2F_1}(\Delta_n+\ell-J,\Delta_n+J-\ell,2\Delta_n,\chi).\nonumber
\end{align}
Here, $n=0,1,2,\ldots$ labels the $n$th operator $\mathcal{O}_n$ appearing in the OPE of $ZZ^J$ (and its conjugate $\bar{\mathcal{O}}_n$ appearing in the OPE of $\bar{Z}\bar{Z}^J$), $\Delta_n$ is the dimension of $\mathcal{O}_n$ (and $\bar{\mathcal{O}}_n$), $\chi^\Delta{_2F_1}(\chi)$ is the conformal block in one dimension and $p_n$ is the $n$th conformal block coefficient.

The operators $\mathcal{O}_n$ in the OPE of $Z$ and $Z^J$ have dimensions of order $J$ or higher. Therefore, two of the three parameters of the hypergeometric function are large, in which case it takes a simplified form \cite{temme2003large}:
\begin{align}\label{eq:SzjtypJR7L}
    {_2F_1}(\alpha,L+\beta,L+\gamma,\chi)&=\frac{1}{(1-\chi)^\alpha}\left(1+O\left(\frac{1}{L}\right)\right).
\end{align}
Letting $L=2J$, $\alpha=\Delta_n+\ell-J$, $\beta=\Delta_n-\ell-J$, $\gamma=2(\Delta_n-J)$, we see that the leading behavior of the conformal block in \eqref{eq:conformal block expansion} is $\chi^{\Delta_n}/(1-\chi)^{\Delta_n+\ell-J}$. Then, matching Eqs.~(\ref{eq:four-point line}) and (\ref{eq:conformal block expansion}), we find that there is only one operator, $\mathcal{O}_0$, contributing to the conformal block expansion in the large charge regime, and that its dimension and conformal block coefficient are:
\begin{align}\label{eq:CFT data}
    \Delta_0&=J+\ell, &p_0&=(2gc)^{2(J+\ell)}e^{-S_{\rm cl}(c)}.
\end{align}
We identify $\mathcal{O}_0=\bar{Z}^{J+\ell}$, $\bar{\mathcal{O}}_0=Z^{J+\ell}$, which have the right dimension. Additionally, the conformal block coefficient $p_0$ is related to the OPE coefficient for the three-point function $ZZ^J\bar{Z}^{J+\ell}$ and the normalization of the two-point function $Z^{J+\ell}\bar{Z}^{J+\ell}$ , which up to factors of $2$ coming from the polarization vectors of $Z$ and $\bar{Z}$ are given by $c_{J+\ell,J,\ell}$ and $n_J$. More precisely, $p_0=2^{J+\ell} c_{J+\ell,J,\ell}^2/n_{J+\ell}$. Then, combining \eqref{eq:CFT data} with \eqref{eq:HH topological final} and \eqref{eq:n_ell finite charge} we find,
\begin{align}\label{eq:extremal OPE coeff leading}
    \frac{c_{J+\ell,J,\ell}^2}{n_{J+\ell}n_Jn_\ell}=\frac{p_0}{2^{J+\ell}n_{J}n_{\ell}}=\frac{(g\pi c^2)^\ell}{\ell!}.
\end{align}
This agrees at leading order with \eqref{eq:extremal OPE coeff subleading}. 

\subsection{Small and large \texorpdfstring{$\mathcal{J}$}{J/g}}\label{sec:small and large J/g}
In this final section, we discuss the behavior of some of the large charge correlators we have computed when $\mathcal{J}\ll 1$ or $\mathcal{J}\gg 1$. We also compare them to the large $g$ and small $g$ perturbative results for the planar, finite charge correlators from \cite{giombi2017half,giombi2018exact,Kiryu:2018phb}. 

\paragraph{Small $\mathcal{J}$.} When $\mathcal{J}\ll 1$, $\mathcal{J}$ and $c^2$ are related by the Taylor series
\begin{align}
    \mathcal{J}&=\pi c^2+\frac{3\pi}{8}c^4+O(c^6), &c^2&=\frac{\mathcal{J}}{\pi}-\frac{3}{8}\frac{\mathcal{J}^2}{\pi^2}+O(\mathcal{J}^3).
\end{align}
The first few terms in the small $\mathcal{J}$ expansions of the two-point function in \eqref{eq:HH topological final} and the normalized extremal three-point function in \eqref{eq:extremal OPE coeff subleading} are:
\begin{align}
    \dbraket{\tilde{\Phi}^J\tilde{\Phi}^J}&=(-g^2)^J \text{exp}\left[g\left(\mathcal{J}\log{\mathcal{J}}-\mathcal{J}(1+\log{\pi})-\frac{3\mathcal{J}^2}{16\pi}+O(\mathcal{J}^3)\right)+O(g^0)\right]\label{eq:two-point small J/g},\\
    \frac{c_{J+\ell,J,\ell}^2}{n_{J+\ell}n_Jn_\ell}&=g^\ell\left(\frac{\mathcal{J}^\ell}{\ell!}-\frac{3}{8\pi}\frac{\mathcal{J}^{\ell+1}}{(\ell-1)!}+O(\mathcal{J}^{\ell+2})\right)\nonumber\\&+g^{\ell-1}\left(\frac{\ell+1}{2(\ell-1)!}\mathcal{J}^{\ell-1}-\frac{3}{16\pi}\frac{\ell(\ell+1)}{(\ell-1)!}\mathcal{J}^\ell+O(\mathcal{J}^{\ell+1})\right) +O(g^{\ell-2}).\label{eq:three-point small J/g}
\end{align}
We recall that our matrix model in Sections \ref{sec:matrix model}-\ref{sec:large J analysis} and the dual string analysis in Section \ref{sec:higher-pt from semistring} only determine the $g^\ell$ term in \eqref{eq:three-point small J/g}. The $g^{\ell-1}$ term comes from the Bremsstrahlung analysis in \ref{sec:bremsstrahlung}.

The expansion of the large charge correlators in small $\mathcal{J}$ pushes results that are valid when $1\ll J\sim g$ into the regime $1\ll J\ll g$. This should agree with the expansion of finite charge correlators in large $J$, which pushes results that are valid when $1\sim J \ll g$ also into the regime $1\ll J \ll g$. More specifically, we expect the coefficient of the $g^{-m} \mathcal{J}^n$ term in the double expansion in $1/g$ and $\mathcal{J}$ to match the coefficient of the $g^{-m-n}J^n$ term in the double expansion in $1/g$ and $1/J$.\footnote{This is equivalent to saying that the $m$th sub-leading term in the $1/g$ expansion with finite  $\mathcal{J}$ should equal the resummation of the $(m+1)$th largest $J$ terms at each order in the $1/g$ expansion with finite $J$.}

Therefore, we can use the $1/g$ expansion of the finite $J$ correlators, determined using localization in \cite{giombi2018exact}, to test Eqs.~(\ref{eq:two-point small J/g})-(\ref{eq:three-point small J/g}). The two-point function (modulo \eqref{eq:topo 2-pt function}) is given in \eqref{eq:n_ell finite charge} and the three-point function is
\begin{align}
    \frac{c_{J+\ell,J,\ell}^2}{n_{J+\ell}n_Jn_\ell}&=\frac{(J+\ell)!}{J!\ell!}\left(1-\frac{3}{8\pi}\frac{J\ell}{g}+O(1/g^2)\right).\label{eq:3-pt finite J}
\end{align}
Putting the two-point function in a form parallel to \eqref{eq:two-point small J/g} and then expanding in large $J$, we find
\begin{align}\label{eq:finite charge two point}
    \dbraket{\tilde{\Phi}^J\tilde{\Phi}^J}&=(-g^2)^J\text{exp}\left[\log\frac{J!}{(g\pi)^J}-\frac{3}{32\pi g}(2J^2+J)+O(1/g^2)\right]\\&=(-g^2)^J \text{exp}\left[\left(J\log{\frac{J}{g}}-J(1+\log{\pi})+O(J^0)\right)-\frac{1}{g}\left(\frac{3J^2}{16\pi }+O(J)\right)+O(1/g^2)\right]\nonumber
\end{align}
which matches \eqref{eq:two-point small J/g} in the overlapping terms. Likewise, expanding the three-point function in large $J$ yields
\begin{align}
    \frac{c_{J+\ell,J,\ell}^2}{n_{J+\ell}n_Jn_\ell}&=\left(\frac{J^\ell}{\ell!}+\frac{(\ell+1)J^{\ell-1}}{2(\ell-1)!}+O(J^{\ell-2})\right)\nonumber\\&-\frac{1}{g}\left(\frac{3}{8\pi}\frac{J^{\ell+1}}{(\ell-1)!}+\frac{3}{16\pi g}\frac{\ell(\ell+1)J^\ell}{(\ell-1)!}+O(J^{\ell-1})\right)+O(1/g^2),
\end{align}
which matches \eqref{eq:three-point small J/g} in the overlapping terms.

Eqs.~(\ref{eq:n_ell finite charge}) and (\ref{eq:3-pt finite J}) follow from localization, but may also be understood in terms of the EAdS$_2$ string dual to the Wilson loop without insertions. The leading strong coupling behavior of the finite charge correlators is determined by a free theory with boundary-to-boundary propagators \cite{giombi2017half}
\begin{align}
    \braket{\Phi_I(\varphi_i)\Phi_J(\varphi_j)}&=\frac{2g}{\pi}\frac{\delta_{IJ}}{d_{ij}^2},
\end{align}
where $I,J=1,\ldots,5$. For non-coincident insertions of the topological operators, the boundary propagator is $\braket{\tilde{\Phi}\tilde{\Phi}}=-g/\pi$. For $Z$ and $\bar{Z}$, the boundary propagator is $\braket{Z(\varphi_i)\bar{Z}(\varphi_j)}=\frac{4g}{\pi}\frac{1}{d_{ij}^2}$, $\braket{ZZ}=\braket{\bar{Z}\bar{Z}}=0$. In this language, the leading term in \eqref{eq:finite charge two point} arises from the $J!$ ways to contract the two insertions of $\tilde{\Phi}^J$, and the leading term in $c_{J+\ell,J,\ell}$ in \eqref{eq:3-pt finite J} comes from the $\binom{J+\ell}{J}J!\ell!$ ways to contract $J$ fields in $\tilde{\Phi}^{J+\ell}$ with $\tilde{\Phi}^J$ and the remaining $\ell$ fields in $\tilde{\Phi}^{J+\ell}$ with $\tilde{\Phi}^\ell$.\footnote{It is also possible to reproduce the next-to-leading order terms by considering fluctuations of the EAdS$_2$ string and including the contribution of the four-point contact diagram with an interaction in the bulk that was studied in \cite{giombi2017half}. However, we stick to the free theory analysis both for simplicity and because it is sufficient for the purpose of checking leading large charge behavior.}

\begin{figure}[t]
\centering
\includegraphics[clip,height=6cm]{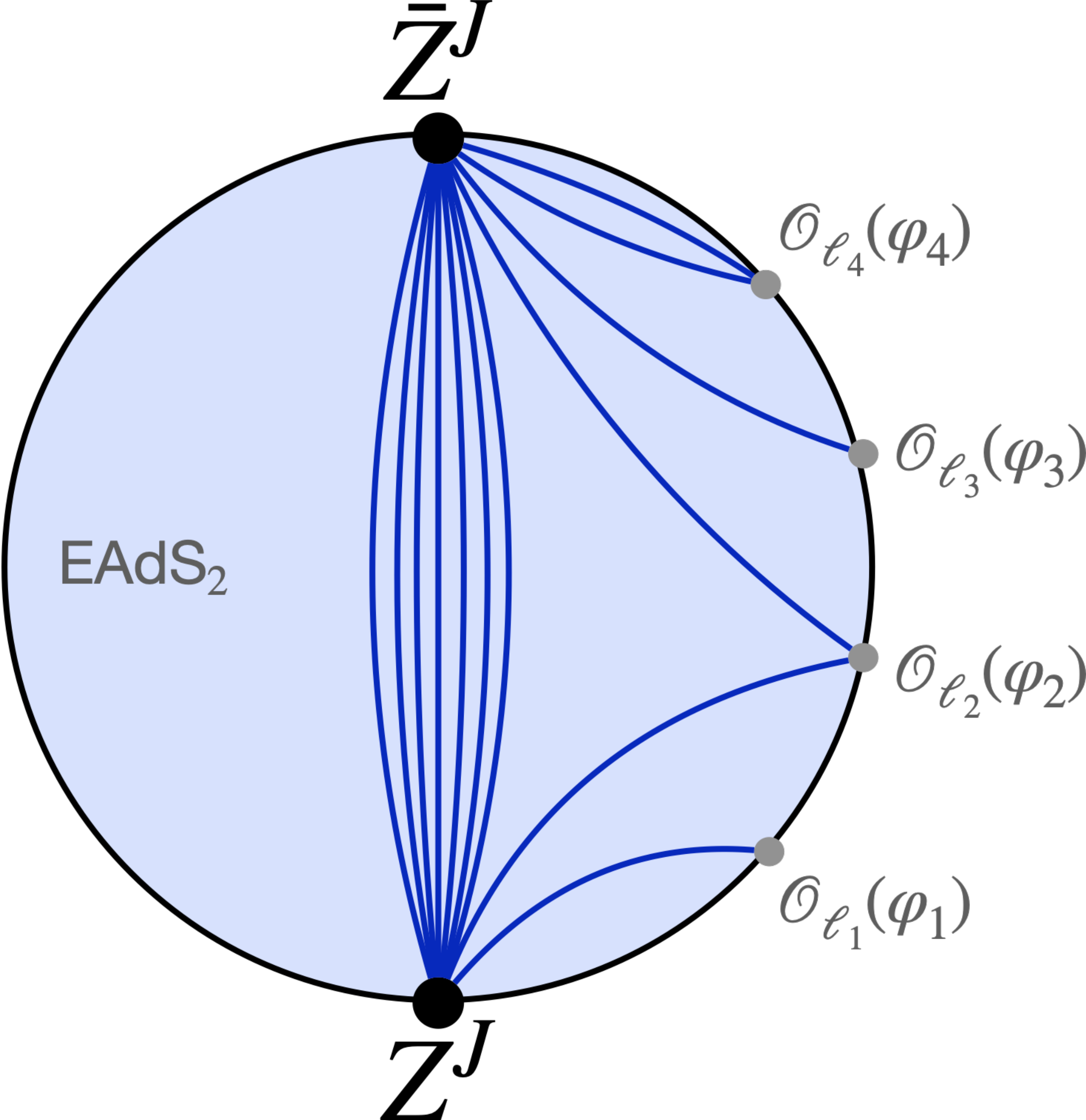}
\caption{In the planar and strongly coupled regime, the leading behavior (in $1/g$) of the defect correlators of operators with parametrically small charges can be analyzed using the free theory on the EAdS$_2$ classical string dual to the Wilson loop without insertions. When two of the operators have a large charge, $1\ll J\ll g$, the leading contributions (in $1/J$) come from diagrams in which the light operators are fully contracted with the large charges, and the remaining fields of the large charges are contracted with each other.}
\label{fig:boundary-to-boundary-contractions}
\end{figure}

Using the EAdS$_2$ free theory, we can similarly determine the leading strong coupling behavior of \eqref{eq:HH LLL correlator} when $J$ is finite. This serves as a check of the large charge result, including the position dependence. The general expression resulting from summing all possible boundary-to-boundary contractions between the $Z$ and $\bar{Z}$ insertions in the numerator of \eqref{eq:HH LLL correlator} is rather cumbersome. However, to compare with the large charge result in \eqref{eq:HH LLL correlator}, we only need the contribution with the largest power of $J$, which comes from the diagrams where the light operators are contracted only with the heavy operators, and the remaining fields of the heavy operators are contracted with each other; see Figure~\ref{fig:boundary-to-boundary-contractions}. The number of such contractions is $J!^2/(J-\ell_{\rm tot})!$. After dividing by $\braket{Z^J\bar{Z}^J}=J!\left(\frac{4g}{\pi d^2}\right)^J$, and noting $J!/(J-\ell_{\rm tot})!=J^{\ell_{\rm tot}}+O(J^{\ell_{\rm tot}-1})$, we find:
\begin{align}
    &\frac{\dbraket{Z^J(-\varphi_L)\bar{Z}^J(\varphi_L)\prod_{i=1}^n Z^{\ell_i}(\varphi_i)\prod_{j=1}^{m}\bar{Z}^{\bar{\ell}_j}(\bar{\varphi}_j)}}{\dbraket{Z^J(-\varphi_L)\bar{Z}^J(\varphi_L)}}\nonumber\\&=g^{\ell_{\rm tot}}\left(J^{\ell_{\rm tot}}\frac{\left(\frac{4}{\pi}\right)^{\ell_{\rm tot}}d^{2\ell_{\rm tot}}}{\prod_{i=1}^n d(\varphi_L,\varphi_i)^{2\ell_i}\prod_{j=1}^m d(-\varphi_L,\varphi_j)^{2\ell_j}}\delta_{\ell_{\rm tot},\bar{\ell}_{\rm tot}}+O(J^{\ell_{\rm tot}-1})\right)+O(g^{\ell_{\rm tot}-1}).
\end{align}
This matches the leading term in the small $\mathcal{J}$ expansion of \eqref{eq:HH LLL correlator}, which is given by replacing $c^2\to \mathcal{J}/\pi$. One consequence of this comparison is that we may view the light vertex operators on the $\mathcal{J}\ll 1$ semiclassical string as arising from the boundary-to-boundary contractions on the EAdS$_2$ string between the light insertions and the charge $J\gg 1$ insertions. The factors of $c^2\sim J$ in the vertex operators reflect the fact that there are $J$ fields in the large insertions that the light insertions can contract with.

\paragraph{Large $\mathcal{J}$.}
When $\mathcal{J}\gg 1$, $\mathcal{J}$ and $c^2$ are related by the asymptotic series
\begin{align}
    \mathcal{J}&=-2\log(1-c^2)+4(2\log{2}-1)+O((1-c^2)\log(1-c^2)),\\c^2&=1-\frac{16}{e^2}e^{-\frac{\mathcal{J}}{2}}+O(\mathcal{J}e^{-\mathcal{J}}).
\end{align}
We have already noted the leading large $\mathcal{J}$ behavior of the two-point function in \eqref{eq:large J/g limit}. Meanwhile, the leading behavior of the three-point function is especially simple:
\begin{align}\label{eq: 3-pt large J/g}
    \frac{c_{J+\ell,J,\ell}^2}{n_{J+\ell}n_Jn_\ell}=g^\ell\left[\frac{\pi^\ell}{\ell!}+O(e^{-\frac{\mathcal{J}}{2}})\right]+O(g^{\ell-1}).
\end{align}
It is perhaps interesting to compare Eqs.~(\ref{eq:three-point small J/g}) and (\ref{eq: 3-pt large J/g}) with the general charge dependence of the heavy-heavy-light OPE coefficient that one would expect based on an effective field theory analysis expanding in inverse powers of the large charge. As in \cite{Monin:2016jmo,Jafferis:2017zna}, the OPE coefficient typically grows as a power of the large charge. Though the extremal OPE coefficient indeed has this behavior at small $\mathcal{J}$ (see \eqref{eq:extremal OPE coeff leading}), we see from \eqref{eq: 3-pt large J/g} that it instead saturates to a constant at large $\mathcal{J}$. 

Finally, as noted in \cite{Zarembo:2002ph}, if $\mathcal{J}$ were the only parameter governing the correlators when $\mathcal{J}$ is large, then the $1\ll g\ll J$ regime probed by Eqs.~(\ref{eq:large J/g limit}) and (\ref{eq: 3-pt large J/g}) should match the $g\ll 1\sim J$ regime accessible to weakly coupled gauge theory, thus creating a bridge between two regimes that a priori seem very different. This agreement was verified in \cite{Zarembo:2002ph}, at least schematically, for the correlator of the half-BPS Wilson loop and a single large charge insertion off the Wilson loop. Likewise, in \cite{drukker2006small} the anomalous dimensions of ``words'' composed of many copies of $Z$ interspersed with orthogonal scalars inserted on the half-BPS Wilson loop were found to match in both the small $g$, large $J$ regime and the large $g$, large $\mathcal{J}$ regime. However, we observe that this agreement between the two regimes does not extend to the correlators we have been studying. From either localization \cite{giombi2018exact} or gauge theory \cite{Kiryu:2018phb}, one finds that that the two and three-point functions in the weakly coupled limit are
\begin{align}
    \dbraket{\tilde{\Phi}^J\tilde{\Phi}^J}&=(-g^2)^J \left(1-\frac{2\pi^2}{3}g^2+O(g^4)\right),\\
    \frac{c_{J+\ell,J,\ell}^2}{n_{J+\ell}n_Jn_\ell}&=1+\frac{2\pi^2 g^2}{3}+O(g^4).
\end{align}
These are manifestly different from Eqs.~(\ref{eq:large J/g limit}) and (\ref{eq: 3-pt large J/g}).

\section*{Acknowledgments}

We are grateful to Nadav Drukker for useful correspondence. The work of SG and BO is supported in part by the US NSF under Grant No.~PHY-1914860. 

 \appendix
 \section{Derivative of the quasi-momentum}\label{ap:quasimomentum}
 In this appendix, we derive the expression for the derivative of the quasi-momentum, \eqref{eq:derpcl}. 
 
 For this purpose, we use the fact that the one form $p_{\rm cl}du$ has poles with residue $J$ at $x=0$ and $x=\infty$, while its integral along two branch cuts gives $-J$. From these properties, we immediately conclude that its derivative $\del_{J}p_{\rm cl}du$ satisfies
 \begin{enumerate}
 \item $\oint_{\mathcal{C}_{\pm}}\del_Jp_{\rm cl}du=-1$.
 \item $\oint_{x=0}\del_Jp_{\rm cl}du=\oint_{x=\infty}\del_Jp_{\rm cl}du=1$.
 \end{enumerate}
 In addition, one can check that $p_{\rm cl}du$ flips a sign upon $x\to -x^{-1}$. This follows from the definition of the (quantum) quasi-momentum, \eqref{eq:sumexpressionp}. 
 
These conditions are sufficient for fixing the one form up to a constant $\alpha$:
\beq\label{eq:uptoconstant}
\del_{J}p_{\rm cl}du=\frac{(x+x^{-1}+\alpha) dx}{\sqrt{(x-e^{i\theta_0})(x-e^{-i\theta_0})(x+e^{i\theta_0})(x+e^{-i\theta_0})}}\period
\eeq
To determine $\alpha$, we expand the quantum quasi-momentum \eqref{eq:sumexpressionp} around $x=0$:
\beq
p(x)du\sim \frac{J}{x}dx+\sum_{k=1}^{J}(x_k-x_k^{-1})dx+\mathcal{O}(x)\period
\eeq
In the classical limit, $x_k$'s are distributed symmetrically with respect to the real axis (see Figure \ref{fig:quasi1}), we have
\beq
\sum_{k=1}^{J}(x_k-x_k^{-1})\overset{J\to \infty}{=}0\period
\eeq
Comparing this with the expansion of \eqref{eq:uptoconstant} around $x=0$, one can show that $\alpha=0$ and reproduce \eqref{eq:derpcl}.

\section{Supersymmetries of the classical string}\label{sec:semistring susy}
This appendix demonstrates that the string solution identified in \eqref{eq:semistring global coords 1} has the same supersymmetries as the string solution analyzed in \cite{drukker2006small}, which corresponds to $c=1$, and therefore has the same supersymmetries as the Wilson operator with $Z^J$ and $\bar{Z}^J$ inserted. More precisely, we will impose the supersymmetry conditions identified in Sec. 4.2 of \cite{drukker2006small} on a string ansatz and will show that \eqref{eq:semistring global coords 1} give the general solution to the resulting pair of differential equations. We closely follow the approach of \cite{drukker2006small}. To avoid confusion, it should be noted that some of the notation in this appendix differs from that in the rest of the text.

Letting $\kappa,\lambda$ denote the spacetime indices on AdS$_5\times S^5$ and run over $t,\rho,\theta,\phi,\ldots$, and letting $X^\kappa=(\rho,t,\theta,\phi,\ldots)$, the relevant part of the metric is
\begin{align}\label{eq:YFcqMM6d2r}
    ds^2=G_{\kappa\lambda}dX^\kappa dX^\lambda&= -\cosh^2{\rho}dt^2+d\rho^2+d\theta^2+\sin^2{\theta}d\phi^2.
\end{align}
We introduce the vielbeins $e^a=\tensor{e}{^a_\kappa}dX^\kappa$ satisfying $ds^2=\eta_{ab}e^ae^b$ where $\eta_{ab}=\text{diag}(-1,1,\ldots,1)$ and $a=0,1,2\ldots,9$. The relevant vielbeins are $e^0=\cosh{\rho}dt$, $e^1=d\rho$, $e^5=d\theta$, and $e^6=\sin{\theta}d\phi$. We introduce $32\times 32$ real constant Dirac matrices $\Gamma_a$ on $10$ dimensional Minkowski space; they satisfy the Clifford algebra $\{\Gamma_a,\Gamma_b\}=2\eta_{ab}$. The Dirac matrices on AdS$_5\times S^5$ are then $\gamma_\kappa\equiv\tensor{e}{^a_\kappa}\Gamma_a$, and satisfy $\{\gamma_\kappa,\gamma_\lambda\}=2G_{\kappa\lambda}$. The relevant $\gamma$ matrices are $\gamma_t=\cosh{\rho}\Gamma_0$, $\gamma_\rho=\Gamma_1$, $\gamma_\theta=\Gamma_5$, and $\gamma_\phi =\sin{\theta} \Gamma_6$. Finally, we define the chirality matrix $\Gamma_\star=\Gamma_0\Gamma_1\Gamma_2\Gamma_3\Gamma_4$, which satisfies $[\Gamma_\star,\Gamma_a]=0$ for $a=0,\ldots,4$, $\{\Gamma_\star,\Gamma_b\}=0$ for $b=5,\ldots,9$, and $\Gamma_\star^2=-1$.

The supersymmetries of a string $X^\kappa(\sigma,\tau)$ embedded in AdS$_5\times S^5$, with $\sigma$ and $\tau$ as worldsheet coordinates, are determined by the linearly independent solutions to the Killing spinor equation,
\begin{align}
    (D_\kappa+\frac{i}{2}\Gamma_\star \gamma_\kappa )\epsilon=0,
\end{align}
that also satisfy the projection condition imposing kappa symmetry, $\Gamma\epsilon=\epsilon$.
Here, $D_\kappa\equiv \partial_\kappa+\frac{1}{4}\omega_\kappa^{ab}\Gamma_{ab}$ is the spinor covariant derivative and the relevant components of the spin connection are $\omega_t^{01}=\sinh{\rho}$ and $\omega_\phi^{56}=-\cos{\theta}$. Further, the kappa symmetry projector is
\begin{align}
    \Gamma\equiv \frac{1}{\sqrt{-h}}\partial_\tau x^\kappa \partial_\sigma x^\lambda \gamma_\kappa \gamma_\lambda K,
\end{align}
where $h$ is the metric induced on the string worldsheet and $K$ acts by complex conjugation to the right. As noted in \cite{drukker2006small}, the general $t,\rho,\theta,\phi$ dependence of the Killing spinors may be written
\begin{align}\label{eq:EBVg1tqKUi}
    \epsilon&=e^{-\frac{i}{2}\left(\rho \Gamma_\star \Gamma_1+\theta\Gamma_\star \Gamma_5\right)}e^{-\frac{i}{2}\left(t\Gamma_\star \Gamma_0+i\phi \Gamma_5\Gamma_6\right)}\epsilon_0,
\end{align}
where $\epsilon_0$ does not depend on the relevant coordinates.

Finally, let us consider the string ansatz of Sec. 4.1 of \cite{drukker2006small}: $t=\omega \tau$, $\phi=w_1\tau$, $\rho=\rho(\sigma)$, $\theta=\theta(\sigma)$. When $w_1=\omega$ and $\sin{\theta}=\sech{\rho}=\tanh(\omega \sigma)$, \cite{drukker2006small} showed that imposing the projection condition on \eqref{eq:EBVg1tqKUi} reduces to two compatible conditions on $\epsilon_0$:
\begin{align}
    (\Gamma_\star \Gamma_0 \Gamma_5 \Gamma_6-i)\epsilon_0=0,\label{eq:bxXm1kXcLO}\\
    (\Gamma_0\Gamma_1K+1)\epsilon_0=0.\label{eq:oKSPQrqESB}
\end{align}
These each halve the degrees of freedom of $\epsilon_0$ and imply that the string solution, like the dual half-BPS Wilson operator with $Z^J$ and $\bar{Z}^J$ inserted, is quarter-BPS. We now derive a more general string configuration, specified by $\theta(\sigma)$ and $\rho(\sigma)$, with these supersymmetries. 

To begin, the metric induced on the worldsheet is
\begin{align}
    h_{\alpha\beta}d\sigma^\alpha d\sigma^\beta=(-\omega^2\cosh^2{\rho}+\sin^2{\theta}w_1^2)d\tau^2+((\rho')^2+(\theta')^2)d\sigma^2,
\end{align}
where the prime denotes differentiation with respect to $\sigma$, and the projector becomes
\begin{align}
    \Gamma&=\frac{\omega\rho'\cosh{\rho}\Gamma_{0}\Gamma_1+\omega\theta'\cosh{\rho}\Gamma_{0}\Gamma_5+w_1\rho'\sin{\theta}\Gamma_{6}\Gamma_1+w_1\theta'\sin{\theta}\Gamma_{6}\Gamma_5}{\sqrt{(\omega^2\cosh^2{\rho}-w_1^2\sin^2{\theta})((\rho')^2+(\theta')^2)}}K.
\end{align}
$\Gamma$ is a function only of $\sigma$, and the $\tau$ dependence of $\epsilon$ is contained in the right exponent in \eqref{eq:EBVg1tqKUi}. Thus, for $\Gamma\epsilon=\epsilon$ to hold for all $\tau$, we follow \cite{drukker2006small} and impose\footnote{One could try to impose more generally that $(\omega \Gamma_\star \Gamma_0 + i w_1 \Gamma_5 \Gamma_6)\epsilon_0=\lambda\epsilon_0$ for some complex $\lambda$. Then, because of the action of $K$ in $\Gamma$, $\lambda$ must be imaginary in order for $\Gamma \epsilon=\epsilon$ to hold for all $\tau$. But since $\Gamma_\star\Gamma_0$ and $\Gamma_5\Gamma_6$ can be simultaneously diagonalized (because they commute) and have eigenvalues $\pm1$ and $\pm i$, respectively, $\lambda$ must also be real. Therefore, $\lambda=0$.}
\begin{align}\label{eq:z6s8CO0EZR}
    (\omega \Gamma_\star \Gamma_0 + i w_1 \Gamma_5 \Gamma_6)\epsilon_0=0.
\end{align}
Acting on the left of the above condition with $-i\Gamma_\star \Gamma_0$, we see that \eqref{eq:z6s8CO0EZR} is equivalent to \eqref{eq:bxXm1kXcLO} only if $w_1=\omega$, as in the Drukker-Kawamoto solution.

More non-trivially, to get the analog of \eqref{eq:oKSPQrqESB}, we commute $\Gamma$ past $e^{-\frac{i}{2}(\rho \Gamma_\star \Gamma_1+\theta\Gamma_\star \Gamma_5)}$ in $\Gamma\epsilon=\epsilon$ and arrive at 
\begin{align}\label{eq:l4WSxnPGFE}
    \frac{(\omega \rho' \cosh{\rho}\cos{\theta}-\omega \theta'\sin{\theta}\sinh{\rho})\Gamma_0\Gamma_1+(\omega \theta' \cosh{\rho}\cos{\theta}+\omega\rho' \sin{\theta}\sinh{\rho})\Gamma_0 \Gamma_5}{\sqrt{(\omega^2\cosh^2{\rho}-w_1^2\sin^2{\theta})((\rho')^2+(\theta')^2)}}K\epsilon_0&=\epsilon_0.
\end{align}
In order for \eqref{eq:l4WSxnPGFE} to be equivalent to \eqref{eq:oKSPQrqESB}, $\theta$ and $\rho$ must satisfy the following two differential equations.
\begin{align}
    \theta' \cosh{\rho}\cos{\theta}+ \rho'\sin{\theta}\sinh{\rho}&=0,
    &&\frac{\rho' \cosh{\rho}\cos{\theta}-\theta'\sin{\theta}\sinh{\rho}}{\sqrt{(\cosh^2{\rho}-\sin^2{\theta})((\rho')^2+(\theta')^2)}}=-1,\label{eq:ZdJz0dp9VR}
\end{align}
where we have assumed $\omega>0$. The first equation is separable and its general solution is
\begin{align}\label{eq:2362815679}
    \sin{\theta}&=\frac{c}{\cosh{\rho}}.
\end{align}
Since $\theta\in[0,\pi]$, it follows that $0\leq c\leq 1$. The second differential equation is then automatically satisfied. Thus, along with $t=\phi=\omega \tau$, \eqref{eq:2362815679} defines the more general string configuration with the supersymmetries specified by Eqs.~(\ref{eq:bxXm1kXcLO}) and (\ref{eq:oKSPQrqESB}).

\newpage 
\bibliographystyle{ssg}
\bibliography{mybib}

\begingroup\raggedright\begin{thebibliography}{10}

\bibitem{Berenstein:2002jq}
D.~E. Berenstein, J.~M. Maldacena, and H.~S. Nastase, ``{Strings in flat space
  and pp waves from N=4 superYang-Mills},'' {\em JHEP} {\bf 04} (2002) 013,
  \href{http://xxx.lanl.gov/abs/hep-th/0202021}{{\tt hep-th/0202021}}.

\bibitem{Gubser:2002tv}
S.~Gubser, I.~Klebanov, and A.~M. Polyakov, ``{A Semiclassical limit of the
  gauge / string correspondence},'' {\em Nucl. Phys. B} {\bf 636} (2002)
  99--114, \href{http://xxx.lanl.gov/abs/hep-th/0204051}{{\tt hep-th/0204051}}.

\bibitem{Beisert:2010jr}
N.~Beisert {\em et.~al.}, ``{Review of AdS/CFT Integrability: An Overview},''
  {\em Lett. Math. Phys.} {\bf 99} (2012) 3--32,
  \href{http://xxx.lanl.gov/abs/1012.3982}{{\tt 1012.3982}}.

\bibitem{Hellerman:2015nra}
S.~Hellerman, D.~Orlando, S.~Reffert, and M.~Watanabe, ``{On the CFT Operator
  Spectrum at Large Global Charge},'' {\em JHEP} {\bf 12} (2015) 071,
  \href{http://xxx.lanl.gov/abs/1505.01537}{{\tt 1505.01537}}.

\bibitem{Monin:2016jmo}
A.~Monin, D.~Pirtskhalava, R.~Rattazzi, and F.~K. Seibold, ``{Semiclassics,
  Goldstone Bosons and CFT data},'' {\em JHEP} {\bf 06} (2017) 011,
  \href{http://xxx.lanl.gov/abs/1611.02912}{{\tt 1611.02912}}.

\bibitem{Alvarez-Gaume:2016vff}
L.~Alvarez-Gaume, O.~Loukas, D.~Orlando, and S.~Reffert, ``{Compensating strong
  coupling with large charge},'' {\em JHEP} {\bf 04} (2017) 059,
  \href{http://xxx.lanl.gov/abs/1610.04495}{{\tt 1610.04495}}.

\bibitem{Jafferis:2017zna}
D.~Jafferis, B.~Mukhametzhanov, and A.~Zhiboedov, ``{Conformal Bootstrap At
  Large Charge},'' {\em JHEP} {\bf 05} (2018) 043,
  \href{http://xxx.lanl.gov/abs/1710.11161}{{\tt 1710.11161}}.

\bibitem{Gaume:2020bmp}
L.~A. Gaum\'e, D.~Orlando, and S.~Reffert, ``{Selected Topics in the Large
  Quantum Number Expansion},'' \href{http://xxx.lanl.gov/abs/2008.03308}{{\tt
  2008.03308}}.

\bibitem{drukker2006small}
N.~Drukker and S.~Kawamoto, ``{Small deformations of supersymmetric Wilson
  loops and open spin-chains},'' {\em JHEP} {\bf 07} (2006) 024,
  \href{http://xxx.lanl.gov/abs/hep-th/0604124}{{\tt hep-th/0604124}}.

\bibitem{giombi2017half}
S.~Giombi, R.~Roiban, and A.~A. Tseytlin, ``{Half-BPS Wilson loop and
  AdS$_2$/CFT$_1$},'' {\em Nucl. Phys.} {\bf B922} (2017) 499--527,
  \href{http://xxx.lanl.gov/abs/1706.00756}{{\tt 1706.00756}}.

\bibitem{Giombi:2020amn}
S.~Giombi, J.~Jiang, and S.~Komatsu, ``{Giant Wilson loops and
  AdS$_{2}$/dCFT$_{1}$},'' {\em JHEP} {\bf 11} (2020) 064,
  \href{http://xxx.lanl.gov/abs/2005.08890}{{\tt 2005.08890}}.

\bibitem{giombi2018exact}
S.~Giombi and S.~Komatsu, ``{Exact Correlators on the Wilson Loop in
  $\mathcal{N}=4$ SYM: Localization, Defect CFT, and Integrability},'' {\em
  JHEP} {\bf 05} (2018) 109, \href{http://xxx.lanl.gov/abs/1802.05201}{{\tt
  1802.05201}}. [Erratum: JHEP11,123(2018)].

\bibitem{Giombi:2018hsx}
S.~Giombi and S.~Komatsu, ``{More Exact Results in the Wilson Loop Defect CFT:
  Bulk-Defect OPE, Nonplanar Corrections and Quantum Spectral Curve},'' {\em J.
  Phys. A} {\bf 52} (2019), no.~12 125401,
  \href{http://xxx.lanl.gov/abs/1811.02369}{{\tt 1811.02369}}.

\bibitem{liendo2018bootstrapping}
P.~Liendo, C.~Meneghelli, and V.~Mitev, ``{Bootstrapping the half-BPS line
  defect},'' {\em JHEP} {\bf 10} (2018) 077,
  \href{http://xxx.lanl.gov/abs/1806.01862}{{\tt 1806.01862}}.

\bibitem{Ferrero:2021bsb}
P.~Ferrero and C.~Meneghelli, ``{Bootstrapping the half-BPS line defect CFT in
  N=4 supersymmetric Yang-Mills theory at strong coupling},'' {\em Phys. Rev.
  D} {\bf 104} (2021), no.~8 L081703,
  \href{http://xxx.lanl.gov/abs/2103.10440}{{\tt 2103.10440}}.

\bibitem{Barrat:2021yvp}
J.~Barrat, A.~Gimenez-Grau, and P.~Liendo, ``{Bootstrapping holographic defect
  correlators in $\mathcal{N}=4$ super Yang-Mills},''
  \href{http://xxx.lanl.gov/abs/2108.13432}{{\tt 2108.13432}}.

\bibitem{Drukker:2012de}
N.~Drukker, ``{Integrable Wilson loops},'' {\em JHEP} {\bf 10} (2013) 135,
  \href{http://xxx.lanl.gov/abs/1203.1617}{{\tt 1203.1617}}.

\bibitem{correa2012exact}
D.~Correa, J.~Henn, J.~Maldacena, and A.~Sever, ``{An exact formula for the
  radiation of a moving quark in N=4 super Yang Mills},'' {\em JHEP} {\bf 06}
  (2012) 048, \href{http://xxx.lanl.gov/abs/1202.4455}{{\tt 1202.4455}}.

\bibitem{Kiryu:2018phb}
N.~Kiryu and S.~Komatsu, ``{Correlation Functions on the Half-BPS Wilson Loop:
  Perturbation and Hexagonalization},'' {\em JHEP} {\bf 02} (2019) 090,
  \href{http://xxx.lanl.gov/abs/1812.04593}{{\tt 1812.04593}}.

\bibitem{Grabner:2020nis}
D.~Grabner, N.~Gromov, and J.~Julius, ``{Excited States of One-Dimensional
  Defect CFTs from the Quantum Spectral Curve},'' {\em JHEP} {\bf 07} (2020)
  042, \href{http://xxx.lanl.gov/abs/2001.11039}{{\tt 2001.11039}}.

\bibitem{Cavaglia:2021bnz}
A.~Cavagli\`a, N.~Gromov, J.~Julius, and M.~Preti, ``{Integrability and
  Conformal Bootstrap: One Dimensional Defect CFT},''
  \href{http://xxx.lanl.gov/abs/2107.08510}{{\tt 2107.08510}}.

\bibitem{Polchinski:2011im}
J.~Polchinski and J.~Sully, ``{Wilson Loop Renormalization Group Flows},'' {\em
  JHEP} {\bf 10} (2011) 059, \href{http://xxx.lanl.gov/abs/1104.5077}{{\tt
  1104.5077}}.

\bibitem{Beccaria:2017rbe}
M.~Beccaria, S.~Giombi, and A.~Tseytlin, ``{Non-supersymmetric Wilson loop in $
  \mathcal{N} $ = 4 SYM and defect 1d CFT},'' {\em JHEP} {\bf 03} (2018) 131,
  \href{http://xxx.lanl.gov/abs/1712.06874}{{\tt 1712.06874}}.

\bibitem{Beccaria:2019dws}
M.~Beccaria, S.~Giombi, and A.~A. Tseytlin, ``{Correlators on
  non-supersymmetric Wilson line in $ \mathcal{N}=4 $ SYM and
  AdS$_{2}$/CFT$_{1}$},'' {\em JHEP} {\bf 05} (2019) 122,
  \href{http://xxx.lanl.gov/abs/1903.04365}{{\tt 1903.04365}}.

\bibitem{Correa:2019rdk}
D.~H. Correa, V.~I. Giraldo-Rivera, and G.~A. Silva, ``{Supersymmetric mixed
  boundary conditions in AdS$_{2}$ and DCFT$_{1}$ marginal deformations},''
  {\em JHEP} {\bf 03} (2020) 010,
  \href{http://xxx.lanl.gov/abs/1910.04225}{{\tt 1910.04225}}.

\bibitem{Cuomo:2021rkm}
G.~Cuomo, Z.~Komargodski, and A.~Raviv-Moshe, ``{Renormalization Group Flows on
  Line Defects},'' \href{http://xxx.lanl.gov/abs/2108.01117}{{\tt 2108.01117}}.

\bibitem{Beccaria:2021rmj}
M.~Beccaria, S.~Giombi, and A.~A. Tseytlin, ``{Higher order RG flow on the
  Wilson line in $\mathcal{N}=4$ SYM},''
  \href{http://xxx.lanl.gov/abs/2110.04212}{{\tt 2110.04212}}.

\bibitem{Gromov:2012eu}
N.~Gromov and A.~Sever, ``{Analytic Solution of Bremsstrahlung TBA},'' {\em
  JHEP} {\bf 11} (2012) 075, \href{http://xxx.lanl.gov/abs/1207.5489}{{\tt
  1207.5489}}.

\bibitem{giombi2021}
S.~Giombi, S.~Komatsu, and B.~Offertaler, ``{Large Charges on the Wilson loop
  in $\mathcal{N} =4$ SYM II}.'' In preparation.

\bibitem{Miwa2006HolographyOW}
A.~Miwa and T.~Yoneya, ``Holography of Wilson-loop expectation values with
  local operator insertions,'' {\em Journal of High Energy Physics} {\bf 2006}
  (2006) 060--060.

\bibitem{drukker1999wilson}
N.~Drukker, D.~J. Gross, and H.~Ooguri, ``{Wilson loops and minimal
  surfaces},'' {\em Phys. Rev.} {\bf D60} (1999) 125006,
  \href{http://xxx.lanl.gov/abs/hep-th/9904191}{{\tt hep-th/9904191}}.

\bibitem{yang2021d}
P.~Yang, Y.~Jiang, S.~Komatsu, and J.-B. Wu, ``D-branes and Orbit Average,''
  {\em arXiv preprint arXiv:2103.16580} (2021).

\bibitem{Bajnok:2014sza}
Z.~Bajnok, R.~A. Janik, and A.~Wereszczy\'nski, ``{HHL correlators, orbit
  averaging and form factors},'' {\em JHEP} {\bf 09} (2014) 050,
  \href{http://xxx.lanl.gov/abs/1404.4556}{{\tt 1404.4556}}.

\bibitem{Grassi:2019txd}
A.~Grassi, Z.~Komargodski, and L.~Tizzano, ``{Extremal correlators and random
  matrix theory},'' {\em JHEP} {\bf 04} (2021) 214,
  \href{http://xxx.lanl.gov/abs/1908.10306}{{\tt 1908.10306}}.

\bibitem{Beccaria:2020azj}
M.~Beccaria, F.~Galvagno, and A.~Hasan, ``{$\mathcal N=2$ conformal gauge
  theories at large R-charge: the $SU(N)$ case},'' {\em JHEP} {\bf 03} (2020)
  160, \href{http://xxx.lanl.gov/abs/2001.06645}{{\tt 2001.06645}}.

\bibitem{Gromov:2008ec}
N.~Gromov, S.~Schafer-Nameki, and P.~Vieira, ``{Efficient precision
  quantization in AdS/CFT},'' {\em JHEP} {\bf 12} (2008) 013,
  \href{http://xxx.lanl.gov/abs/0807.4752}{{\tt 0807.4752}}.

\bibitem{Vicedo:2008jy}
B.~Vicedo, ``{Semiclassical Quantisation of Finite-Gap Strings},'' {\em JHEP}
  {\bf 06} (2008) 086, \href{http://xxx.lanl.gov/abs/0803.1605}{{\tt
  0803.1605}}.

\bibitem{Zarembo:2002ph}
K.~Zarembo, ``{Open string fluctuations in AdS(5) x S**5 and operators with
  large R charge},'' {\em Phys. Rev. D} {\bf 66} (2002) 105021,
  \href{http://xxx.lanl.gov/abs/hep-th/0209095}{{\tt hep-th/0209095}}.

\bibitem{giombi2010correlators}
S.~Giombi and V.~Pestun, ``{Correlators of local operators and 1/8 BPS Wilson
  loops on S**2 from 2d YM and matrix models},'' {\em JHEP} {\bf 10} (2010)
  033, \href{http://xxx.lanl.gov/abs/0906.1572}{{\tt 0906.1572}}.

\bibitem{giombi2013correlators}
S.~Giombi and V.~Pestun, ``{Correlators of Wilson Loops and Local Operators
  from Multi-Matrix Models and Strings in AdS},'' {\em JHEP} {\bf 01} (2013)
  101, \href{http://xxx.lanl.gov/abs/1207.7083}{{\tt 1207.7083}}.

\bibitem{drukker2007more}
N.~Drukker, S.~Giombi, R.~Ricci, and D.~Trancanelli, ``{More supersymmetric
  Wilson loops},'' {\em Phys. Rev.} {\bf D76} (2007) 107703,
  \href{http://xxx.lanl.gov/abs/0704.2237}{{\tt 0704.2237}}.

\bibitem{Drukker:2007yx}
N.~Drukker, S.~Giombi, R.~Ricci, and D.~Trancanelli, ``{Wilson loops: From
  four-dimensional SYM to two-dimensional YM},'' {\em Phys. Rev. D} {\bf 77}
  (2008) 047901, \href{http://xxx.lanl.gov/abs/0707.2699}{{\tt 0707.2699}}.

\bibitem{Drukker:2007qr}
N.~Drukker, S.~Giombi, R.~Ricci, and D.~Trancanelli, ``{Supersymmetric Wilson
  loops on S**3},'' {\em JHEP} {\bf 05} (2008) 017,
  \href{http://xxx.lanl.gov/abs/0711.3226}{{\tt 0711.3226}}.

\bibitem{Hellerman:2017sur}
S.~Hellerman and S.~Maeda, ``{On the Large $R$-charge Expansion in ${\mathcal
  N} = 2$ Superconformal Field Theories},'' {\em JHEP} {\bf 12} (2017) 135,
  \href{http://xxx.lanl.gov/abs/1710.07336}{{\tt 1710.07336}}.

\bibitem{erickson2000wilson}
J.~K. Erickson, G.~W. Semenoff, and K.~Zarembo, ``{Wilson loops in N=4
  supersymmetric Yang-Mills theory},'' {\em Nucl. Phys.} {\bf B582} (2000)
  155--175, \href{http://xxx.lanl.gov/abs/hep-th/0003055}{{\tt
  hep-th/0003055}}.

\bibitem{drukker42exact}
N.~Drukker and D.~J. Gross, ``{An Exact prediction of N=4 SUSYM theory for
  string theory},'' {\em J. Math. Phys.} {\bf 42} (2001) 2896--2914,
  \href{http://xxx.lanl.gov/abs/hep-th/0010274}{{\tt hep-th/0010274}}.

\bibitem{pestun2007localization}
V.~Pestun, ``{Localization of gauge theory on a four-sphere and supersymmetric
  Wilson loops},'' {\em Commun. Math. Phys.} {\bf 313} (2012) 71--129,
  \href{http://xxx.lanl.gov/abs/0712.2824}{{\tt 0712.2824}}.

\bibitem{Cooke:2017qgm}
M.~Cooke, A.~Dekel, and N.~Drukker, ``{The Wilson loop CFT: Insertion
  dimensions and structure constants from wavy lines},'' {\em J. Phys. A} {\bf
  50} (2017), no.~33 335401, \href{http://xxx.lanl.gov/abs/1703.03812}{{\tt
  1703.03812}}.

\bibitem{Kim:2017sju}
M.~Kim, N.~Kiryu, S.~Komatsu, and T.~Nishimura, ``{Structure Constants of
  Defect Changing Operators on the 1/2 BPS Wilson Loop},'' {\em JHEP} {\bf 12}
  (2017) 055, \href{http://xxx.lanl.gov/abs/1710.07325}{{\tt 1710.07325}}.

\bibitem{Qiao:2017xif}
J.~Qiao and S.~Rychkov, ``{A tauberian theorem for the conformal bootstrap},''
  {\em JHEP} {\bf 12} (2017) 119,
  \href{http://xxx.lanl.gov/abs/1709.00008}{{\tt 1709.00008}}.

\bibitem{Kazakov:2004qf}
V.~A. Kazakov, A.~Marshakov, J.~A. Minahan, and K.~Zarembo,
  ``{Classical/quantum integrability in AdS/CFT},'' {\em JHEP} {\bf 05} (2004)
  024, \href{http://xxx.lanl.gov/abs/hep-th/0402207}{{\tt hep-th/0402207}}.

\bibitem{Dorey:2006zj}
N.~Dorey and B.~Vicedo, ``{On the dynamics of finite-gap solutions in classical
  string theory},'' {\em JHEP} {\bf 07} (2006) 014,
  \href{http://xxx.lanl.gov/abs/hep-th/0601194}{{\tt hep-th/0601194}}.

\bibitem{Eynard:2014zxa}
B.~Eynard, ``{A short overview of the ''Topological recursion''},''
  \href{http://xxx.lanl.gov/abs/1412.3286}{{\tt 1412.3286}}.

\bibitem{Eynard:2007kz}
B.~Eynard and N.~Orantin, ``{Invariants of algebraic curves and topological
  expansion},'' {\em Commun. Num. Theor. Phys.} {\bf 1} (2007) 347--452,
  \href{http://xxx.lanl.gov/abs/math-ph/0702045}{{\tt math-ph/0702045}}.

\bibitem{Sizov:2013joa}
G.~Sizov and S.~Valatka, ``{Algebraic Curve for a Cusped Wilson Line},'' {\em
  JHEP} {\bf 05} (2014) 149, \href{http://xxx.lanl.gov/abs/1306.2527}{{\tt
  1306.2527}}.

\bibitem{Maldacena:1998im}
J.~M. Maldacena, ``{Wilson loops in large N field theories},'' {\em Phys. Rev.
  Lett.} {\bf 80} (1998) 4859--4862,
  \href{http://xxx.lanl.gov/abs/hep-th/9803002}{{\tt hep-th/9803002}}.

\bibitem{Rey:1998ik}
S.-J. Rey and J.-T. Yee, ``{Macroscopic strings as heavy quarks in large N
  gauge theory and anti-de Sitter supergravity},'' {\em Eur. Phys. J. C} {\bf
  22} (2001) 379--394, \href{http://xxx.lanl.gov/abs/hep-th/9803001}{{\tt
  hep-th/9803001}}.

\bibitem{Drukker:2006ga}
N.~Drukker, ``{1/4 BPS circular loops, unstable world-sheet instantons and the
  matrix model},'' {\em JHEP} {\bf 09} (2006) 004,
  \href{http://xxx.lanl.gov/abs/hep-th/0605151}{{\tt hep-th/0605151}}.

\bibitem{Enari:2012pq}
T.~Enari and A.~Miwa, ``{Semi-classical correlator for 1/4 BPS Wilson loop and
  chiral primary operator with large R-charge},'' {\em Phys. Rev. D} {\bf 86}
  (2012) 106004, \href{http://xxx.lanl.gov/abs/1208.0821}{{\tt 1208.0821}}.

\bibitem{Dolan:2011dv}
F.~A. Dolan and H.~Osborn, ``{Conformal Partial Waves: Further Mathematical
  Results},'' \href{http://xxx.lanl.gov/abs/1108.6194}{{\tt 1108.6194}}.

\bibitem{temme2003large}
N.~M. Temme, ``Large parameter cases of the Gauss hypergeometric function,''
  {\em Journal of computational and applied mathematics} {\bf 153} (2003),
  no.~1-2 441--462.

\end{thebibliography}\endgroup


\begingroup\raggedright\endgroup
\end{document}